\documentclass[12pt]{article}

\usepackage{geometry} 	% \usepackage[a4paper]{geometry}
\usepackage{setspace}   	% to use \setstretch{1.5} (e.g. here: 1.5-spacing)
\usepackage{amsmath,amssymb,amsfonts,amsthm}		% math
\usepackage{bm}                                 % bold math

\usepackage{enumitem}
\setlist{nosep} 

\usepackage{graphicx}
\usepackage{subcaption}

\usepackage{fancyhdr}			% headers, footers

\usepackage{xcolor}				% color definitions
\usepackage{soul} 				% to highlight text in colors

\usepackage{hyperref}			% hyperlinks
	\hypersetup{colorlinks=true, citecolor = black, linkcolor = black, urlcolor=black}

\usepackage{tocloft}			% dots for sections in table of contents

\usepackage{natbib}
\usepackage[USenglish]{babel}
\usepackage[style=iso]{datetime2} %set date as yyyy-mm-dd

\usepackage[capitalise]{cleveref}

\usepackage{dutchcal} % supports lowercase caligrahic letters and has nice "l"s

\usepackage[textsize=tiny, disable]{todonotes} % disable in [] for notes to disappear
\usepackage{marginnote}
\newcommand{\mytodo}[2][]{{%		\mytodo{} puts notes on left margin, \todo{} on right
 \let\marginpar\marginnote
 \reversemarginpar
 \todo[#1]{#2}}}

\usepackage{booktabs} % all for tables
\usepackage{multicol}
\usepackage{multirow}

\usepackage{array}
\newcolumntype{N}{c@{}S}
\newcolumntype{H}{>{\setbox0=\hbox\bgroup}c<{\egroup}@{}}
\newcolumntype{L}[1]{>{\raggedright\let\newline\\\arraybackslash\hspace{0pt}}m{#1}}
\newcolumntype{C}[1]{>{\centering\let\newline\\\arraybackslash\hspace{0pt}}m{#1}}
\newcolumntype{R}[1]{>{\raggedleft\let\newline\\\arraybackslash\hspace{0pt}}m{#1}}

\usepackage[round-mode=places, table-number-alignment=center]{siunitx}

\usepackage{longtable}

\usepackage{lscape} % landscape format

\usepackage[raggedright]{titlesec} %to prevent titles in end of page
\makeatletter
\renewcommand*{\@seccntformat}[1]{\csname the#1\endcsname\hspace{20pt}} %space after number in section title
\makeatother

\definecolor{darkpowderblue}{rgb}{0.0, 0.2, 0.6}
\definecolor{carmine}{rgb}{0.59, 0.0, 0.09}
\definecolor{cadmiumgreen}{rgb}{0.0, 0.42, 0.24}
\definecolor{burntorange}{rgb}{0.8, 0.33, 0.0}

\newcommand{\colW}[1]{{\color{white}{#1}}}

\newcommand{\colDB}[1]{{\color{darkpowderblue}{#1}}}

\newtheorem{corollary}{Corollary}

\newtheorem{proposition}{Proposition}
\newtheorem{definition}{Definition}
\newtheorem{lemma}{Lemma}
\newtheorem{algo}{Algorithm}

\newtheorem{runexx}{Running Example}

\makeatletter
\newtheorem*{rep@theorem}{\rep@title}
\newcommand{\newreptheorem}[2]{%
\newenvironment{rep#1}[1]{%
 \def\rep@title{#2 \ref{##1}}%
 \begin{rep@theorem}}%
 {\end{rep@theorem}}}
\makeatother

\newreptheorem{theorem}{Theorem}
\newreptheorem{proposition}{Proposition}
\newreptheorem{example}{Example}

\newcommand{\myproof}[1]{
    \noindent {\bf Proof:} #1 
}

\newcommand{\myproposition}[3]{
    \vspace*{5pt}
    \noindent %\begin{minipage}{1\textwidth}
    \begin{proposition}[#1] \label{#2} \hspace*{0cm}\\[3pt]
        #3
    \end{proposition} %\end{minipage}
}

\newcommand{\myrepproposition}[3]{
    \vspace*{5pt}
    \noindent %\begin{minipage}{1\textwidth}
    \begin{repproposition}{#2}[#1] \hspace*{0cm}\\[3pt]
        #3
    \end{repproposition}%\end{minipage}
}

\newcommand{\myrunningexample}[3]{
    \vspace*{5pt}
    \noindent %\begin{minipage}{1\textwidth}
        \begin{runexx}[#1] \label{#2} \hspace*{1cm}\\
            #3
        \end{runexx}%\end{minipage}
}

\newcommand{\mylemma}[3]{
    \vspace*{5pt}
    \noindent %\begin{minipage}{1\textwidth}
        \begin{lemma}[#1] \label{#2} \hspace*{1cm}\\
        	#3
        \end{lemma} %\end{minipage}
}

\newcommand{\mycorollary}[3]{
    \vspace*{5pt}
    \noindent %\begin{minipage}{1\textwidth}
        \begin{corollary}[#1] \label{#2} \hspace*{1cm}\\
        	#3
        \end{corollary} %\end{minipage}
}

\newcommand{\myremark}[2]{
    \vspace*{5pt}
    \noindent %\begin{minipage}{1\textwidth}
    {\bf Remark (#1):} #2
    \vspace*{5pt}  %\end{minipage}
}

\newcommand{\cbr}[1]{\left\{ {#1} \right\}}
\newcommand{\sbr}[1]{\left[ {#1} \right]}
\newcommand{\br}[1]{\left( {#1} \right)}

\newcommand{\myquote}[1]{``{#1}"}

\newcommand{\mean}{\mathbb{E}}
\newcommand{\var}{\mathbb{V}}

\newcommand{\reals}{\mathbb{R}}

\newcommand{\bmat}[1]{\begin{bmatrix} #1 \end{bmatrix}}

\newcommand{\calF}{\mathcal{F}}

\newcommand{\calM}{\mathcal{M}}

\newcommand{\calE}{\mathcal{E}}

    {\begin{minipage}[t]{1\linewidth}\begin{itemize}\setlength\itemsep{0pt}}
    {\end{itemize}\end{minipage}}

\newcommand*{\plotPathOne}{FiguresCausal}%

\usepackage{multicol}
\usepackage{comment}
\usepackage{longtable}

\begin{document}

%% -------------------------------------------------------------------------- %%
%% Title Page & Table of Contents
%% -------------------------------------------------------------------------- %%

\title{ \fontsize{19}{25}\selectfont \textbf{Origins and Nature of Macroeconomic Instability \\ in Vector Autoregressions}}

\author{%
    \begin{normalsize}
        \begin{minipage}{0.3\textwidth}
        	\centering
        	{Pooyan Amir-Ahmadi\\
            \emph{\small Amazon \\ \colW{.}}
            }
        \end{minipage}
        \begin{minipage}{0.3\textwidth}
        	\centering
        	{Marko Mlikota\\
            \emph{\small Geneva Graduate Institute \\ \colW{.}}
            }
        \end{minipage}
        \begin{minipage}{0.3\textwidth}
        	\centering
        	{Dalibor Stevanovi\'c\footnote{
				\setlength{\baselineskip}{4mm} \fontsize{9}{11}\selectfont Corresponding author:  dstevanovic.econ@gmail.com. Pooyan contributed to this work prior to joining Amazon. For helpful discussions we thank Jean Boivin, Frank Diebold, Simon Freyaldenhoven, Daniel Lewis, Alexey Onatski, Giorgio Primiceri, Frank Schorfheide and participants at the Frank Diebold 65+35 Conference and EC$^2$ Conference 2025.} \\
            \emph{\small Université du Québec \\ à Montréal}
            }
    	\end{minipage}
    \end{normalsize}
}

% Acknowledgements:
% Giorgio Primiceri (Diebold conference)
% Daniel Lewis (Diebold conf.)
% Simon Freyaldenhoven (EC^2 Lugano)
% Jean-Marie Dufour (EC^2 Lugano)

% Conferences:
% - 65+35 Diebold
% - EC^2 Lugano 

\date{\vspace*{10pt} {\normalsize This Version: \today \\[0pt] {\footnotesize [latest version: click \href{https://www.stevanovic.uqam.ca}{\colDB{here}}]}
% PRELIMINARY and INCOMPLETE \\
% Do not circulate without authors' permission
    } \\[10pt] 
} %{\footnotesize \href{https://markomlikota.github.io}{\colDB{[latest version: click here]}}} }
\maketitle

\todo{Dalibor needs to maintain link to arXiv on his website}

%% -------------------------------------------------------------------------- %%
%% Abstract:
%% -------------------------------------------------------------------------- %%

\vspace*{-20pt}

\begin{abstract}

	\noindent %We provide theoretical results on the origins and nature of time-varying parameters (TVPs) in vector-autoregressions (VARs). 
	For a general class of dynamic and stochastic structural models, we show that 
    (i) non-linearity in economic dynamics is a necessary and sufficient condition for time-varying parameters (TVPs) in the reduced-form VARMA process followed by observables, and 
    (ii) all parameters' time-variation is driven by the same, typically few sources of stochasticity: the structural shocks. 
	Our results call into question the common interpretation that TVPs are due to \myquote{structural instabilities}.
	Motivated by our theoretical analysis, we model a set of macroeconomic and financial variables as a TVP-VAR with a factor-structure in TVPs.
	This reveals that most instabilities are driven by a few factors, which comove strongly with measures of macroeconomic uncertainty and the contribution of finance to real economic activity, commonly emphasized as important sources of non-linearities in macroeconomics.
	Furthermore, our model yields improved forecasts relative to the standard TVP-VAR where TVPs evolve as independent random walks.
	\todo{REMEMBER: DISABLE TODONOTES \& DELETE ToC BEFORE SHARING }

\end{abstract}



\thispagestyle{empty}

%% -------------------------------------------------------------------------- %%
%% Table of Contents:
%% -------------------------------------------------------------------------- %%

% \newpage

% \thispagestyle{empty}

% 	\renewcommand{\cftsecleader}{\cftdotfill{\cftdotsep}}

% 	\tableofcontents
% 	\thispagestyle{empty}
% 	\vfill
% 	\pagebreak

%% -------------------------------------------------------------------------- %%
%% Main Part
%% -------------------------------------------------------------------------- %%

\newpage

\clearpage
\setcounter{page}{1}

%% -------------------------------------------------------------------------- %%
%% -------------------------------------------------------------------------- %%

\newpage

\clearpage
\setcounter{page}{1}

%% -------------------------------------------------------------------------- %%
%% -------------------------------------------------------------------------- %%
%% -------------------------------------------------------------------------- %%

\section{Introduction}
\label{sec_intro}

% \item 1st INTRO PARAGRAPH: 
% \begin{itemize}
%     \item VARs are THE tool in empirical macro; in forecasting and causal inference (from v1)
%     \item the mapping between DSGE models and (CP-)VARs is well-established (linearized DSGE leads to VAR (well, VARMA); people test restrictions from DSGE using VAR, develop priors for VAR using DSGE or vice versa, ...; cite JFV's ABCDs paper, An \& Schorf (2007), Del Negor \& Schorf, ... )
% \end{itemize}

Since their introduction by \citet{Sims1980}, vector autoregressions (VARs) have arguably become the most widely used tool for empirical work in macroeconomics, both for forecasting and for causal inference (see \citet{Clark-McCracken2013} and \citet{Ramey2016} for surveys). 
Since the mapping between linearized dynamic stochastic general equilibrium (DSGE) models and standard, constant parameter (CP)-VARs is well established (see e.g. \citet{Sims2001,FernandezVillaverde-RubioRamirez-Sargent-Watson2007}),
these structural and empirical frameworks are often used in conjunction: researchers use VARs to test restrictions implied by DSGEs\todo{cite}, develop priors for DSGE models based on VARs and vice versa \citep{delnegroschorfheide2004ier}, and estimate DSGE parameters using VAR impulse-responses \citep{Christianoetal2005}.

% \item 2nd INTRO PARAGRAPH:
% \begin{itemize}
%     \item * Numerous studies allow for instability of VAR parameters, in various shapes and forms, i.e. using various methodologies (split-sample/break point searches, TVP-VAR, TVP-LP, ML-TVP, regime-switching models, model-averaging and -mixtures, ...)
%     \item * Among them, many associate instability of VAR parameters with instability of economic environment agents operate in.
%     \item For example: Boivin \& Giannoni (2006): split-sample analysis of IRF to MP shocks in VAR and in DSGE model pre- and post-1980
%     \item For example: globalization or European integration used as motivation for allowing for TVPs in VAR (Ciccarelli \& Rebucci (2002), Bianchi \& Civelli (2015),  Gambetti \& Musso (2012))
%     \item For example: good luck vs good policy debate associated with the Great moderation (papers mentioned in Benati \& Surico (2009))
%     \item However, this relation is not obvious:
%     \begin{itemize}
%         \item 1. Benati \& Surico (2009) showed with simulated data that TVP-SVARs are not suitable to detect structural changes. 
%         \item 2. From theory point of view; how to reconcile forward-looking, rational agents with TVPs? 
%     \end{itemize}
%     \item Conclusion: we only have a vague idea what TVPs in VARs represent (how to interpret them) or how to best model them.
% \end{itemize}

Many studies allow for instability of VAR parameters using a wide range of methodologies: break-point or split-sample approaches \citep{andrews1993tests,boivingiannoni2006}, time-varying parameter (TVP) models with or without stochastic volatility \citep{cogley2005,Primiceri2005}, regime-switching models \citep{Hamilton1989,KimNelson1999,sims2006}, machine-learning techniques \citep{GOULETCOULOMBE2025982,Hauzenberger02012025} as well as model averaging strategies \citep{ArtisGalvaoMarcellino2007,koop2012forecasting}. 
Among them, many associate instability of VAR parameters with instability of the economic environment that agents operate in (\myquote{structural instabilities}).
For example, a range of papers use time-varying VARs to explore whether the Great Moderation is due to a change in the conduct of monetary policy \citep{stock2003,Primiceri2005,cogley2005,sims2006,boivingiannoni2006,canovagambettipappa2007} 
or to study the effects of globalization or European integration
\citep{CICCARELLI2006737,GambettiMusso2017,BIANCHI2015406}.

However, the mapping between structural models and time-variation in VARs remains unclear. 
From a practical point of view, \cite{BenatiSurico2009} show that TVP-VARs are not suitable to detect a shift in monetary policy in simulated data.
From a theoretical angle, it is unclear how to reconcile rational, forward-looking agents with drifting VAR representations.\footnote{
    See also the discussion in \citet{FernandezVillaverde-RubioRamirez2007}.
} 
In the end, we are left with an imprecise understanding of the structural origins and nature of TVPs in VARs.

% \item WHAT WE DO : THEORY 1: ORIGINS
% \begin{itemize}
%     \item We provide theoretical results on the origins of TVPs in VARs
%     \item We depart from a general D+S macro model (henceforth: DSGE), represented by system of equations that are either linear or some finite-order polynomial in variables\\
%     + rational agents\\
%     + exogenous process follow a CP-VAR
%     \item In this context, we show that TVPs in VARMA for observables = nonlinearity. This establishes that, under RE + smooth exogenous processes, smoothly changing params in the linear emipirical model correspond to smooth nonlinearities in the true DGP.
%     \item * (BRIEF:) Our result seriously calls into question any association of TVPs with structural changes.
%     \item * (BRIEF:) More specifically, we show that, regardless of the presence or absence of time-variation in DSGE-parameters normally assumed to be fixed and referred to as "structural", we obtain a CP-VARMA under linear model and a TVP-VARMA under non-linear model
% \end{itemize}

We narrow this gap. 
First, we characterize the origins of TVPs in the linear, reduced-form model for observables generated by a general class of dynamic and stochastic (macroeconomic) models. 
Under continuous exogenous variables with stable dynamics, 
we show that non-linearity of the structural equations is both a necessary and sufficient condition for time-variation in the parameters of the VARMA process followed by observables.
For this result, we restrict our attention to smooth non-linearities,\footnote{
    Structural equations are a finite-order polynomial in endogenous and exogenous variables.
}
and we assume 
that agents form rational expectations. 
Under more general exogenous dynamics, TVPs are obtained regardless of the (non-)linearity of structural dynamics.
% and that exogenous variables follow an autoregressive process with CPs.
%
In line with \citet{BenatiSurico2009}, our analysis calls into question the interpretation of TVPs as evidence of evolving policy rules or regime shifts. %; TVPs arise only in non-linear economies, and they occur regardless of which objects are specified as invariant parameters and which are assumed to exhibit time-variation.
Instead, it suggests that the linear relationships among variables in reduced-form models appear time-varying because they are, in fact, not linear.\todo{This can be regarded as model misspecification, no?}

% \item WHAT WE DO : THEORY 2: NATURE
% \begin{itemize}
%     \item Furthermore, our results also characterize the nature of TVPs in VARs
%     \item Specifically,  our analysis shows that TVPs are non-linear functions of the DSGE model variables. Since all these variables vary due to same few shocks, TVPs have RR structure.
%     \item * (BRIEF:) This rationalizes the ex-post observed correlation in TVPs even under the widespread assumption of independent RWs
%     \item * (BRIEF:) Also, it provides a theoretical justification to econometrically motivated RR approaches
% \end{itemize}

Our second theoretical contribution concerns the nature of time-variation of parameters in observables' reduced-form processes.
We show that a non-linear structural model leads to TVPs that are (non-linear) functions of the model's endogenous and exogenous variables.
Since the dynamics of these variables are ultimately driven by a small set of structural shocks, the time-variation of all parameters originates from a few common sources of stochasticity, which suggests reduced-rank variation.
This provides a theoretical explanation for the 
recent success of machine-learning approaches that model TVPs as general functions of observables like \citet{GOULETCOULOMBE2025982},
it rationalizes the strong ex‑post correlations among estimated TVPs documented in empirical work like \citet{cogley2005},
and it justifies econometric approaches that -- motivated by this evidence and by a desired dimensionality-reduction -- impose a reduced‑rank structure or cross-equation restrictions on the evolution of TVPs, like \citet{CanovaCiccarelli2009} and \citet{Grassi-VanDerWel2013}.
% By showing that such structures arise naturally from standard macroeconomic environments with nonlinear dynamics, our analysis bridges the gap between theoretical foundations and practical modelling choices.
\todo{add that params change smoothly; here and in theory (main text)?}
% In other words, smoothly changing params in the linear emipirical model correspond to smooth nonlinearities in the true DGP.

% \item WHAT WE DO : MODEL \& APPLICATION \todo{adding here nonlinearity-informed Factor-TVP-VAR (+ relevant empirical contribution) would be huge additional contribution!}
% \begin{itemize}
%     \item In line with our theory, we build a TVP-VAR model where TVPs follow factor structure
%     \item ... sentence on estimation ...
%     \item We apply it to ... sentence on application ...
%     \item This reveals several interesting results:
%     \begin{enumerate}
%         \item We get interpretable factors that indeed, in macro-finance theory,  (typically) induce nonlinearities
%         \item We get changing IRF to credit conditions
%         \item We get better forecasting performance compared to CP-VAR and TVP-VAR of Primiceri
%     \end{enumerate}
% \end{itemize}

Guided by these theoretical insights, we propose an empirical framework that embeds the reduced‑rank nature of parameter-instability directly into the specification of a TVP‑VAR. In contrast to the standard approach of modelling each time‑varying parameter as an independent random walk \citep{Primiceri2005}, we allow all intercepts, autoregressive coefficients, contemporaneous relationships, and stochastic volatilities to evolve according to a small number of latent factors. This Factor‑TVP‑VAR specification sharply reduces the dimensionality of the state space and aligns with the theoretical result that only a few underlying forces drive all parameter changes. We further develop a grouped‑factor variant that distinguishes between factors driving the propagation mechanism, the contemporaneous covariance structure, and stochastic volatilities, improving interpretability. Applied to U.S. macro‑financial data, this approach yields economically meaningful factors—such as those linked to the role of finance in real activity—uncovers structural changes in transmission mechanisms, and delivers improved forecasting performance relative to both CP- and standard TVP‑VAR benchmarks.

\paragraph*{Related Literature}

    % \item LONG-STANDING PARAMETER INSTABILITY LITERATURE
    % \begin{itemize}
    %     \item characterize literature; unpack first sentence of second paragraph from above (Cogley \& Sargent (2005), Primiceri (2005), ...)
    %     \item contribution: theory + corresponding model + application
    %     \item cite here Granger (2008) (our results apply his insight that any non-linear process can be written as linear process with TVPs) + Aruoba et al. (2017) (they explicitly write non-linear process)
    %     \item \colDB{cite here also Aruoba et al. (2021,2022), where they (we) show how ZLB in DSGE translates into Markov-switching VAR (two sets of parameters, depending on whether rate is above or below zero); in contrast, we are more general (no paticular model) and we consider smooth TVPs}
    % \end{itemize}

A large literature has examined parameter-instability in VARs using a variety of approaches.
The influential contributions of \cite{cogley2005} and \cite{Primiceri2005} establish the now‑standard TVP‑VAR framework with stochastic volatility.
They also show that allowing for time-variation improves empirical fit and that ignoring changes in volatility can produce spurious time-variation in other parameters.
Subsequent refinements include 
improved Bayesian estimation algorithms \citep{delnegroprimiceri2015}, 
extensions to local projections \citep{INOUE2024105726}, 
and the incorporation of high‑dimensional or machine‑learning methods to inform TVPs with richer information sets \citep{GOULETCOULOMBE2025982,Hauzenberger02012025}. 
Parallel strands have pursued specifications with more specific time-variation such as regime‑switching models \citep{Hamilton1989,KimNelson1999}, 
though \citet{BaumeisterPeersman2013} show that TVP-models can capture both gradual and abrupt changes in parameters.\todo{and so TVP-models are dominant?}
Within this broad literature, we contribute a structural foundation for when and why TVPs arise, 
and we develop a theory-consistent modelling strategy that addresses the dimensionality challenges identified in earlier empirical studies.\todo{do we address "interpretability challenges"?}
Our treatment builds on the insight of \citet{Granger2008} that non-linear processes can be written as linear processes with TVPs,
and it contrasts with the approach of \citet{AruobaBocolaSchorfheide2017}, who proceed with explicitly non-linear (quadratic) processes for observables derived under a DSGE model linearized to second order.
Our focus on continuous endogenous variables and smooth non-linearities differs from \citet{Aruoba-CubaBorda-HigaFlores-Schorfheide-Villalvazo2021}, who derive a regime-switching VAR under a DSGE model with an occasionally binding constraint, variants of which are analyzed in \citet{Mavroeidis2021,AruobaMlikotaSchorfheideVillalvazo2022,DuffyMavroeidisWycherley2024,DuffyMavroeidisWycherley2025}.

    % \item ORIGINS $\neq$ STRUCTURAL CHANGES: IMPLICATIONS FOR MACRO INFERENCE
    % \begin{itemize}
    %     \item Recall: papers use TVP-SVAR for inference (debate about great moderation in particular), and Benati \& Surico (2009) show TVP-SVAR not suitable.
    %     \item They simulate data .... of a small New Keynesian economy and time-varying Taylor-rule parameters. ((and they assume abrupt change))           
    %     \item Our theory calls into question link of TVPs with structural changes in a general DSGE model, ((assuming smooth non-linearities and exogenous processes.)) 
    % \end{itemize}
    
Our theoretical results have important implications for the economic interpretation of time-variation in VARs.
In response to a range of studies that used TVP-VARs to analyze whether the Great Moderation is explained by good policy (a change in the conduct of monetary policy) or good luck (a reduced volatility of shocks), \citet{BenatiSurico2009} point to an unclear mapping between a simulated shift in monetary policy and the instability of VAR-parameters.
Specifically, they simulate data from a linearized New Keynesian economy under a baseline specification and under an increased responsiveness of the interest rate to inflation -- with agents unaware of this shift --,
and they show that VARs estimated on the two samples display differing autoregressive and error-covariance-parameters in all equations.
This questions the applicability of TVP-VARs for inference on \myquote{structural instabilities}.
It also renders SVAR-based counterfactuals, which only change the autoregressive parameters in the interest rate-equation, ill-motivated -- a point further developed in \citet{Benati2010}.
Our analysis casts doubt on the link between TVPs in VARs and \myquote{structural instabilities} for a general class of structural models:
under typical dynamics of exogenous processes, we establish a correspondence between TVPs in VARMAs and non-linearities in the underlying structural model.
This relation of TVPs to non-linearities is preserved even under more general exogenous dynamics, including discrete changes in objects typically specified as time-invariant, such as Taylor-rule parameters. 
In line with \cite{Hurwicz1962}, we define structural parameters as objects invariant to shocks, which prompts us to cast \myquote{structural instabilities} as exogenous processes, with respect to which agents form rational expectations.

Our theoretical analysis also concerns the nature of time-variation in VAR parameters, with consequences for the methodology used to model them.
By pointing to non-linearities as sources of time-variation, our theory is in line with the recent success of machine-learning approaches for modelling TVPs, like \citet{GOULETCOULOMBE2025982,Hauzenberger02012025}.
By pointing to reduced-rank variation in TVPs, our theory rationalizes the ex-post correlation found in estimated TVPs that are specified ex-ante as independent processes \citep{cogley2005,stevanovic2016,Renzetti2024}.
Also, it justifies the methodological approaches of \citet{CanovaCiccarelli2009,Grassi-VanDerWel2013,dewind2014reduced,stevanovic2016,CarrieroClarkMarcellino2016,CHAN2020105}, who impose a factor structure or cross‑equation restrictions on TVPs.
In line with \citet{stevanovic2016}, our subsequent empirical analysis points to forecasting gains of this reduced-rank structure relative to the typical specification of TVPs as independent processes. % this constrasts with \citet{Grassi-VanDerWel2013}.
Like \citet{CHAN2020105}, we also get more precisely estimated impulse-responses. % and unlike \citet{CarrieroClarkMarcellino2016}.
\todo{Carriero et al. should be 2012! And we have it double.}

The rest of the paper is structured as follows. 
\cref{sec_theory} presents our theoretical results,
based on which \cref{sec_model} introduces our proposed Factor-TVP-VAR.
Structural and predictive applications are discussed in \cref{sec_app_structural,sec_app_fcst}.
\cref{sec_conclusion} concludes.

\section{Theory: TVPs in VARs}
\label{sec_theory}

In this section, we present our theoretical results. 
Proofs are in \cref{appsec_theory}.
\cref{appsec_illustrativemodels} develops in more detail the example models we use to illustrate our setup and results.

%% -------------------------------------------------------------------------- %%

Consider a vector of macroeconomic observables $y^o_t$.
Motivated by the Wold representation, in empirical work, researchers predominantly stipulate that $y^o_t$ evolves as a Constant Parameter (CP-)VARMA process, as set up in \cref{def_CPVARMA}.
A second popular class of models are VARMA processes with time-varying parameters (TVPs), as specified by \cref{def_TVPVARMA} for a generic type of time-variation.
\citet{Granger2008} provides an econometric justification for this choice, arguing that any non-linear process, i.e. any process outside the class of CP-VARMA models, can be written as a linear process with TVPs, i.e. a TVP-VARMA. 
Researchers, however, commonly allow for TVPs based on an economic rationale that intuitively relates TVPs to \myquote{structural instabilities}.
Once the choice is made in favor of a TVP-VARMA, there is a vast array of possibilities how to set up the time-variation of parameters.\todo{cite again, as in intro? (reasoning for TVPs + types of TVPs)}
A particularly common approach is to specify a regime-switching (RS) variation, leading to a RS-VARMA process, as stated in \cref{def_RSVARMA}.

Our theoretical analysis aims to characterize the mapping between the structural environment that generates $y^o_t$ and the stability of parameters in its linear reduced-form process.
By inverting this mapping, then, 
we can determine which kind of inference on the properties of the structural environment is permissible based on the presence of TVPs in observables' VARMA process.
The analysis also allows us to investigate salient properties of TVPs that can be exploited for selecting among the plethora of methods proposed to model them.

\begin{definition}[CP-VARMA] \label{def_CPVARMA}
    We say $x_t$ follows a CP-VARMA($p,q$) process if
    \begin{align*}
        x_t = \Phi_0 + \sum_{l=1}^{p} \Phi_{l}x_{t-l} + \sum_{k=1}^{q} \Theta_{k} u_{t-k} + u_t \; , \quad u_t \sim WN(0,\Sigma) \; .
    \end{align*}
\end{definition}

\begin{definition}[TVP-VARMA] \label{def_TVPVARMA}
    We say $x_t$ follows a TVP-VARMA($p,q$) process if
    \begin{align*}
        x_t = \Phi_{0,t} + \sum_{l=1}^{p} \Phi_{l,t}x_{t-l} + \sum_{k=1}^{q} \Theta_{k,t} u_{t-k} + u_t \; , \quad u_t \sim WN(0,\Sigma_t) \; ,
    \end{align*}    
    and $\exists$ at least one $(r,s)$ and one element $\calE_t$ of $\{\Phi_{l,t}\}_{l=0}^p$, $\{\Theta_{k,t}\}_{k=1}^q$ or $\Sigma_t$ s.t. $\calE_r \neq \calE_s$.
\end{definition}

\begin{definition}[RS-VARMA] \label{def_RSVARMA}
    We say $x_t$ follows a RS-VARMA($p,q$) process if 
    \begin{align*}
        x_t = \Phi_{0}(s_t) + \sum_{l=1}^{p} \Phi_{l}(s_t)x_{t-l} + \sum_{k=1}^{q} \Theta_{k}(s_t) u_{t-k} + u_t \; , \quad u_t \sim WN(0,\Sigma(s_t)) 
    \end{align*}   
    % it follows a TVP-VARMA($p,q$) with 
    % \begin{align*}
    %     \Phi_{l,t} = \Phi_l(s_t) \; , \; l=0:p \; , \quad 
    %     \Theta_{k,t} = \Theta_k(s_t) \; , \; k=1:q \; , \quad 
    %     \text{and} \quad
    %     \Sigma_t = \Sigma (s_t) 
    % \end{align*} 
    for $s_t \sim$ $n_s$-state Markov chain with transition matrix $T$,
    and $\exists$ at least one pair $(s,s')$ and one element $\calE(s_t)$ of $\Phi_{0}(s_t)$, $\{\Phi_{l}(s_t)\}_{l=1}^p$, $\{\Theta_{k}(s_t)\}_{k=1}^q$ or $\Sigma(s_t)$ s.t. $\calE(s) \neq \calE(s')$.
\end{definition}

%% -------------------------------------------------------------------------- %%

To trace out this mapping, we set up a general structural environment and derive the resulting process followed by observables under different structural properties.
Let $y_t$ be an $n_y \times 1$ vector of endogenous variables whose dynamics are determined by the set of equations
\begin{align}
    \mean_t \sbr{ F(y_t,y_{t+1},y_{t-1},e_t,e_{t+1};\theta) } &= 0 \; , \label{eq_DSGEeqconds}
\end{align}
given the dynamics of an $n_e \times 1$ vector of exogenous processes $e_t$. 
We assume $e_t = (e^{c\prime}_t,e^{d\prime}_t)'$ is composed of an $n^c_e$-dimensional continuous component $e^c_t$ and an $n^d_e$-dimensional discrete-valued component $e^d_t$, which are mutually independent\footnote{
    As can be verified easily, our conclusions are unchanged if $G$ and $\Sigma$ in the equation for $e^c_t$ depend on $e^d_t$.
} and evolve as:
\begin{align}
    e^c_{t+1} &= G_t(\theta)e^c_{t} + \varepsilon_{t+1} \; , \quad \varepsilon_t \sim WN(0,\Sigma_t(\theta)) \; , \quad \Sigma_t(\theta) \; p.d. \; , \label{eq_DSGEexoprocesses_cont} \\
    e^d_{t+1} &\sim \; \text{Markov Chain with $n_s$ states and transition matrix $T_t(\theta)$} \; . \label{eq_DSGEexoprocesses_disc}
\end{align}
$F$ is a vector-valued function with an $n_y$-dimensional domain. 
The function $F$ and matrices $G_t$, $\Sigma_t$ and $T_t$ are indexed by a vector of time-invariant parameters $\theta$.
We define $\varepsilon_t = \Sigma_{t,tr}\epsilon_t$, where $\Sigma_{t,tr}$ is the Cholesky-factor of $\Sigma_t$ and $\epsilon_t$ is a vector of shocks with $\var[\epsilon_t] = I$.
$\mean_t[\cdot] = \mean[\cdot|\calF_t]$ is the expectation taken with the information set $\calF_t = \{y_{t-l},e_{t-l}\}_{l=0}^\infty$ and with full knowledge of the model's structure embodied by \cref{eq_DSGEeqconds,eq_DSGEexoprocesses_cont,eq_DSGEexoprocesses_disc} and the laws of motion of $G_t$, $\Sigma_t$ and $T_t$, which we leave unspecified.\footnote{
    For technical reasons, we assume that $\partial y_{i,t+h}/\partial e^c_{jt} \neq 0$ for at least one $h\geq0$, $i=1:n_y$ and $j=1:n_e$, and -- if $n^d_e > 0$ -- $\mean_t \sbr{ F(\cdot) |e^d_t = s} \neq \mean_t \sbr{ F(\cdot) |e^d_t = s'}$ for at least one pair $(s,s')$. 
}

\myrunningexample{Neoclassical Growth (NCG) Model}{runex_basicsetup}{
    The canonical NCG model can be characterized in terms of two non-linear equations:
    \begin{align*}
        c_t^{-\tau} - \beta \mean_t \sbr{ c_{t+1}^{-\tau} (1-\delta+\alpha e^{z_{t+1}} k_{t+1}^{\alpha-1}) } &= 0 \; ,\\
        c_t + k_{t+1}-(1-\delta)k_t -e^{z_t} k_t^\alpha &= 0 \; .
    \end{align*}
    They describe the dynamics of endogenous variables $y_t=(c_t,k_{t+1})'$ in terms of the exogenous Total Factor Productivity (TFP) process $e_t = z_t$, typically assumed to follow an AR(1): 
    $$ z_t = \rho_z z_{t-1} + \sigma \epsilon_t \; , \quad \epsilon_t \sim N(0,1) \; . $$
    The structural parameters are $\theta=(\alpha,\beta,\tau,\delta,\rho_z,\sigma)'$. 
}

\cref{eq_DSGEeqconds} is general enough to accommodate any dynamic and stochastic macroeconomic structure of interest.
Nevertheless, for now we restrict our attention to linear and smoothly non-linear functions $F$ by requiring $F$ to be a finite-order polynomial, and -- relatedly -- we focus on endogenous variables $y_t$ with continuous support.
\todo{mention Weierstrass or Taylor theorems?}
$F$ can be the set of equilibrium conditions derived from agents' optimization problems under some notion of equilibrium, as in \cref{runex_basicsetup}.
In that case, our definition of $\mean_t$ amounts to assuming that agents form rational expectations (REs).
Our definition of $\theta$ as time-invariant renders those parameters structural in the sense of \citet{Hurwicz1962}; they are invariant to shocks. 
% %
% For simplicity, we refer to the structural model characterized by \cref{eq_DSGEeqconds,eq_DSGEexoprocesses_cont,eq_DSGEexoprocesses_disc} as a Dynamic Stochastic General Equilibrium (DSGE) model.\todo{could also just always say structural model (except in intro)}

The literature typically considers structural models where all exogenous processes in $e_t$ are continuous ($n^d_e = 0$) and evolve based on an autoregressive law of motion with CPs ($G_t = G$ and $\Sigma_t = \Sigma$). 
We refer to this environment as \myquote{typical exogenous dynamics}.

Our setup in \cref{eq_DSGEexoprocesses_cont,eq_DSGEexoprocesses_disc} is considerably more general.
It can accommodate both continuous and discrete exogenous variables, whose dynamics can exhibit arbitrary instabilities over time, i.e. we allow for arbitrary laws of motion of $G_t$, $\Sigma_t$ and $T_t$.
In particular, our setup can accommodate \myquote{structural instabilities}.
For example, suppose that -- following a policy reform -- the depreciation rate $\delta$ in \cref{runex_basicsetup} undergoes a one-time increase from $\delta_L$ to $\delta_H$ at some specific period.
Then it ceases to be structural in the sense of \citet{Hurwicz1962}.
Instead, we include it as $\delta_t$ in $e^d_t$, specify $\{\delta_L,\delta_H\}$ as its support, and stipulate in $T_t$ some probabilities of transitioning between those two values.
Similarly, a smoothly time-varying discount factor $\beta_t$ would be included in $e^c_t$ and $G_t$ and $\Sigma_t$ would be specified accordingly.
Our definition of $\mean_t$ requires agents to form REs with regard to any such instabilities.
However, we allow for arbitrary dynamics of $G_t$, $\Sigma_t$ and $T_t$ and a wide range of low probability-realizations of $e_{t+1}$.
Ultimately, we only require agents to be aware of any structural instabilities.
\todo{would be great if we can be more precise about that; RE or only awareness? is bounded rationality ok?}

We emphasize this aspect of our setup because researchers often disagree on which parameters are structural, and some use TVPs in observables' VARMA process to conduct inference on structural instabilities.
A prominent example are Taylor-rule parameters, which characterize the interest rate response to changes in inflation and output in the New Keynesian (NK) model.
While they are typically specified as constant -- and therefore structural --, a range of studies explores an exogenous increase in the responsiveness of the interest rate to inflation as a possible explanation for the Great Moderation \citep{BenatiSurico2009}.
Analogously, in open economy models, the discount factor or the world-interest rate (as perceived by domestic agents) are often specified as functions of domestic asset holdings in order to induce stationarity of endogenous variables \citep{SchmittGrohe-Uribe2003}.
Therefore, we are particularly interested in the relation between TVPs and discrete structural instabilities (breaks). As discussed above, we include the latter in $e^d_t$ and we assume that agents are aware of their time-variation.\footnote{
    We refrain from referring to them as \emph{time-varying} or \emph{unstable structural parameters} due to our definition of \emph{structural} borrowed from \citet{Hurwicz1962} and \citet{FernandezVillaverde-RubioRamirez2007}.
    Disregarding semantics, a pragmatic view on our setup is the following: $\theta$ contains all time-invariant structural objects, $e_t$ all time-varying objects determined outside the macroeconomic structure in \cref{eq_DSGEeqconds}, and $y_t$ all endogenously time-varying objects.
}

%% -------------------------------------------------------------------------- %%
%% -------------------------------------------------------------------------- %%
%% -------------------------------------------------------------------------- %%

Under the typical exogenous dynamics referenced above, it has been established that a linear(ized) DSGE model leads to a finite-order CP-VARMA process for observables $y^o_t \subseteq y_t$.\footnote{
    See e.g. \citet{FernandezVillaverde-RubioRamirez-Sargent-Watson2007}. 
    In the proof of \cref{prop_linearDSGE}, we combine results from \citet{Sims2001} and \citet{lutkepohl2005} to arrive at this result. 
}
We summarize this result by \cref{prop_linearDSGE},
and we illustrate it with \cref{runex_linearized}.

%Stability of VARMA-Parameters under Linear DSGE Model
\myproposition{}{prop_linearDSGE}{
    Suppose the dynamics of $y_t$ are non-explosive and uniquely generated by \cref{eq_DSGEeqconds,eq_DSGEexoprocesses_cont,eq_DSGEexoprocesses_disc}, and
    \begin{enumerate}
        \item \label{prop_linearDSGE_cond_typicalexo} 
        $n^c_e > 0$ and $n^d_e = 0$ (all $\{e_{jt}\}_{j=1}^{n_e}$ are continuous), 
        and $G_t=G$ and $\Sigma_t = \Sigma$ are constant;

        \item \label{prop_linearDSGE_cond_linear} 
        $F$ is linear in $(y_t',y_{t+1}',y_{t-1}',e^{\prime}_t,e^{\prime}_{t+1})'$.
    \end{enumerate}
    Then, any $y^o_t\subseteq y_t$ follows a CP-VARMA($p',q'$) with $p',q' < \infty$.
}

\myrunningexample{NCG Model, Linear Dynamics}{runex_linearized}{
    Consider the following linear dynamics of $y_t = (c_t,k_{t+1})'$ given $e_t = z_t$:
    \begin{align*}
        \varrho^1_1 c_t + \varrho^1_2 k_{t+1} + \varrho^1_3 k_t + \varrho^1_4 k_t + \varrho^1_5 z_t
        &= 0 \; ,\\
        \varrho^2_1 c_t + \varrho^2_2 \mean_t[c_{t+1}] + \varrho^2_3 \mean_t[ z_{t+1}] + \varrho^2_4 k_{t+1}
        &= 0\; ,
    \end{align*}
    where $\{\varrho^1_i\}_{i=1}^5$ and $\{\varrho^2_i\}_{i=1}^4$ are known functions of $\theta=(\alpha,\beta,\tau,\delta,\rho_z,\sigma_z)'$.
    This is a linear DSGE model; $F$ is linear in $(y_t,y_{t+1},y_{t-1},e_t,e_{t+1})'$.\footnote{
        Such a system is obtained by linearizing the two equations from \cref{runex_basicsetup} around the steady state (see \cref{appsec_illustrativemodels}). The origin of $F$ is, however, irrelevant to our analysis.
    }
    Inserting $\mean_t[z_{t+1}]=\rho_z z_t$ and augmenting the two equations with the law of motion of $z_t$ and with expectational errors $\eta_t = c_t - \mean_{t-1}[c_t]$ yields a linear RE-system with CPs for $x_t = (y_t',z_t,\mean_{t}[c_{t+1}])'$:
    \begin{align*}
        \Gamma_0(\theta) x_t = \gamma(\theta) + \Gamma_1(\theta) x_{t-1} + \Psi(\theta)\epsilon_t + \Pi \eta_t \; .
    \end{align*}
    Its non-explosive solution yields a CP-VAR for $x_t$, which implies a CP-VARMA for any $y^o_t \subseteq y_t \subset x_t$.
    %
    % The same conclusion is reached when the two equations from \cref{runex_setupExoTVtau} are linearized.
    % In that case, $x_t = (y_t',e_t',\mean_{t}[c_{t+1}])'$, with $e_t = (z_t,\hat\tau_t)'$.
    % %
    % If instead we substitute $\tau_t$ for $\tau$ in the equations above -- which in the optimization problems underlying the NCG model amounts to assuming that agents are oblivious to the time-variation in $\tau$ --, then the model ceases to be linear in $(y_t',e_t')'=(c_t, k_{t+1}, z_t,\tau_t)'$ and \cref{prop_linearDSGE} no longer applies.
    % % If instead we substitute $\tau_t$ for $\tau$ in the equations above, then the model ceases to be linear in $(y_t',e_t')'=(c_t, k_{t+1}, z_t,\tau_t)'$ and \cref{prop_linearDSGE} no longer applies.

    % % This is done, e.g., by \citet{BenatiSurico2009} in the specific context of a small NK economy and time-varying Taylor-rule parameters. 
}

%% -------------------------------------------------------------------------- %%

Motivated by \citet{Granger2008} and the prevalent modeling choices in the literature, we restrict the reduced-form processes followed by $y^o_t$ to two classes: CP-VARMA ($\calM_0$) and TVP-VARMA processes ($\calM_1$).
This allows us to regard TVP-VARMA models as the complement of CP-VARMA models. 
Then, the contrapositive form of \cref{prop_linearDSGE} states that -- under typical exogenous dynamics (Condition \ref{prop_linearDSGE_cond_typicalexo} of \cref{prop_linearDSGE}) --  nonlinear structural dynamics are a necessary condition for the parameters in the VARMA process for $y^o_t$ to exhibit time-variation; only if $F$ is nonlinear, does the VARMA process for $y^o_t$ feature TVPs. 
This is summarized by \cref{cor_nonlinearDSGEnecessary}.

%DSGE-Nonlinearity Necessary for TVPs in VARMA
 \mycorollary{}{cor_nonlinearDSGEnecessary}{
    Suppose the dynamics of $y_t$ are non-explosive and uniquely generated by \cref{eq_DSGEeqconds,eq_DSGEexoprocesses_cont,eq_DSGEexoprocesses_disc}, and
    \begin{enumerate}
        \item \label{cor_nonlinearDSGEnecessary_cond_typicalexo} 
        $n^c_e > 0$ and $n^d_e = 0$ (all $\{e_{jt}\}_{j=1}^{n_e}$ are continuous), 
        and $G_t=G$ and $\Sigma_t = \Sigma$ are constant;
        
        \item \label{cor_nonlinearDSGEnecessary_cond_twomodelclasses} 
        the researcher considers the following two classes of models for $y^o_t \subseteq y_t$: 
        $$
            \calM_0 : \; \text{CP-VARMA} \quad \text{and} \quad \calM_1 : \; \text{TVP-VARMA} \; .
        $$
    \end{enumerate}
    Then, if $y^o_t$ follows a TVP-VARMA, $F$ must be nonlinear.
}

% do we have to add a presumption that the researcher has a way to perfectly tell the truth, i.e. whether $y^o_t \sim \calM_0$ or $y^o_t \sim \calM_1$? 
% I don't like the sound of "rejecting". It suggests that this is about finite samples, but it's not!

%% -------------------------------------------------------------------------- %%
%% -------------------------------------------------------------------------- %%

\cref{prop_nonlinearDSGE} establishes that non-linear structural dynamics are also a sufficient condition for observables to follow a TVP-VARMA, at least when attention is restricted to smooth non-linearities and when typical exogenous dynamics are considered.
In its proof, we write \cref{eq_DSGEeqconds} as a function that features time-varying coefficients and is linear in $(y_t, y_{t-1}, e_t)$ and expectations of their first- and higher-order interactions.
\cref{runex_2ndOrderLin} illustrates.
Our approach echoes the treatment of non-linear processes in \citet{Granger2008} and contrasts with that in \citet{AruobaBocolaSchorfheide2017}, who translate a second-order linearized DSGE model into an explicitly non-linear (quadratic) process for observables.
In this sense, \cref{prop_nonlinearDSGE} does not claim that the process followed by observables is uniquely represented as a TVP-VARMA, and \emph{follows} should be read as \emph{can be written as}. 

%Instability of VARMA-Parameters under Nonlinear DSGE Model
\myproposition{}{prop_nonlinearDSGE}{
    Suppose the dynamics of $y_t$ are non-explosive and uniquely generated by \cref{eq_DSGEeqconds,eq_DSGEexoprocesses_cont,eq_DSGEexoprocesses_disc}, and
    \begin{enumerate}
        \item \label{prop_nonlinearDSGE_cond_typicalexo} 
        $n^c_e > 0$ and $n^d_e = 0$ (all $\{e_{jt}\}_{j=1}^{n_e}$ are continuous), 
        and $G_t=G$ and $\Sigma_t = \Sigma$ are constant;

        \item \label{prop_nonlinearDSGE_cond_poly} 
        $F$ is a $p$th-order polynomial in $(y_t',y_{t+1}',y_{t-1}',e^{\prime}_t,e^{\prime}_{t+1})'$ for $p \geq 2$.
    \end{enumerate}
    Then, any $y^o_t \subseteq y_t$ follows a TVP-VARMA($p',q'$) with $p',q' < \infty$.
}

\myrunningexample{NCG Model, Second-Order Linearized Dynamics}{runex_2ndOrderLin}{
    Consider the following non-linear dynamics of $y_t = (c_t,k_{t+1})'$ given $e_t = z_t$:
    \begin{align*}
        \varrho^1_1 c_t + \varrho^1_2 k_{t+1} + \varrho^1_3 k_t 
        + \varrho^1_4 k_t + \varrho^1_5 k_t^2 + \varrho^1_6 z_t + \varrho^1_7 z_t^2 + \varrho^1_8 k_t z_t
        &= 0 \; ,\\
        \varrho^2_1 c_t + \varrho^2_{2} c_t^2 
        +\varrho^2_{3} \mean_t[c_{t+1}] + \varrho^2_{4} \mean_t[c_{t+1}^2] + \varrho^2_{5} \mean_t[z_{t+1}] 
        + \varrho^2_{6} \mean_t[z_{t+1}^2] + \varrho^2_{7} k_{t+1} + \varrho^2_{8}  k_{t+1}^2  & \\
        + \varrho^2_{9} k_{t+1}\mean_t[z_{t+1}] + \varrho^2_{10} \mean_t[c_{t+1} z_{t+1}] + \varrho^2_{11} \mean_t[c_{t+1}] k_{t+1}
        &= 0 \; ,
    \end{align*}
    where $\{\varrho^1_i\}_{i=1}^{8}$ and $\{\varrho^2_i\}_{i=1}^{11}$ are known functions of $\theta$ (unchanged relative to \cref{runex_basicsetup}).
    Then, $F$ is a second-order polynomial in $(y_t',y_{t+1}',y_{t-1}',e_t',e_{t+1}')'$.
    Inserting the known expressions for $\mean_t[z_{t+1}]$ and $\mean_t[z_{t+1}^2]$, 
    both equations above can be written as linear in $x_t = (y_t', z_t,\mean[c_{t+1}], \mean[c_{t+1}^2],\mean[c_{t+1}z_{t+1}])'$.
    Augmenting the system with the law of motion of $z_t$ and with expectational errors $\eta_t = (c_t, c_t^2 , c_tz_t)'
    - \mean_{t-1}\sbr{ (c_t, c_t^2 , c_tz_t)' }$
    yields a linear RE system with TVPs:
    \begin{align*}
        \Gamma_{0t}(\theta) x_t = \gamma(\theta) + \Gamma_{1t}(\theta) x_{t-1} + \Psi(\theta)\epsilon_t + \Pi \eta_t \; ,
    \end{align*}
    whereby elements of $\Gamma_{0t}$ and $\Gamma_{1t}$ are linear functions of $(y_t',y_{t-1}',e_t')'$.
    Its non-explosive solution yields a TVP-VAR for $x_t$, which implies a TVP-VARMA for any $y^o_t \subseteq y_t \subset x_t$.
}

Taken together, \cref{cor_nonlinearDSGEnecessary,prop_nonlinearDSGE} establish nonlinearity of $F$ as both a necessary and a sufficient condition for TVPs in the VARMA process for $y^o_t$ under the typical dynamics of $e_t$.
This is summarized by \cref{cor_nonlinearDSGEnecessaryandsufficient}.

%DSGE-Nonlinearity Necessary and Sufficient for TVPs in VARMA
\mycorollary{}{cor_nonlinearDSGEnecessaryandsufficient}{
    Suppose the dynamics of $y_t$ are  non-explosive and uniquely generated by \cref{eq_DSGEeqconds,eq_DSGEexoprocesses_cont,eq_DSGEexoprocesses_disc}, and
    \begin{enumerate}
        \item \label{cor_nonlinearDSGEnecessaryandsufficient_cond_typicalexo} 
        $n^c_e > 0$ and $n^d_e = 0$ (all $\{e_{jt}\}_{j=1}^{n_e}$ are continuous), 
        and $G_t=G$ and $\Sigma_t = \Sigma$ are constant;

        \item \label{cor_nonlinearDSGEnecessaryandsufficient_cond_poly} 
        $F$ is a $p$th-order polynomial in $(y_t',y_{t+1}',y_{t-1}',e^{\prime}_t,e^{\prime}_{t+1})'$ for $p \geq 1$ (including linearity);

        \item \label{cor_nonlinearDSGEnecessaryandsufficient_cond_twomodelclasses} 
        we consider the two classes of models $\calM_0$ and $\calM_1$ from \cref{cor_nonlinearDSGEnecessary} for $y^o_t \subseteq y_t$.
    \end{enumerate}
    Then, $y^o_t$ follows a TVP-VARMA if and only if $p\neq 1$ (i.e. $F$ is nonlinear).
  }

%% -------------------------------------------------------------------------- %%

Under more general dynamics of exogenous processes $e_t$, observables follow a TVP-VARMA regardless of (non-)linearity of $F$. 
Under time-varying dynamics of $e^c_t$ and still excluding discrete-valued exogenous variables $e^d_t$, this is established by \cref{prop_DSGE_TVPexo}.
Going a step further, \cref{prop_DSGE_discexo} presumes the presence of discrete-valued exogenous processes $e^d_t$.
Under linear structural dynamics and stable exogenous dynamics, the presence of $e^d_t$ with time-invariant transition probabilities $T$ leads to an RS-VARMA process for observables, whereby the regimes are determined by $e^d_t$. 
In more general cases, however, this regime switching-nature is subsumed into the generic time-variation due to non-linear structural dynamics or, possibly, time-varying dynamics for $e^c_t$ or $e^d_t$.

\myproposition{}{prop_DSGE_TVPexo}{
    Suppose the dynamics of $y_t$ are non-explosive and uniquely generated by \cref{eq_DSGEeqconds,eq_DSGEexoprocesses_cont,eq_DSGEexoprocesses_disc}, and
    \begin{enumerate}
        \item \label{prop_DSGE_TVPexo_cond_TVPVARcontexo} 
        $n^c_e > 0$ and $n^d_e = 0$ (all $\{e_{jt}\}_{j=1}^{n_e}$ are continuous), 
        and $G_t$ and $\Sigma_t$ vary over time: $\exists$ at least one $(r,s)$ and one element $\calE_t$ of $G_t$ or $\Sigma_t$ s.t. $\calE_r \neq \calE_s$;

        \item \label{prop_DSGE_TVPexo_cond_poly} 
        $F$ is a $p$th-order polynomial in $(y_t',y_{t+1}',y_{t-1}',e^{\prime}_t,e^{\prime}_{t+1})'$ for $p\geq 1$ (including linearity).
    \end{enumerate}
    Then, any $y^o_t \subseteq y_t$ follows a TVP-VARMA($p',q'$) with $p',q' < \infty$.
}

\myproposition{}{prop_DSGE_discexo}{
    Suppose the dynamics of $y_t$ are non-explosive and uniquely generated by \cref{eq_DSGEeqconds,eq_DSGEexoprocesses_cont,eq_DSGEexoprocesses_disc}, and 
    \begin{enumerate}
        \item \label{prop_DSGE_discexo_cond_nodiscexo} 
        $n^c_e > 0$ and $n^d_e > 0$ (there are both continuous and discrete exogenous processes);

        \item \label{prop_DSGE_discexo_cond_poly} 
        $F$ is a $p$th-order polynomial in $(y_t',y_{t+1}',y_{t-1}',e^{c\prime}_t,e^{c\prime}_{t+1})'$ for $p\geq 1$ (including linearity).
    \end{enumerate}
    Then, if $p=1$ and $G_t = G$, $\Sigma_t = \Sigma$ and $T_t = T$ are constant, any $y^o_t \subseteq y_t$ follows a RS-VARMA($p',q'$) with $p',q' < \infty$ and regimes determined by $e^d_t$.
    Else, $y^o_t \subseteq y_t$ follows a TVP-VARMA($p',q'$) with $p',q' < \infty$.
}

%% -------------------------------------------------------------------------- %%

% To recapitulate, under typical exogenous dynamics, \cref{cor_nonlinearDSGEnecessaryandsufficient} establishes a correspondence between the time-variation of parameters in observables' linear reduced-form process and the non-linearity of the data-generating DSGE model.
% Under more general exogenous dynamics, we obtain TVPs in the VARMA process followed by observables regardless of the (non-)linearity of the underlying DSGE model.
% %
Overall, our analysis shows that the existence of TVPs in the reduced-form process is not related to \myquote{structural breaks}.
As the true structure underlying the dynamics of macroeconomic observables is likely non-linear, their linear reduced-form process features TVPs regardless of the exogenous, discrete shocks $e^d_t$ affecting the system.
Whether such shocks magnify the time-variation in reduced-form parameters or impact their nature in a way that can be exploited for inference is -- necessarily -- an empirical and structure-specific question.
However, even in the specific context of a small NK economy and time-varying Taylor-rule parameters, and with agents oblivious to changes in the monetary policy regime,  \citet{BenatiSurico2009} point to a unintuitive mapping between this structural instability and the TVPs in the reduced-form process estimated for simulated observables.
% \footnote{
%     They depart from a linearized NK economy and then assume that Taylor-rule parameters change following the change of the Fed's chairman.
%     As discussed in \cref{runex_linearized}, this yields a non-linear DSGE model and amounts to agents oblivious in the time-variation of the Taylor-rule parameter.
%     In line with \cref{prop_nonlinearDSGE}, their VAR model for observables simulated under that environment features TVPs.
% } 
\todo{LATER ON, I'D PUT THIS PARAGRAPH AS START OF SECTION 3}

%% -------------------------------------------------------------------------- %%

While our analysis so far concerns the origins of TVPs in VARMA models for macroeconomic observables, the proof of \cref{prop_nonlinearDSGE} and the derivative proofs of \cref{prop_DSGE_TVPexo,prop_DSGE_discexo} point to a salient aspect of their nature. 
Under stable exogenous dynamics, 
if $F$ is a $p$th-order polynomial, 
the TVPs in the linear RE-system are $(p-1)$th-order polynomials of $(y_t',y_{t-1}',e_t')'$.
This is illustrated by \cref{runex_3rdOrderLin} for $p=3$.
Since the stochasticity of these variables is entirely due to the shocks to exogenous processes $e_t$, this implies that the time-variation of all parameters in the matrices of the RE-system is due to structural shocks.
Subsequent transformations that turn these RE-system-parameters into parameters of the VARMA process for $y^o_t$ alter the exact mapping between $e_t$ and TVPs, but do not change this fundamental origin of stochasticity.
Presuming that the dimensionality of VARMA parameters exceeds the dimensionality of structural shocks, our analysis points to reduced-rank variation of VARMA parameters.
This rationalizes the high correlation (ex-post) observed empirically in estimated TVPs even under TVP-VAR specifications that ignore their reduced rank-nature ex-ante (see e.g. \citet{cogley2005,stevanovic2016,Renzetti2024}).

Time-variation in exogenous dynamics are another source of stochasticity for the TVPs in observables' VARMA process. 
Provided that this time-variation is not pervasive, the dynamics of TVPs are nevertheless mainly driven by the time-varying shocks -- rather than time-variation in their probabilistic structure -- and have reduced rank.

\myrunningexample{NCG Model, Third-Order Linearized Dynamics}{runex_3rdOrderLin}{
    Suppose we replace the first equation in \cref{runex_2ndOrderLin} by 
    \begin{align*}
        \varrho^1_1 c_t + \varrho^1_2 k_{t+1} + \varrho^1_3 k_t 
        &= \varrho^1_4 k_t +  \varrho^1_5 k_t^2 + \varrho^1_6 z_t + \varrho^1_7 z_t^2 + \varrho^1_8 k_t z_t \\
        &\colW{=} + \varrho^1_{9} k_t^3 
        + \varrho^1_{10} z_t^3 
        + \varrho^1_{11} k_t z_t^2 
        + \varrho^1_{12} k_t^2 z_t  \; ,
    \end{align*}
    which results in $F$ being a third-order polynomial.
    This equation can be written as 
    \begin{align*}
        \varrho^1_1 c_t + \varrho^1_2 k_{t+1} + \psi^c_{1t} k_t + \psi^c_{2t} z_t
        = 0   \; ,
    \end{align*}
    where $\psi^c_{1t}$ and $\psi^c_{2t}$ are second-order polynomials of $(y_t',y_{t-1}',z_t')'$ and appear in the resulting linear RE system with TVPs. 
}

Note that all preceding statements concern population dynamics. 
They do not make predictions about which class of models is preferred in finite samples. 
This choice depends on the exact specification of the considered CP- and TVP-VARMA models.
Due to the dense parameterization of standard TVP-VARMA approaches, it is likely that a CP-VARMA model is preferred even under non-linear data-generating processes (DGPs).
This holds in particular as the DGP approaches linearity and exogenous dynamics exhibit little time-variation; by \cref{cor_nonlinearDSGEnecessaryandsufficient} and continuity, one can reason that close-to-linear DSGE models lead to negligible time-variation in VARMA parameters even in population.\todo{ADD SOME OUTLOOK ON SECTION 3?}

Also, note that we trace out the mapping between TVPs and the underlying economic structure under the assumption that structural equations are smooth and -- relatedly -- endogenous variables $y_t$ have continuous support.
Specifically, due to our requirement that $F$ is a finite-order polynomial, 
our analysis accounts for smoothly non-linear dynamics as captured by continuously differentiable structural equations, 
but we can only imperfectly account for kinked and censored dynamics, as discussed e.g. in \citet{Aruoba-CubaBorda-HigaFlores-Schorfheide-Villalvazo2021} under an effective lower-bound (ELB) on interest rates.
The extent of this shortcoming is determined by the extent to which dynamics evolve around such censoring-points. To illustrate, consider a toy-model defined by 
\begin{align*}
    y_t - \max\{0, a + b e_t \} = 0 \; , \quad e_{t+1} = \rho e_t + \sigma \epsilon_t \; , \; \; \epsilon_t \sim N(0,1) \; ,
\end{align*}
with $\theta = (a,b,\rho,\sigma)'$.
The larger $a$ and the smaller $(b,\sigma)$, the better can our analysis approximate the dynamics of $y_t$, since the censoring-point becomes less relevant.
This is the case for the typical evolutions of aggregate consumption, working hours and capital stock, all of which are -- at least in theory -- censored at zero.
On the other hand, the censored nature of dynamics is important in periods where the interest rate is at (or close to) the ELB or around sudden stops.

%% -------------------------------------------------------------------------- %%
%% -------------------------------------------------------------------------- %%

%% -------------------------------------------------------------------------- %%
%% -------------------------------------------------------------------------- %%
%% -------------------------------------------------------------------------- %%

\section{Factor-TVP-VAR}
\label{sec_model}

In line with the theoretical results from \cref{sec_theory}, we propose a TVP-VAR that acknowledges the commonalities in parameters' time-variation. \cref{subsec_model_model} presents the model, \cref{subsec_model_est} discusses its estimation.

\subsection{Model}
\label{subsec_model_model}

We consider a TVP-VAR model inspired by \citet{cogley2005} and \citet{Primiceri2005}:
\begin{align}
y_{t} &= c_t + \sum^{p}_{l=1}  B_{l,t} y_{t-l} + A^{-1}_{t}\Sigma_{t}\epsilon_{t} \quad t=1:T \; ,	 \label{eq_TVPVAR}\\
\nonumber
\epsilon_{t} &\sim  N(0,I).
\end{align}
where $y_t$ is an $n \times 1$ vector of observed endogenous variables, $c_t$ is an $n\times 1$ vector of time varying constants, $\{B_{l,t}\}_{l=1:p}$ are $n\times n$ matrices of time varying coefficients, $A_t$ is an $n\times n$ lower triangular matrix capturing contemporaneous responses of the endogenous variables to shocks $\epsilon_{t}$, and $\Sigma_t$ is a diagonal $n\times n$ matrix containing stochastic volatilities. Stacking in a $n(1+np) \times 1$ vector $b_t \equiv vec([c_t \ B_{1,t} \  \ldots \ B_{p,t}])$ all the regression coefficients and defining 
$X_{t}' \equiv I \otimes  [\mathbf{1} \ y_{t-1}',\ldots,y_{t-p}']$, \cref{eq_TVPVAR} can be rewritten as
\begin{align}
Y_{t} = X_{t}' b_{t} + A^{-1}_{t}\Sigma_{t}\epsilon_{t} \; . \label{eq_TVPVAR_LRM}
\end{align}
Let the $m\times 1$ vector $\theta_t = (b_t',a_t',h_t')'$ contain all TVPs in this model, whereby 
$a_t$ is an $(n(n-1)/2) \times 1$ vector containing the non-zero and non-one elements of the matrix $A_t$ (stacked by rows), 
and $h_t$ is the $n\times 1$ vector with the natural logarithm of the diagonal elements of the matrix $\Sigma_t$. 
As a result, $m = n + n^2 p + n(n+1)/2$. 

The traditional approach -- as found, for example, in \citet{Primiceri2005} -- posits that each TVP in $\theta_t$ follows an independent random walk:
\begin{align}
    \theta_t = \theta_{t-1} + \omega_t \; , \quad \omega_t \sim N(0,\Omega) \; , \label{eq_RWTVP}
\end{align}
where $\Omega$ is a diagonal matrix containing the variances of the innovations to the TVP processes $\omega_t$.
This structure is flexible and permits a straightforward elicitation of the prior for a whole TVP process $\theta_{jt}$ given priors for the initial value $\theta_{j0}$ and innovation variance $\Omega_{jj}$. 
However, it is inefficient, as it supposes a-priori the same number of sources of time variation as the number of VAR parameters. The resulting approach is computationally particularly demanding in higher dimensions, i.e. for higher $n$ and $p$.
Henceforth, we refer to the model consisting of \cref{eq_TVPVAR,eq_TVPVAR_LRM,eq_RWTVP} as RW-TVP-VAR.

Based on our theoretical results as well as empirical evidence on strong a-posteriori correlation of TVPs, we propose a new class of TVP-VAR models where the TVPs $\theta_t$ -- i.e. the intercepts, slope coefficients, covariances and stochastic volatilities -- have a factor representation.
\begin{definition}{Common Factor-TVP-VAR}\label{def1} \newline
Let $\theta_t = [b_t^\prime , \ a_t^\prime, \ h_t^\prime]^\prime$ be a 
column-wise  $m$-dimensional vector of all the TVPs in (\ref{eq_TVPVAR_LRM}), where $m = n + n^2 p + n(n+1)/2$. 
The VAR process (\ref{eq_TVPVAR_LRM}) is said to be in Common Factor-TVP form if $\theta_t$ has a common linear factor representation:
\begin{eqnarray}
\label{eq_FactorTVP_eq1}
	\theta_{t} &=& \theta_{0} +  \Lambda_{\theta} f_{\theta,t} + \omega_{\theta,t}, \quad \omega_{\theta,t} \sim N(0,\Omega_{\theta}),	 \\
\label{eq_FactorTVP_eq2}
	f_{\theta,t} & = & \rho_\theta(L) f_{\theta,t-1} + \eta_{\theta,t}, \quad \eta_{\theta,t} \sim N(0,H_{\theta}),
\end{eqnarray}
where $f_{\theta,t}$ is a vector of $q$  factors, with $q << m$, $\Lambda_{\theta}$ is an $m \times q$ matrix of factor loadings, $\rho_\theta(L)$ is a finite (matrix) polynomial,
$\omega_{\theta,t}$ and $\eta_{\theta,t}$ are uncorrelated at all leads and lags.
\end{definition}

%\begin{align}
%        \theta_t &= \gamma +  \Lambda f_{t} + \zeta_{t} \; , \quad \zeta_t \sim N(0,Z),	\label{eq_FactorTVP_eq1} \\
%        f_t &= \rho(L) f_{t-1} + \eta_{t} \; , \quad \eta_{t} \sim N(0,H) \; . \label{eq_FactorTVP_eq2}
%\end{align}
%Thereby, $f_{t}$ is a $q \times 1$ vector of factors, with $q << m$, $\Lambda$ is an $m \times q$ matrix of factor loadings, $\rho(L)$ is a finite (matrix) polynomial, and the factor-innovations $\eta_{t}$ and idiosyncratic TVP-errors $\zeta_{t}$ are uncorrelated at all leads and lags.
This model acknowledges the commonalities in the dynamics of $\theta_t$ and ameliorates the proliferation of state variables as $n$ and $p$ increase. 
We label the model consisting of \cref{eq_TVPVAR,eq_TVPVAR_LRM,eq_FactorTVP_eq1,eq_FactorTVP_eq2} as Common Factor-TVP-VAR.
It reflects our motivation to reduce the dimensionality of the state space compared to the traditional TVP-VAR setup. For 
instance, the standard TVP-VAR requires filtering $m$ states in (\ref{eq_RWTVP}) while our 
Common Factor-TVP model has only $q$ state variables in (\ref{eq_FactorTVP_eq2}). In addition, commonalities within TVPs 
follow the idea that there are only few sources of structural changes that can affect the stability of parameters in 
empirical macroeconomic models. On one hand, the VAR relationships may vary with changes in monetary policy rule 
(e.g. adoption of inflation targeting),  sectoral changes (e.g. service sector became more important) and regulation 
(e.g. financial deregulation since 80s). On the other hand, the uncertainty (or time-varying predictability) can affect the variance of the forecast 
errors, and hence imply common stochastic volatility, as motivated by \cite{jurado2015}. Despite this list is not 
exhaustive, the number of these changes is still very small compared to the number of parameters in a typical VAR model.

The  above Common Factor-TVP representation suggests that all conditional means of endogenous
variables, the VAR autoregressive coefficients (hence the propagation mechanism) as well as the second moments load 
on the same underlying factors. Despite its very general nature, the potential drawbacks of this representation are 
the lack of interpretation of the latent factors and the existence of block-level factors that can alter the estimation of the common 
component and contaminate the idiosyncratic errors. Hence, grouping  the elements of $\theta_t$ and 
restricting the parameter space can produce interpretable factors and provide a more efficient estimation of 
time instability. %A subset of these restrictions that are likely to occur in typical macroeconomic applications is discussed 
%in following model variations.  

%One can argue that conditional means could present low-frequency variations, the propagation mechanism may vary  with changes in policy rules while the variances of  the shocks hitting economy may load only on few stochastic volatility factors. Several exemples can serve to rationalize this  hypothesis. For the conditional mean, \cite{Blinder-Watson(2016)} highlight the down-slopping trend in the growth rate of  US real GDP. Similar trend is present in inflation rate since Volcker period as well as in the short interest rate.  In the case of the transmission mechanism and orthogonalized shocks, the literature on identifying causes of the Great  Moderation has postulated two hypothesis. The first, called `Good Policy', states that the VAR autoregressive coefficients  vary possibly due to underlying changes in the monetary policy rule. The second, `Good Luck', suggest simply the shocks  hitting economy became less important during post-1985 period. Of course, all three sources of time variation can  occur simultaneously but affect in different way specific components of the TVP-VAR model. 

The following representation, that we call \emph{Grouped-Factor TVP-VAR}, takes into account these considerations 
by grouping the VAR parameters into three categories: VAR regressions coefficients ($b_t$), 
covariance states ($a_t$) and stochastic volatilities ($h_t$).

\begin{definition}{Grouped Factor-TVP-VAR}\label{def2} \newline
The VAR process (\ref{eq_TVPVAR_LRM}) is said to be in 
Grouped Factor-TVP form if $\theta_t$ has the following restricted linear factor representation:
\begin{eqnarray}
\label{gftvp_eq1}
	b_{t} &=& b_0 + \Lambda_{b} f_{b,t} + \omega_{b,t}, \quad \omega_{b,t} \sim N(0,\Omega_{b}),	 \\
\label{gftvp_eq2}
	a_{t} &=& a_0 + \Lambda_{a} f_{a,t} + \omega_{a,t}, \quad \omega_{a,t} \sim N(0,\Omega_{a}),	 \\
\label{gftvp_eq3}
	h_{t} &=& h_0 + \Lambda_{h} f_{h,t} + \omega_{h,t}, \quad \omega_{h,t} \sim N(0,\Omega_{h}),
\end{eqnarray}
\newline
with the following VAR(1) processes for each group of factors:
\begin{eqnarray}
\label{gftvp_eq4}
	f_{b,t} & = & \rho_{b} f_{b,t-1} + \eta_{b,t}, \quad \eta_{b,t} \sim N(0,H_{b}),	 \\
\label{gftvp_eq5}
	f_{a,t} & = & \rho_a f_{a,t-1} + \eta_{a,t}, \quad \eta_{a,t} \sim N(0,H_{a}),	 \\
\label{gftvp_eq6}
	f_{h,t} & = & \rho_h f_{h,t-1} + \eta_{h,t}, \quad \eta_{h,t} \sim N(0,H_{h}),
\end{eqnarray}
where $f_{i,t}$ contains $q_i$ group-specific factors, $\Lambda_i$ are the corresponding factor loading matrices, 
$\omega_{i,t}$ and $\eta_{i,t}$ are uncorrelated at all lead and lags, with $i \in (b, a, h)$. The covariance matrices 
$\Omega_i$ are assumed diagonal.
\end{definition}
This specification is obviously a restricted version of the general model (\ref{eq_FactorTVP_eq1})-(\ref{eq_FactorTVP_eq2}). %with 
%$\Lambda_{\theta} = \mathrm{diag}(\Lambda_i)$, $f_{\theta,t} = [f_{b,t}^\prime, f_{a,t}^\prime, f_{h,t}^\prime ]^\prime$, 
%$\omega_{\theta,t} = [\omega_{b,t}^\prime, \omega_{a,t}^\prime, \omega_{h,t}^\prime]^\prime$, 
%$\eta_{\theta,t} = [\eta_{b,t}^\prime, \eta_{a,t}^\prime, \eta_{h,t}^\prime]^\prime$, $\Omega_{\theta} = \mathrm{diag}(\Omega_i)$
%and $H_{\theta} = \mathrm{diag}(H_i)$ with $i \in (b, a, h)$. 
This structure is appealing since it allows to pin down the  
sources of time variation specific to a group of coefficients. In addition, the group-level factors can be easier to interpret than 
the genuinely common shocks. Moreover, imposing a lot of zeros in factor loadings implies more efficient estimates 
if the restrictions are likely to hold. Hence, this more parsimonious specification may provide a better forecasting 
performance than the Common Factor-TVP form.

A few modeling decisions are required to make the Factor-TVP-VAR model operational. 
First, we must assume the existence of an approximate (exact) factor 
structure in case of principal component (likelihood-based) estimation. 
Second, the number of factors, $q$, has to be specified ax-ante 
or selected using a statistical procedure.
Our estimation approach is discussed in \cref{subsec_model_est}.\footnote{Note that our Factor-TVP-VAR nests the RW-TVP-VAR. The latter is obtained under $q=m$ and restricting $\gamma=0$, $\Lambda=I$, $Z=0$ (which implies degenerate $\zeta_t$), $\rho(L)=I$ as well as parameterizing $H=\Omega$.}

\subsection{Estimation}
\label{subsec_model_est}

For now, we employ a two-step estimation approach. In the first step, we estimate a RW-TVP-VAR. Following \citet{Primiceri2005}, we inform our prior for $\theta_0$ by the preliminary OLS estimation of a constant parameter (CP)-VAR.
For a prior on $\Omega$ -- and therefore the degree of time-variation in $\theta_t$ --, we follow \citet{AmirAhmadi-Matthes-Wang2020} and estimate those hyperparameters jointly with all other parameters in the model.
In a second step, then, we apply estimate a factor structure for the posterior mean of $\theta_t$ using principal components analysis.\footnote{
    Estimation in a single step is work in progress. It is analogous to the estimation of the RW-TVP-VAR in \citet{Primiceri2005}, with one additional step in the Gibbs sampling iterations. More concretely, relative to that paper, the state space models used to sample $B_{1:T}$, $A_{1:T}$ and $\Sigma_{1:T}$ conditional on all other parameters feature a different linear-Gaussian transition equation, reflecting the fact that these TVPs evolve according to a factor structure and not as independent RWs. In the additional step, conditional on $\theta_{1:T}$, the parameters in \cref{eq_FactorTVP_eq1,eq_FactorTVP_eq2} are sampled using standard factor model estimation.
}
%\cref{subsec_app_priors} contains details.

%% -------------------------------------------------------------------------- %%
%% -------------------------------------------------------------------------- %%

%% -------------------------------------------------------------------------- %%
%% -------------------------------------------------------------------------- %%

%% -------------------------------------------------------------------------- %%
%% -------------------------------------------------------------------------- %%
%% -------------------------------------------------------------------------- %%

\section{Application: Analyzing Real-Economic and Financial Interactions with the Factor-TVP-VAR}
\label{sec_app_structural}

\subsection{Data}

We apply our Factor TVP framework proposed in the previous section to estimate a model for the US economy with a role for the financial sector. Specifically we want to model the joint dynamics and its evolution over time for the following vector of observables

\begin{align*}
y_{t}= \left(GDP_{t}, GDPDEF_{t}, FFR_{t}, BUSLOANS_{t}, CS_{t}\right)' %\label{Dataeq}%
\end{align*}

%method to the following 5-variable TVP-VAR(2):
%\begin{equation}\label{VAR}
	%y_t = c_{t} + B_{1t} y_{t-1} + B_{2t} y_{t-2} + A_t ^{-1}\Sigma_t \varepsilon_t
%\end{equation}

where $GDP_{t}$ and  $GDPDEF_t$ are GDP growth rate and GDP deflator inflation respectively. The first two variables constitute the non-policy block in the system. A role for monetary policy is given via the third variable $FFR_t$, namely the Federal Funds Rate. A financial block is included via the last two variables, total business loans growth rate and Credit spread which is the Moody's BAA Corporate Bond Yield minus 10-year Treasury Constant Maturity Rate.\footnote{Similar specifications have been used by \citet{BoivinGiannoniStevanovic2020} to study the dynamic macroeconomic effects of credit shocks.}

%XXX WHY THIS VAR IS INTERESTING? WHY THE FINANCIAL BLOCK IS OF INTEREST? WHAT KIND OF SHOCKS DO WE IDENTIFY? SIMILAR 
%MODELS HAVE BEEN USED BY OTHERS TO ESTIMATE THE EFFECTS OF CREDIT/FINANCIAL SHOCKS ON REAL ECONOMY: we should list 
%them. XXX

We use quarterly data from the FRED database (Federal Reserve Economic Data of the Federal Reserve Bank of St-Louis). The time span is 1954Q1 - 2013Q2. In our benchmark specification we use a lag length of 2. Hence, the vector of VAR time varying coefficients, $b_t = [ c_t^{\prime} \ \mathrm{vec}(B_{t})^{\prime}]^{\prime}$,  contains 55 elements. The vector $a_t$ contains 10 unique contemporaneous states of $A_t$.  Finally, 5 stochastic volatilities from $\Sigma_t$ are stacked in $h_t$. Accordingly, the complete set of time varying parameters is contained in $\theta_t = [b_t^{\prime} \ a_t^{\prime} \ h_t^{\prime}]^{\prime}$. 

We apply the \emph{Grouped-Factor TVP-VAR} model (\ref{gftvp_eq1})-(\ref{gftvp_eq6}). The objective to group the parameters in this way is to potentially 
disentangle the sources of common time variation affecting the transmission mechanism from the ones causing instability in the second moments. On one hand, 
changes in economic policies, financial deregulation or in the behaviour of economic agents are likely to affect mainly the VAR coefficients. On the other 
hand, the (financial) risk and (macroeconomic) uncertainty are potential determinants of the instability in the covariance matrix of the VAR residuals.

%MORE DETAILS ON ESTIMATION HERE DEPENDING ON ESTIMATION SECTION ABOVE...

\subsection{Priors}

We follow the standard approach of \citet{Primiceri2005}  to parameterize the prior distributions of the model based on a training sample. We fit a fixed coefficient VAR to the training sample covering 1954Q4 - 1964Q4. The corresponding mean and the variance of our time varying AR coefficients (coefficients states, covariance states and log volatility states) are chosen to be the OLS point estimates and six times its variance in a time invariant VAR, estimated on the small
initial subsample. Summarizing, the priors take the form:

\begin{eqnarray*}
b_0 &\sim& N\left(\hat{b}_{OLS},4\times V(\hat{b}_{OLS})\right)\\
a_0 &\sim& N\left(\hat{a}_{OLS},4\times V(\hat{a}_{OLS})\right)\\
h_0 &\sim& N\left(h_{OLS},I_M\right)\\
\Omega^b &\sim& IW\left(\kappa_b^2\times 40\times V(\hat{b}_{OLS}),40\right)\\
\Omega^a_{i} &\sim& IW\left(\kappa_a^2\times (i+1)\times V(\hat{a}_{i,OLS}),(i+1)\right)\\
\Omega^h &\sim& IW\left(\kappa_h^2\times 4 ,4\right)
\end{eqnarray*}
\newline
where $\Omega^a_{i}$ denotes the corresponding block of $\Omega^a$, while $\hat{a}_{i,OLS}$ stand for the two correspondent blocks of $\hat{a}_{OLS}$. The specific choices for $\kappa_b$,$\kappa_a$ and $\kappa_h$ are important. To remain comparable with the literature we choose our benchmark specification follow \citet{Primiceri2005}  setting $\kappa_b=0.01$,$\kappa_h=0.01$ and $\kappa_a=0.1$.

\subsection{Exploring Common Sources of Instability in Macroeconomic Dynamics}

We start by studying the factorability of the estimated time varying parameters (scree plots and explanatory power of factors). Then, we present the estimated factors and interpret them in terms of correlations with observable macroeconomic variables.

\begin{figure}[t!]
	% \centering
	\begin{center}
	\includegraphics[width=1.00\textwidth]{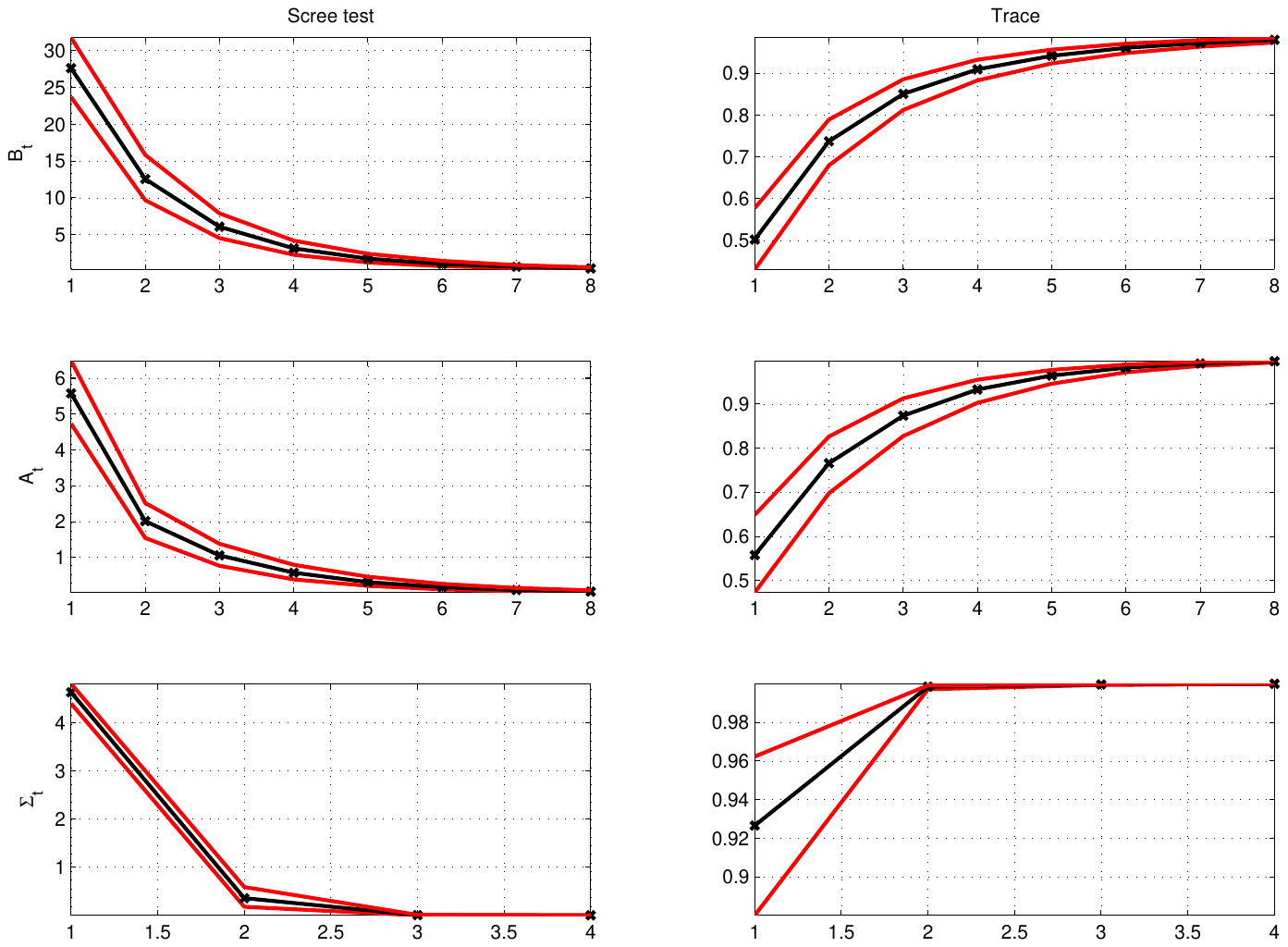}
	\end{center}
	\vspace{-2.0cm}
	\begin{flushleft}
	\caption{Factorability in case of Grouped-Factor-Structure representation.} \label{fig:factorability2}
	This figure shows the posterior median and 68\% bands of eigenvectors, cumulative product of eigenvectors and trace in the case of Grouped-Factor-Structure model. For each group of parameters these quantities are calculated separately.
 \end{flushleft}
\end{figure}

% \paragraph{Factorability}
\subsubsection{Do Macroeconomic Dynamics Change due to Common Sources?}

We start by evaluating the strength of the factor structure among all TVPs  for the benchmark case. Figure~\ref{fig:factorability2} shows the scree plots,  trace and the corresponding $68\%$ posterior bands for each group of coefficients: $\hat b_t$, $\hat a_t$ and $\hat h_t$. These results suggest 2-3 factors for VAR coefficients, 2 for covariances and 2 factors for stochastic volatilities.

\begin{figure}[t!]
	% \centering
	\begin{center}
	\includegraphics[width=1.00\textwidth]{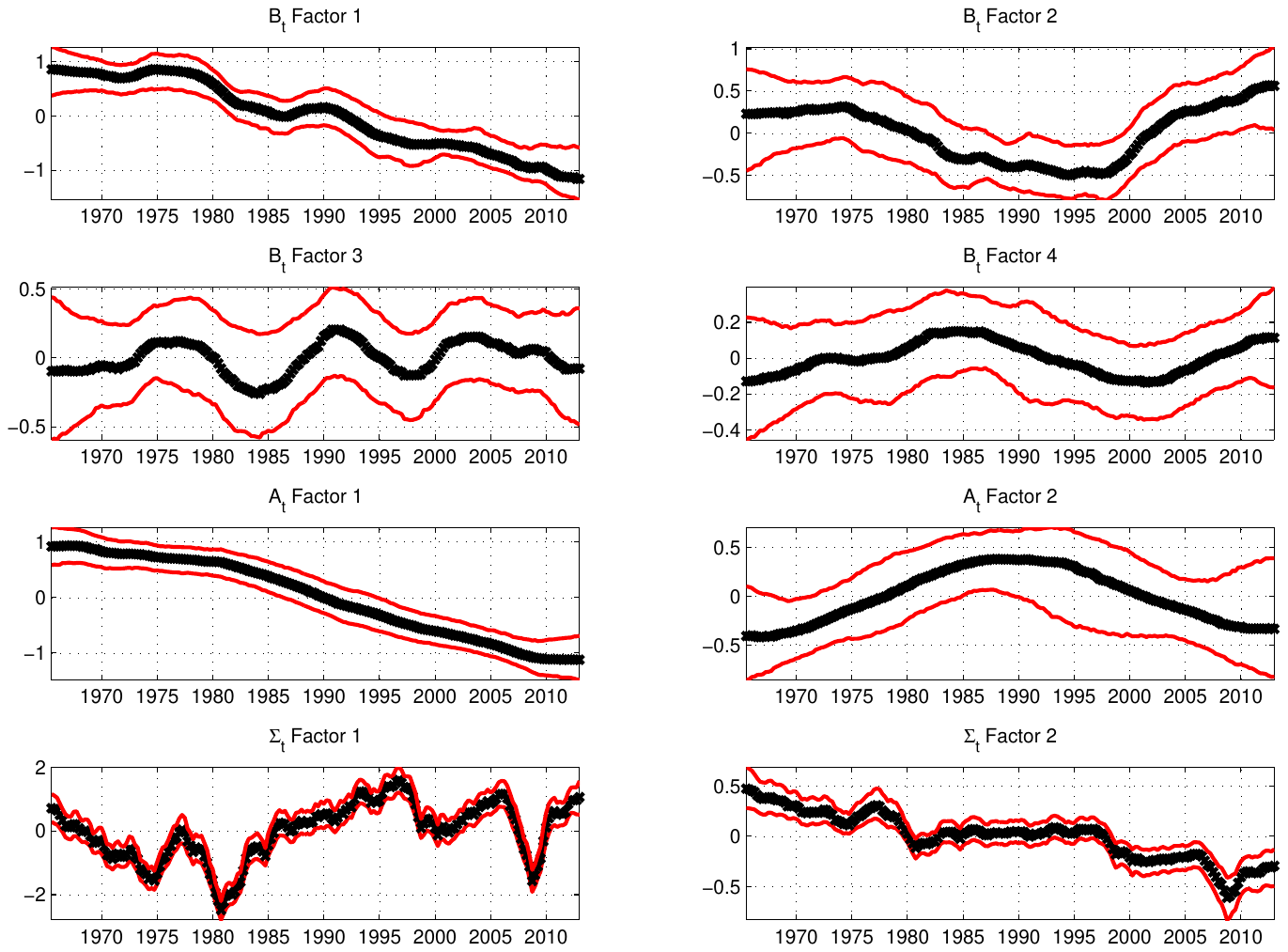}
	\end{center}
	\vspace{-2.0cm}
	\begin{flushleft}
	\caption{Factors from the Grouped-Factor-Structure model.} \label{fig:TVPfactorsCase2}
	{\raggedright This figure shows the posterior median and 68\% bands of the sampled factors from the corresponding drifting parameters and volatilities. The $B_t$ factors are estimates of elements of $f_t^b$ and are obtained as the first 3 principal components of $\hat b_t $. The $A_t$ factors are constructed as the first 2 principal components of $\hat a_t $. Finally, the $\Sigma_t$ factors are the estimates of stochastic volatility factors, $f_t^h$, and are obtained as the first 2 principal components of $\hat h_t$.}
 \end{flushleft}
\end{figure}

Following factorability results, we now estimate the specific common factors underlying the drifting parameters and stochastic volatilities. Figure \ref{fig:TVPfactorsCase2} plot the median of sampled factors and the corresponding posterior $68\%$ error bands. The first $B_t$ factor is precisely estimated while the rest of coefficient factors are less significant and even not very relevant.   In the case of covariance states, the first factor is highly persistent and significant, while the is  imprecise. Finally, both stochastic volatility factors is very significant and precisely estimated. These findings are also supported by explanatory power analysis in terms of $R^2$s.

\begin{figure}[t!]
	% \centering
	\begin{center}
	\includegraphics[width=1.00\textwidth]{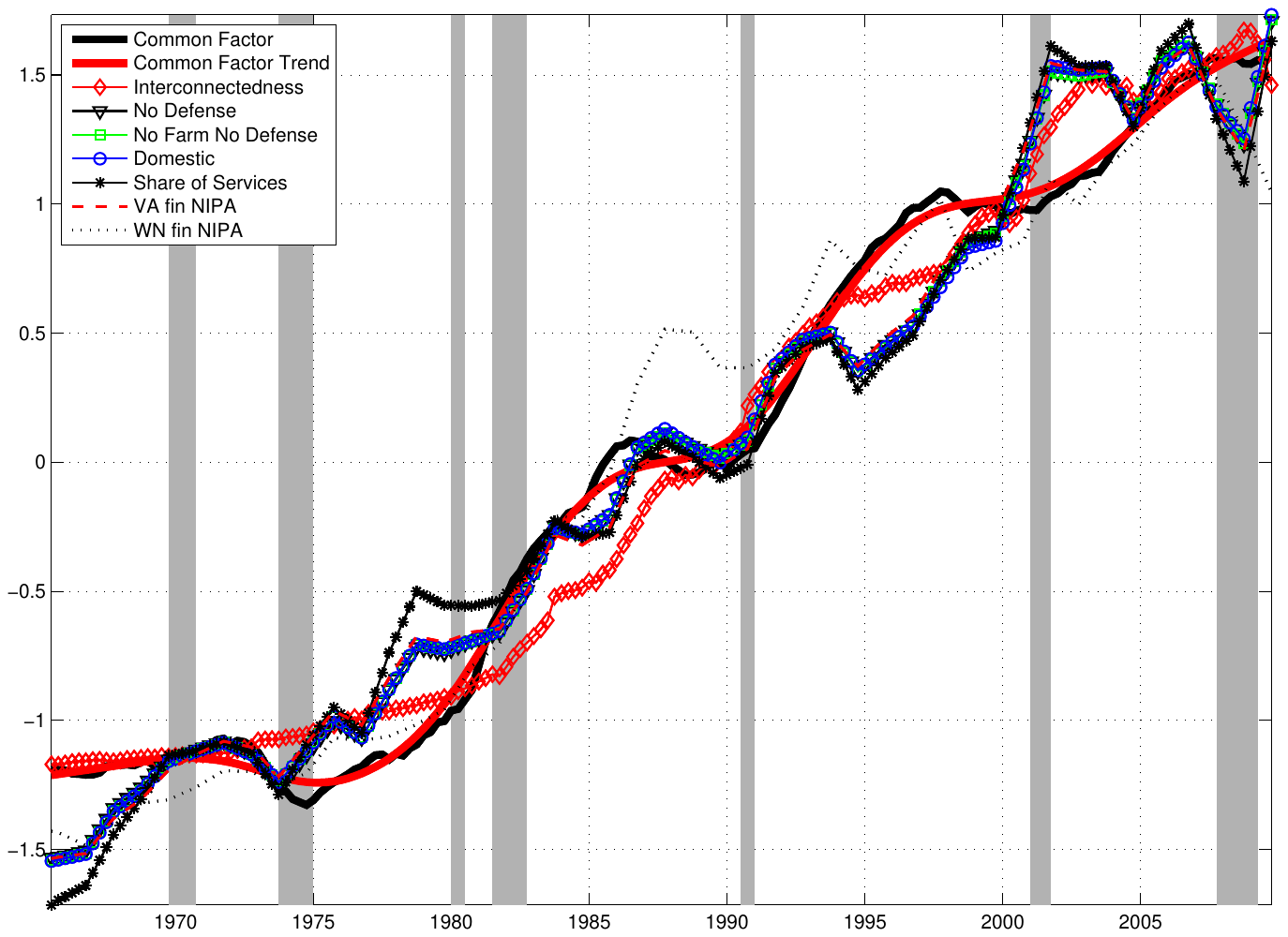}
	\end{center}
	\vspace{-2.0cm}
	\begin{flushleft}
	\caption{Common TVP factor and various measures of the share of finance.} \label{fig:CommonFactorFinance}
	The thick black line, labeled $Common Factor$ corresponds to $\hat f_{1t}^b$, the thick red line, labeled $Common Factor Trend$ is the trend from $\hat f_{1t}^b$ obtained by HP filter. The blue line is the measure of \emph{interconnectedness} between financial and real sectors from \cite{barattieri2019}. The rest are various measures of the share of finance in GDP from \cite{Philippon2012}.
 \end{flushleft}
\end{figure}

\subsubsection{The common factor and the share of finance}

The Figure \ref{fig:CommonFactorFinance} compares the first $B_t$ factor, (labelled \emph{Common Factor}) and its HP-filter trend (labelled \emph{Common Factor Trend}) with several measures of the share of financial sector in economy. The \emph{Interconnectedness} is one minus the measure of \emph{Direct Connectedness} between financial and real sectors from \cite{barattieri2019} that is constructed as the share of the credit to the non-financial sectors over the total credit market instruments. The authors argue that the aggregate measure of connectedness declined by about 27\% in the period 1952-2009, and show that this increase in disconnectedness between the financial sector and the real economy may have dampened the sensitivity of the real economy to monetary shocks. The other series are the measures of the share of finance in U.S. GDP reported by \cite{Philippon2012}. Recall that the common factor captures essentially the co-movements  within the time varying VAR coefficients. Hence, the apparent structural time change in the share of finance is somehow related to the transmission mechanism in our VAR representation of the economy.

\begin{figure}[h!]
	% \centering
	\begin{center}
	\includegraphics[width=1.00\textwidth]{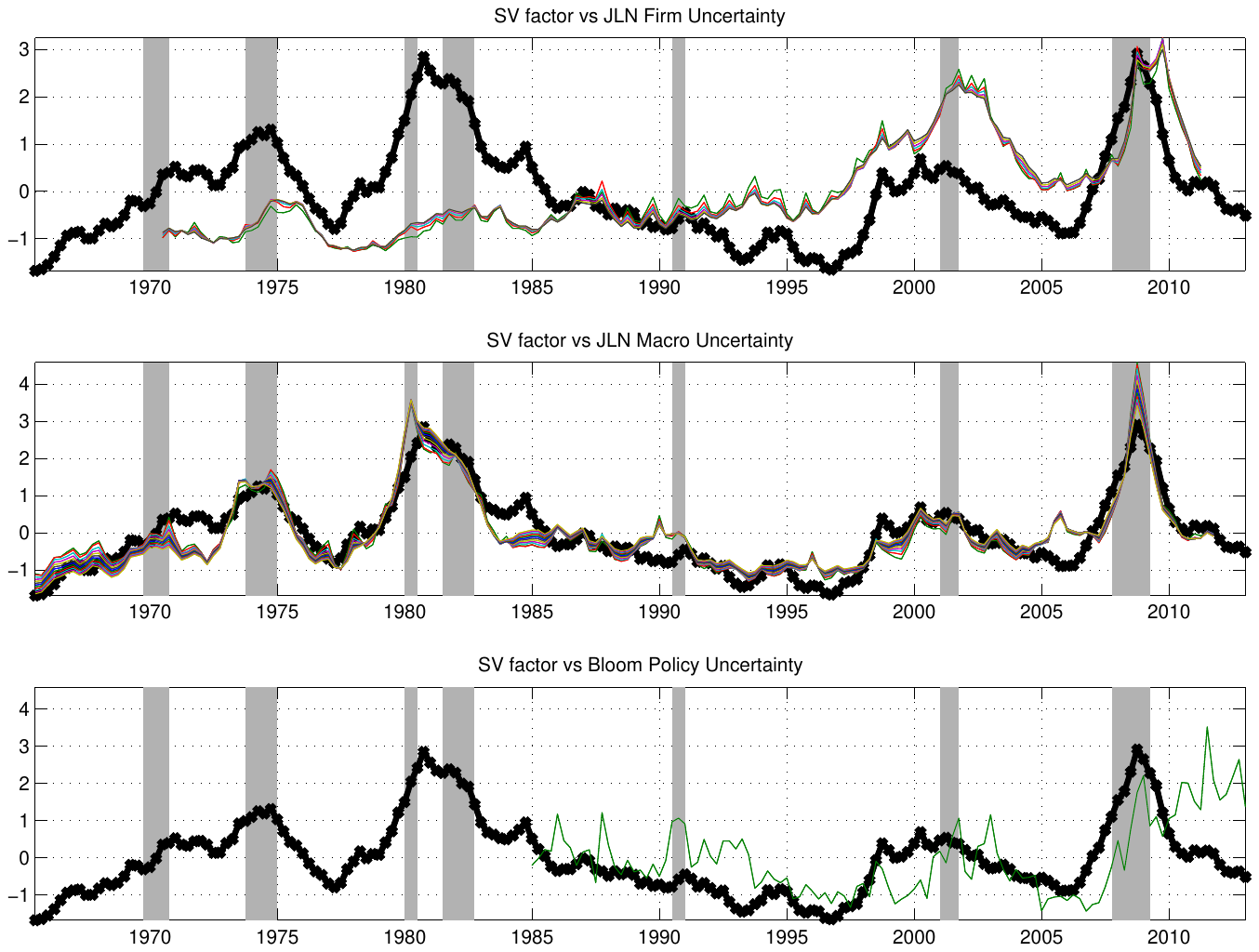}
	\end{center}
	\vspace{-2.0cm}
	\begin{flushleft}
	\caption{Stochastic volatility factor and various uncertainty measures.} \label{fig:SV1}
	The thick black line is the first stochastic volatility factor, $- \hat f_{1t}^h$. The firm uncertainty contains common unforecastable components for horizons of 1 to 6 quarters, and the macro uncertainty is constructed for horizons of 1 to 12 months. Since the macro uncertainties are measured in months in \cite{jurado2015}, we aggregate  them to quarterly frequency. \cite{baker2013} calculate the policy uncertainty in monthly frequency from 1985M01 and we aggregate it to quarters.
 \end{flushleft}
\end{figure}

\subsubsection{Stochastic volatility factor and uncertainty}

The Figure \ref{fig:SV1} compares our first stochastic volatility factor, $\hat f_{1t}^h$, with the firm and the macro uncertainty measures from \cite{jurado2015} and \cite{baker2013} policy uncertainty. The stochastic volatility factor is highly correlated with the common macro uncertainty measure that has been calculated using a factor model applied to hundreds of macroeconomic time series. In particular, \cite{jurado2015} estimate forecasting equations for each series and produce the uncertainty measure as an aggregate stochastic volatility of the forecasting error across all variables. Hence, the uncertainty is defined as the volatility of the unpredictable part of macro series. The advantage of our approach is that we control for time variation in the systematic part of the forecasting model, i.e. VAR coefficients are allowed to vary, so measuring the instability in the unpredictable part is more robust. Compared to \cite{carriero2016}, our stochastic volatility factor is conditional on the first  factor common to all time-varying parameters in the VAR. Hence, it is purged of sources of instability that might  affect the VAR representation of data. The second panel of Figure \ref{fig:SV1} suggests that our stochastic volatility factor is smoother and stays higher, on average, during recessions in 70's and 80's. It is less correlated with the firm-level uncertainty and policy uncertainty measures. The Table \ref{tab:corrSVuncertainty} present the correlations between our first and second stochastic volatility factors and each uncertainty measure from \cite{jurado2015} and \cite{baker2013}. We then investigate the dynamic correlation structure between $\hat f_{1t}^{h}$ and the quarterly aggregate of the uncertainty measure from Jurado, Ludvigson and Ng (2013) calculated from the one quarter ahead forecasting horizon. The results show there is strong evidence on reverse causality\footnote{The VAR and Granger causality test results are presented in the Online Appendix.}. Thus, we are confident that $\hat f_{1t}^h$  can be given the interpretation of an uncertainty measure.%Also, it leads all JLN macro uncertainty measures by 1-2 quarters at the beginning of the last recession.

\begin{figure}[t!]
	% \centering
	\begin{center}
	\includegraphics[width=1.00\textwidth]{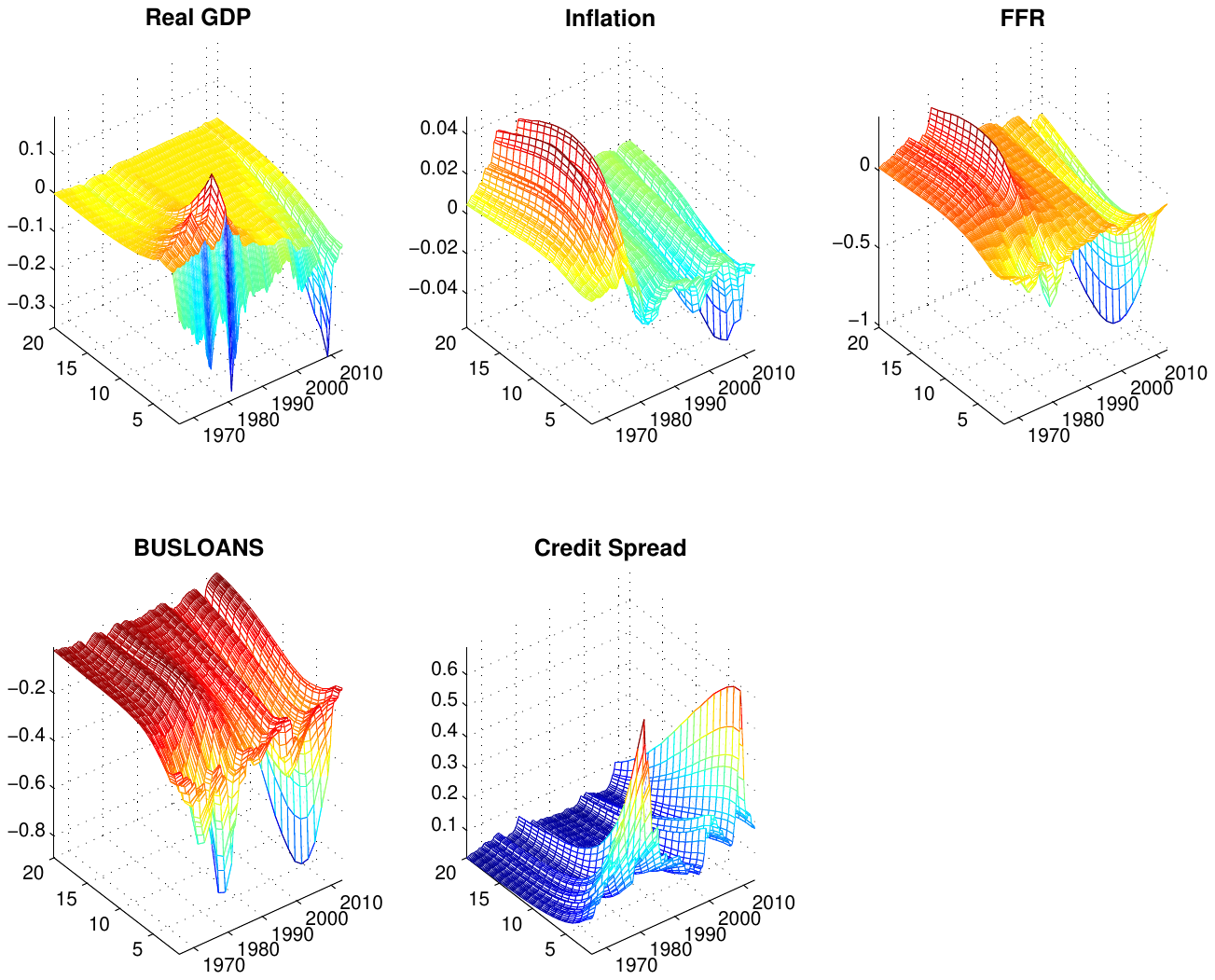}
	\end{center}
	\vspace{-2.0cm}
	\begin{flushleft}
	\caption{Impulse responses to a shock on Credit Spread one standard deviation in size at each point in time across the sample analyzed.} \label{fig:figcs1}
	{ \raggedright This Figure contains impulse responses to a credit spread shock one standard deviation in size identified after recursive Wold causal ordering with credit spread ordered last.}
 \end{flushleft}
\end{figure}

\subsection{Impulse response analysis with common structural changes}

What are the effects of an unanticipated widening in the credit spread on the macroeconomy over time? In order to address this question we employ our benchmark model with group specific common factors in the law of motion of the time varying objects. We report results for two distinct perspectives on the same questions delivering results of distinct interest. First, we ask what are the dynamics effects of credit spread shocks given the historical size of a shock at a given point. Following the vast literature we choose a one standard deviation shock. This approach allows to analyze the importance of the credit spread shocks in historical perspective taking specifically into account the specific size of a shock. We can distinguish between high volatility periods in the credit markets versus low volatility periods. Results are reported in Figure \ref{fig:figcs1}.

Second we analyze evolution and stability of a transmission mechanism from credit market shocks on the macroeconomy. For that we calculate impulse response functions at each point in time normalizing the size of the shock to $1\%$, thus allowing to detect whether the transmission mechanism has truly changed, correcting for the role of stochastic volatilities in changing the size of a shock.

\subsubsection{Credit Spread Shocks historical perspective}

Figure \ref{fig:figcs1} shows that for all variables in the system particularly two periods have been marked with strong dynamic effects following an unanticipated credit spread widening. During the end $70$ies up to the Volcker disinflation period and the Great Recession the size of a typical one-standard deviations shock was much higher than on average. Hence, the real consequences are strongest during those episodes. However, interestingly the magnitude and the sign off the effects during those episodes are very different. The adverse shock during the first episode of large shocks shows a strong reversal in real GDP growth response, and is followed by an increase in the dynamic responses of Inflation and the federal funds rate. The picture is quite different for a typical shock during the Great Recession. Here, the negative response of real GDP remains negative without a reversal, exhibiting a larger persistence. The same hold true for the dynamic responses of the short term interest rate and inflation. This part is very different from typical dynamic responses to credit spread shock during the end $70$ies up to the mid-$80$ies. This results suggest that there is a change in the transmission mechanism of credit market shocks on the macroeconomy. It is also striking, that we get quite smooth and sensical results although the dynamics in the time varying parameters
is based on one common factor only. Also the resulting posterior distribution of the impulse response functions exhibits a lower degree of uncertainty.

\subsubsection{Characterizing the effects of Credit Spread Shocks over time}

\begin{figure}[t!]
	% \centering
	\begin{center}
	\includegraphics[width=1.00\textwidth]{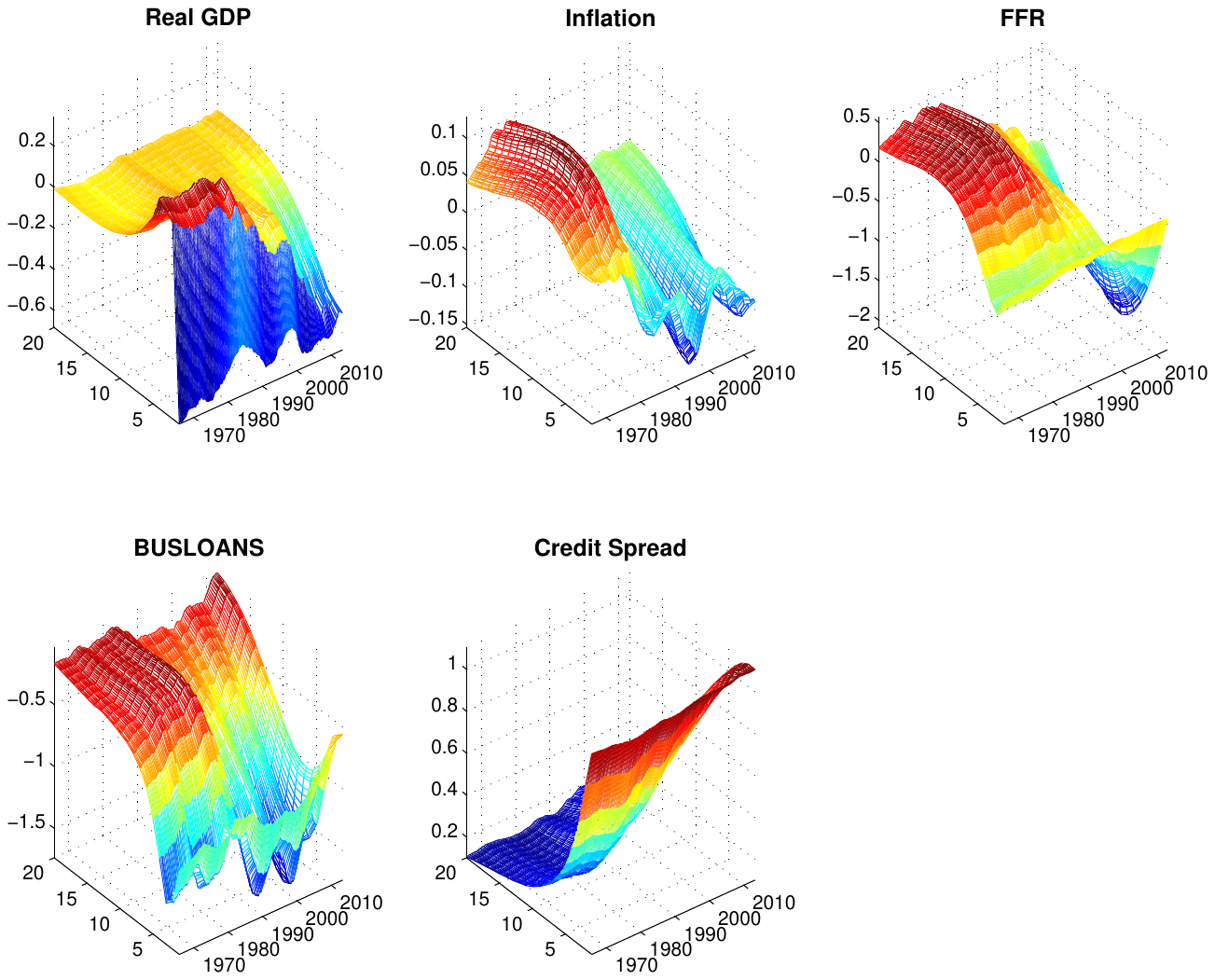}
	\end{center}
	\vspace{-2.0cm}
	\begin{flushleft}
	\caption{Impulse responses to a $1\%$ Credit Spread shock at each point in time across the sample analyzed.} \label{fig:figcs2}
	{ \raggedright This Figure contains impulse responses to $1\%$ Credit Spread shock identified after recursive Wold causal ordering with credit spread ordered last.}
 \end{flushleft}
\end{figure}

The previous section already indicated an important change in the transmission mechanism under study. To further elaborate on that we recalculate at each point in time the impulse responses to a credit spread shock, however normalizing the size of the shock at each point in time to $1\%$. This way we explicitly focus on the changes in the transmission mechanism that are truly due to changes in the underlying dynamics of the system and not due to changes in the size of a shock. Figure \ref{fig:figcs2} further highlights a change in the transmission mechanism. The responses of real GDP growth show a higher persistence during the Great Recession following an adverse credit spread shock. Interestingly, the timing in the change in the trend of the share of finance marks the changing nature of the evolving impulse response functions of real GDP, inflation, the federal funds rate and business loans.

\section{Application: Forecasting with Factor-TVP-VAR}
\label{sec_app_fcst}

\subsection{Data \& Model-Setup}
\label{subsec_app_data}

We apply our Factor-TVP-VAR proposed in \cref{sec_model} to model the following set of macroeconomic and financial variables in the US economy:
\begin{align*}
y_{t}= \br{ GDP_{t}, \ GDPDEF_{t}, \ FFR_{t}, \ CS_{t} }' \; , %\label{eq_data}%
\end{align*}
where $GDP_{t}$ is the GDP growth rate, $GDPDEF_t$ is GDP deflator inflation, $FFR_t$, is the Federal Funds Rate, and $CS_{t}$ is the Moody's BAA Corporate Bond Yield minus 10-year Treasury Constant Maturity Rate (\myquote{credit spread}). We use quarterly data from the Federal Reserve Bank of St-Louis database spanning 1954:Q1 - 2013:Q2. 

\paragraph*{Priors}
We follow the standard approach of \citet{Primiceri2005} and parameterize the prior distribution of $\theta_0$ based on a training sample. Concretely, we fit a fixed coefficient VAR based on data 1954Q4 - 1964Q4. The prior mean and variance of $\theta_0$ are taken to be the OLS point estimates and four times their variance:
$$ \theta_0 \sim N(\hat{\theta}_{OLS}, 4 \var[\hat\theta_{OLS}]) \; .$$
For tuning the degree of time variation we follow \citet{AmirAhmadi-Matthes-Wang2020} and set the prior
$$ \Omega \sim IW(diag(\bar{\kappa}) \nu V diag(\bar{\kappa}),\nu) \; , $$
where $\nu$ are the degrees of freedom and the prior scaling matrix $V$ is initialized using the training sample.
Given posterior estimates from the RW-TVP-VAR, we fit a principal components factor model to the posterior mean of $\theta_{1:T}$.

\subsection{Prediction}
\label{sec:Prediction}

%Here we show how to obtain the sequence of $H$ iterative out-of-sample forecasts of endogenous variables given the information 
%available at the end of sample $I_{t}$. 
We assume  quadratic loss function so the optimal $h$-step-ahead prediction is the conditional  expectation $Y_{T+h \mid T} = {\mathrm{E}}\left[Y_{T+h} \mid Y_{1:T}\right]$. We are interested in constructing a finite sequence of $H$ iterative out-of-sample forecasts
$Y_{T+1:T+H \mid T}$. Following \cite{diebold2017}, draws are generated from the posterior predictive density
\begin{eqnarray}
\label{preddensity}
	p(Y_{T+1:T+H} \mid Y_{1:T}) & = \int p(Y_{T+1:T+H}|\lambda_\theta, \rho_\theta, \theta_0, f_{\theta}^{T}, \theta^T, V)
\end{eqnarray}

The following algorithm describes how the predictive density draws are calculated in the case of general Common Factor-TVP process in 
(\ref{eq_FactorTVP_eq1})-(\ref{eq_FactorTVP_eq2})

\textbf{Algorithm 1: Calculating predictive density draws} \\
For $j=1$ to $n_{sim}$
\begin{enumerate}
	\item Draw $\lambda_{\theta}^{(j)}$, $\theta_0^{(j)}$, $f_{\theta,T}^{(j)}$, $H_{\theta}^{(j)}$, $\Omega_{\theta}^{(j)}$ from the posterior. 
	%as described in Appendix \ref{sec:Estimation and Inference}.
	\item Draw TVPs from $p(\theta_{T+1:T+H} | \lambda_{\theta}^{(j)}, f_{\theta,T}^{(j)}, H_{\theta}^{(j)}, \Omega_{\theta}^{(j)})$ as follows:
	\begin{enumerate}
		\item Draw the sequence of factors' innovations $\eta_{\theta,T+1:T+H}^{(j)} \sim N(0,H_{\theta}^{(j)})$
		\item Starting from $f_{\theta,T}^{(j)}$ use the law of motion (\ref{eq_FactorTVP_eq2}) to obtain the sequence 
		 $f_{\theta,T+1:T+H}^{(j)}$:
		$f_{\theta,t}^{(j)} = \rho_\theta^{(j)} f_{\theta,t-1}^{(j)} + \eta_{\theta,t}$, $t=T+1,\ldots,T+H$.
		\item Draw the sequence of the idiosyncratic innovations $\omega_{\theta,T+1:T+H}^{(j)} \sim 
		N(0,\Omega_{\theta}^{(j)})$ 
		\item Draw the sequence $\theta_{T+1:T+H}^{(j)}$ using the equation (\ref{eq_FactorTVP_eq1}): 
		$\theta_{T+1:T+H}^{(j)} = \theta_0^{(j)} + \lambda_{\theta}^{(j)} f_{\theta,T+1:T+H}^{(j)} + \omega_{\theta,T+1:T+H}^{(j)}$
	\end{enumerate}
	\item Compute the sequence $Y_{T+1:T+H}^{(j)}$ with the VAR equation (\ref{eq_TVPVAR_LRM}) by accordingly 
	splitting $\theta_{T+1:T+H}^{(j)}$ into $c_{T+1:T+H}^{(j)}$, $B_{T+1:T+H}^{(j)}$, $A_{T+1:T+H}^{(j)}$ 
	and $\Sigma_{T+1:T+H}^{(j)}$ and using $Y_{T-p+1:T}$ as starting values.  
	$\epsilon_{T+1:T+H}$ is simulated from $N(0,1)$.
\end{enumerate}
Clearly, the difference with respect to the standard RW-TVP model is in the step 2 where one would use 
the random walk dynamics in (\ref{eq_RWTVP}) to draw the corresponding elements of 
$\theta_{T+1:T+H}^{(j)}$. This would be preceded by drawing from the posterior distributions of $n+n^2 p + n(n+1)/2$ 
states. Instead, we only need to draw $q$ states from the law of motion (\ref{eq_FactorTVP_eq2}). Therefore, if the commonalities 
among TVPs are strong enough, this dimension reduction should result in a more accurate approximation of the true 
data generating process and hence provide more precise predictive ability of the model.

\subsection{Evaluation}

In following, we describe the evaluation metrics for each predictive quantity. These are based on \cite{diebold2017} and \cite{Clark-McCracken-Mertens(2018)}. 

%\begin{comment}
\subsubsection{Point forecast evaluation}

According to the quadratic loss function, the optimal point forecasts are constructed as posterior means. In particular, these are 
computed by Monte Carlo averaging,
\begin{equation}
	\hat Y_{T+h|T} = \frac{1}{n_{sim}} \sum^{n_{sim}}_{j=1} Y_{T+h}^{(j)},
\end{equation}
where $Y_{T+h}^{(j)}$ is the $h^{th}$ element of the sequence $Y_{T+1:T+H}^{(j)}$ generated with Algorithm 1. The models 
are then compared with the standard root mean squared errors (RMSE) computed over the pseudo-out-of-sample 
period. \cite{DieboldMariano1995} tests of equal MSE, against the one-sided alternative that the model with
Factor-TVP structure is more accurate, are performed as well.

\subsubsection{Interval forecast evaluation}

An advantage of having in hand the posterior density (\ref{preddensity}) is that interval forecasts are easy to compute. 
The highest-density $100(1-\alpha)$ percent interval forecast for $Y_{i,T+h}$, $i=1,\ldots n$ 
is computed numerically by looking for the shortest connected interval containing $100(1-\alpha)$ percent of the draws 
$\{ Y_{i,T+h}^j \}_{j=1}^{n_{sim}}$, obtained from Algorithm 1.  

Few evaluation standards are available to compare interval forecasts. Here, we consider relative standards 
(coverage rate and interval length) and the absolute standard (Christoffersen likelihood-ratio tests). The coverage 
rate is the frequency with which the forecasts of each variable in $Y$ fall inside the $100(1-\alpha)$ 
highest posterior density interval. A realized frequency of more (less) than the nominal level means 
that the corresponding predictive density is too wide (narrow). The average prediction interval lengths are 
considered as well. 

The usual absolute evaluation metrics is obtained from \cite{Christoffersen1998}. If interval forecasts are well conditionally 
calibrated, then the hit sequence, $I_t^{(1-\alpha)} = \mathbf{1}_{\{ Y_{i,T+h|T} \ \in \ 
\text{interval} \}}$, should have 
mean $(1-\alpha)$ (coverage) and have at most $h-1$ degree of autocorrelation (independence). We consider individual and 
joint Christoffersen's likelihood-ratio tests for correct coverage and independence.

\subsubsection{Density forecast evaluation}

Density forecast evaluation standard considered is the continuous ranked probability score (CRPS). As discussed in \cite{Clark-McCracken-Mertens(2018)}, the CRPS is preferred to the log score since it rewards better forecasts closer, 
but not equal, to the true outcome, and because it is less sensitive to outliers in the predictive density. It is a generalization of 
the mean absolute error: it does not focus on any specific point of the probability distribution, but considers the distribution of the 
forecasts as a whole. The CRPS is defined as follows
\begin{equation}
	CRPS_T(Y_{i,T+h}) = \int_{-\infty}^{\infty} \left[ F(z) - \mathbf{1}_{\{ Y_{i,T+h} \leq z \}} \right]^2 dz = 
	\mathrm{E}_f \left| Y_{i,T+h|T} - Y_{i,T+h} \right| - 0.5 \mathrm{E}_f \left| Y_{i,T+h|T} - Y_{i,T+h|T}^{\prime} \right|,
\end{equation}
where $F$ denotes the CDF associated with the predictive density $f$, $\mathbf{1}$ takes values 1 if the observation 
$Y_{i,T+h} \leq z$ and zero otherwise, and $Y_{i,T+h|T}$ and $Y_{i,T+h|T}^{\prime}$ are independent random draws from 
the posterior predictive density (\ref{preddensity}). 
The significance of differences in CRPS is computed with $t$-tests for equality of average CRPS.

\subsection{Results}

We conduct a pseudo-out-of-sample forecasting exercise. The validation period covers 1980Q1 - 2013Q2 span. The models 
are estimated recursively. We evaluate the performance of the Factor-TVP relative to the random walk 
TVP with respect to point, interval and density forecasts. %In following, we describe the evaluation metrics for each predictive quantity. These 
\subsubsection{Point forecast accuracy}

We now discuss the forecasting evaluation. We report results for both Common and Grouped Factor-TVP-VAR specifications, with lag lengths of 2 and 5. %The lag length $2$ and $5$ and discuss further specifications and variation in section (XXX). 

In Tables (\ref{fig:RMSFEModel1Prior3Lag2FcstNeval98SamplingMode1DifMode1FacMode1}) and (\ref{fig:RMSFEModel1Prior3Lag5FcstNeval98SamplingMode1DifMode1FacMode1})  we present forecast RMSEs for  a $2$ lag and $5$ lag specification respectively. We show RMSEs for the benchmark standard TVP-VAR-SV model in the first line of each panel of the respective variable, and RMSE ratios in the subsequent lines. Ratios less than one indicate that the forecasts from our corresponding common factor TVP-VAR-SV model variation are more accurate than the benchmark standard TVP-VAR-SV model forecasts. We use the following abbreviations: ‘‘TVP-VAR-SV’’ is the traditional TVP-VAR model (\ref{eq_TVPVAR_LRM}), ‘‘CF-TVP-VAR-SV’’ is the common factor TVP-VAR model (\ref{eq_FactorTVP_eq1}-\ref{eq_FactorTVP_eq2}) and ‘GF-TVP-VAR-SV’’ is the grouped factor TVP-VAR-SV model (\ref{gftvp_eq1}-\ref{gftvp_eq6}). In parentheses we show p-values of \cite{DieboldMariano1995}  tests of equal MSE against the one-sided alternative that the ‘‘TVP-VAR-SV’’ is more accurate. Statistically significance are indicated by ${^{*}}$, ${^{**}}$, and ${^{***}}$ which correspond to $10$, $5$, and $1$ percent significance level, respectively.

Generally, with increasing the VAR lag orders the common factor model specifications tend to produce more accurate forecasts. The highest and statistically significant result emerge for inflation forecasts while federal funds rate  forecast are worse under the common factor specification.  

In particular, Table (\ref{fig:RMSFEModel1Prior3Lag5FcstNeval98SamplingMode1DifMode1FacMode1}) shows that inflation forecasts from the specifications with common and grouped factors are significantly more accurate at all horizons and the accuracy gains increase with the horizon. Forecast accuracy for output and credit spreads are similar, while not statistically significant. In contrast, federal funds rate forecast accuracy is worse across models and horizons, while those losses are not significant. There is no clear ranking across the two common factor time-varying specifications. %, however there is a stronger tendency of the ''CF-TVP-VAR'' model to produce more accurate forecasts than ''GF-TVP-VAR-SV''. 

%The results show that dimension reduction in the space of state variables, by means of factor structure, can improve the point forecast ability especially in the case of parameter proliferation, i.e. when the lag orders are large. This is of particular interest for the TVP-VAR literature that usually consider small systems because of the difficulty to accurately track a large number of TVPs.

% ==========================================================
\subsubsection{Interval forecast accuracy}

In the interval forecast evaluation we report relative standards (i) coverage and length in tables (\ref{fig:InFevalModel1Prior3Lag2FcstNeval98SamplingMode1DifMode1FacMode1})-(\ref{fig:InFevalModel1Prior3Lag5FcstNeval98SamplingMode1DifMode1FacMode1}) and absolute standards (ii) Christoffersen Likelihood-Ratio tests  in tables (\ref{tab:InFevalCTestHorizon1Model1Prior3Lag2FcstNeval98SamplingMode1DifMode1FacMode1})-(\ref{tab:InFevalCTestHorizon8Model1Prior3Lag5FcstNeval98SamplingMode1DifMode1FacMode1}). %We furthermore show the figures on the posterior $68$ percent forecast interval in figures (\ref{fig:2FIHorizon8CRPSModel1Prior3Lag2FcstNeval98SamplingMode1DifMode1FacMode1})-(\ref{fig:2FIHorizon8CRPSModel1Prior3Lag5FcstNeval98SamplingMode1DifMode1FacMode1}). 

\paragraph*{Coverage rates.}
In Tables (\ref{fig:InFevalModel1Prior3Lag2FcstNeval98SamplingMode1DifMode1FacMode1}-\ref{fig:InFevalModel1Prior3Lag5FcstNeval98SamplingMode1DifMode1FacMode1}) (values without square brackets) we report the frequency with which forecast outcomes for output growth, inflation rate, the federal funds rate and credit spreads fall inside %real-time 
68\% intervals. Correct coverage corresponds to frequencies of about 68\%, whereas a frequency of greater than (less than) 68\% means that on average over a given sample, the posterior density is too wide (narrow). The respective ${^{*}}$, ${^{**}}$, and ${^{***}}$ correspond to $10$, $5$, and $1$ percent significance level  p-values of t-statistics of the hypothesis of correct coverage (empirical = nominal coverage of 68\%), calculated using Newey-West standard errors. For all variables and horizons, estimated coverage is quite close to 68\%. At horizons of 1 year  and longer for output and inflation the empirical coverage rates exceed the nominal ones and are too large. In contrast, for the federal funds rate and credit spreads the coverage rates remain closer to $68\%$ across all horizons. Here as well, more complex models tend to benefit more from the common structure modelling of the evolving coefficients. On average the model ''GF-TVP-VAR-SV'' performs best across the models. 

\paragraph*{Interval length.}
Tables (\ref{fig:InFevalModel1Prior3Lag2FcstNeval98SamplingMode1DifMode1FacMode1}-\ref{fig:InFevalModel1Prior3Lag5FcstNeval98SamplingMode1DifMode1FacMode1})  (values in square brackets) also shows average prediction interval lengths which are based on both, the ''CF-TVP-VAR-SV'' and ''GF-TVP-VAR-SV'' models throughout shorter. For output we find significant differences of up to $20\%$ shorter intervals compared to the traditional ''TVP-VAR-SV'' model.  It turns out that the forecasting performance gains of our proposed models, in terms of shorter forecast interval length are increasing with both, the forecast horizon and model complexity (here, higher lag specifications). 

\paragraph*{Conditional calibration.}
The previous subsections evaluated unconditional coverage rates and interval lengths. These measures provide a first indication of forecast accuracy but do not guarantee full calibration. A stricter requirement is conditional calibration, which combines correct coverage with the absence of serial dependence in forecast errors. Christoffersen’s likelihood ratio (LR) tests provide a joint assessment of these conditions. Tables \ref{tab:InFevalCTestHorizon1Model1Prior3Lag2FcstNeval98SamplingMode1DifMode1FacMode1}-\ref{tab:InFevalCTestHorizon8Model1Prior3Lag5FcstNeval98SamplingMode1DifMode1FacMode1} report LR statistics for coverage, independence, and the joint conditional calibration test across horizons and model variants.

At the one-quarter horizon with two lags (Table \ref{tab:InFevalCTestHorizon1Model1Prior3Lag2FcstNeval98SamplingMode1DifMode1FacMode1}), output growth forecasts strongly reject the coverage condition under the benchmark TVP-VAR-SV and GF-TVP-VAR-SV models, while the CF-TVP-VAR-SV partially mitigates these rejections. Inflation forecasts generally attain correct coverage, consistent with the results in Section 5.4.1, while interest rate forecasts often fail the independence test, indicating clustering of forecast misses despite seemingly accurate coverage.

With five lags (Table \ref{tab:InFevalCTestHorizon1Model1Prior3Lag5FcstNeval98SamplingMode1DifMode1FacMode1}), the picture is similar: inflation and credit spreads achieve satisfactory coverage, but output growth remains problematic, and the federal funds rate continues to exhibit dependence in the hit sequence. At the eight-quarter horizon (Tables \ref{tab:InFevalCTestHorizon8Model1Prior3Lag2FcstNeval98SamplingMode1DifMode1FacMode1} and \ref{tab:InFevalCTestHorizon8Model1Prior3Lag5FcstNeval98SamplingMode1DifMode1FacMode1}), deficiencies become more severe. The benchmark TVP-VAR-SV produces particularly poor calibration for output and inflation, with large and highly significant joint LR statistics. The GF and CF extensions sometimes reduce these values, but rejections remain common at conventional significance levels.

Taken together, the conditional calibration results reinforce the earlier findings. While interval coverage for inflation is relatively reliable and forecast bands are reasonably tight, the combination of correct coverage and independence is rarely achieved for output growth and interest rates, especially at longer horizons. The GF- and CF-TVP-VAR-SV models offer improvements in some cases but cannot fully eliminate violations of conditional calibration.

% ==========================================================
\subsubsection{Density Forecast Accuracy}
While calibration evaluates reliability, density forecast accuracy is assessed by the continuous ranked probability score (CRPS), a proper scoring rule that jointly rewards sharpness and calibration of the entire predictive density. Lower CRPS values indicate more accurate density forecasts. Tables \ref{fig:CRPSModel1Prior3Lag2FcstNeval98SamplingMode1DifMode1FacMode1} and \ref{fig:CRPSModel1Prior3Lag5FcstNeval98SamplingMode1DifMode1FacMode1} report CRPS values for horizons of 1, 2, 4, and 8 quarters ahead, based on VAR specifications with two and five lags, respectively.

With two lags (Table \ref{fig:CRPSModel1Prior3Lag2FcstNeval98SamplingMode1DifMode1FacMode1}), the GF- and CF-TVP-VAR-SV models generally outperform the benchmark TVP-VAR-SV, especially for output growth and credit spreads. At short horizons (1–2 quarters), all models deliver comparable CRPS values for inflation, in line with its stronger calibration results. Differences widen at longer horizons (4–8 quarters), where the benchmark tends to generate diffuse densities and higher CRPS, while the common factor specification achieves noticeable gains.

When the lag length is increased to five (Table \ref{fig:CRPSModel1Prior3Lag5FcstNeval98SamplingMode1DifMode1FacMode1}), the same broad patterns emerge, though the relative improvements are somewhat muted. The baseline TVP-VAR-SV again exhibits the weakest performance at medium and long horizons. The CF specification continues to provide the largest accuracy gains for real activity and inflation, while the GF specification yields more mixed results: in some cases improving upon the benchmark, but in others producing CRPS values close to or above those of the baseline.

Taken together, the CRPS results indicate that factor-structured extensions to the TVP-VAR-SV can improve density forecast accuracy, particularly at longer horizons where the benchmark model deteriorates. The benefits are most visible for output growth and credit spreads, while inflation forecasts remain relatively well behaved across models and horizons.

\section{Conclusion}
\label{sec_conclusion}

For a general class of dynamic and stochastic structural models,
we show  
(i) that non-linearity of structural dynamics is necessary and sufficient for the presence of TVPs in the VARMA process followed by observables,
and (ii) that all parameters' time-variation is driven by the same, typically few sources of variation: the shocks in the structural model.
Building on this insight, we develop the Factor-TVP-VAR -- a TVP-VAR with TVPs evolving as a dynamic factor model -- and we apply it to study a set of macroeconomic and financial variables.
Besides an improved forecasting performance relative to a benchmark TVP-VAR, this also yields interpretable factors, which are correlated to commonly emphasized sources of non-linearity in macroeconomics.

%% -------------------------------------------------------------------------- %%
%% Bibliography:
%% -------------------------------------------------------------------------- %%

\bibliography{FactorTVPVAR}

@article{DieboldMariano1995,
   author = {Francis X. Diebold and Roberto S. Mariano},
   issue = {3},
   journal = {Journal of Business \& Economic Statistics},
   month = {7},
   title = {Comparing Predictive Accuracy},
   volume = {13},
   year = {1995},
}

@article{BoivinGiannoniStevanovic2020,
   author = {Jean Boivin and Marc P. Giannoni and Dalibor Stevanović},
   doi = {10.1080/07350015.2018.1497507},
   issn = {0735-0015},
   issue = {2},
   journal = {Journal of Business \& Economic Statistics},
   month = {4},
   pages = {272-284},
   title = {Dynamic Effects of Credit Shocks in a Data-Rich Environment},
   volume = {38},
   year = {2020},
}

@article{AmirAhmadi-Matthes-Wang2020,
   author = {Pooyan Amir-Ahmadi and Christian Matthes and Mu-Chun Wang},
   doi = {10.1080/07350015.2018.1459302},
   issn = {0735-0015},
   issue = {1},
   journal = {Journal of Business \& Economic Statistics},
   month = {1},
   pages = {124-136},
   title = {Choosing Prior Hyperparameters: With Applications to Time-Varying Parameter Models},
   volume = {38},
   year = {2020},
}

@article{DuffyMavroeidisWycherley2025,
	title = {Cointegration with {Occasionally} {Binding} {Constraints}},
	volume = {252},
	issn = {03044076},
	url = {http://arxiv.org/abs/2211.09604},
	doi = {10.1016/j.jeconom.2025.106103},
	language = {en},
	urldate = {2025-12-20},
	journal = {Journal of Econometrics},
	author = {Duffy, James A. and Mavroeidis, Sophocles and Wycherley, Sam},
	month = nov,
	year = {2025},
	pages = {106103},
}

@article{DuffyMavroeidisWycherley2024,
	title = {Stationarity with {Occasionally} {Binding} {Constraints}},
	url = {https://www.ssrn.com/abstract=4819163},
	doi = {10.2139/ssrn.4819163},
	language = {en},
	urldate = {2025-12-20},
	journal = {Manuscript, Oxford University (arXiv: 2307.06190)},
	author = {Duffy, James and Mavroeidis, Sophocles and Wycherley, Sam},
	year = {2024},
}

@article{Mavroeidis2021,
	title = {Identification at the {Zero} {Lower} {Bound}},
	volume = {89},
	copyright = {https://creativecommons.org/licenses/by/4.0/},
	issn = {0012-9682},
	url = {https://www.econometricsociety.org/doi/10.3982/ECTA17388},
	doi = {10.3982/ECTA17388},
	language = {en},
	number = {6},
	urldate = {2025-05-07},
	journal = {Econometrica},
	author = {Mavroeidis, Sophocles},
	year = {2021},
	pages = {2855--2885},
}

@article{AruobaMlikotaSchorfheideVillalvazo2022,
	title = {{SVARs} with occasionally-binding constraints},
	volume = {231},
	issn = {03044076},
	url = {https://linkinghub.elsevier.com/retrieve/pii/S0304407621002487},
	doi = {10.1016/j.jeconom.2021.07.013},
	language = {en},
	number = {2},
	urldate = {2025-05-07},
	journal = {Journal of Econometrics},
	author = {Aruoba, S. Borağan and Mlikota, Marko and Schorfheide, Frank and Villalvazo, Sergio},
	month = dec,
	year = {2022},
	pages = {477--499},
}

@article{Sims2001,
	title = {Solving {Linear} {Rational} {Expectations} {Models}},
	volume = {20},
	journal = {Computational Economics},
	author = {Sims, Christopher A.},
	year = {2001},
	pages = {1--20},
}

@article{Aruoba-CubaBorda-HigaFlores-Schorfheide-Villalvazo2021,
	title = {Piecewise-linear approximations and filtering for {DSGE} models with occasionally-binding constraints},
	volume = {41},
	issn = {10942025},
	url = {https://linkinghub.elsevier.com/retrieve/pii/S1094202520301149},
	doi = {10.1016/j.red.2020.12.003},
	language = {en},
	urldate = {2025-05-07},
	journal = {Review of Economic Dynamics},
	author = {Aruoba, S. Borağan and Cuba-Borda, Pablo and Higa-Flores, Kenji and Schorfheide, Frank and Villalvazo, Sergio},
	month = jul,
	year = {2021},
	pages = {96--120},
}

@article{Philippon2012,
   author = {Thomas Philippon},
   doi = {10.1257/aer.20120578},
   issn = {19447981},
   issue = {4},
   journal = {American Economic Review},
   month = {4},
   pages = {1408-1438},
   publisher = {American Economic Association},
   title = {Has the US finance industry become less efficient? On the theory and measurement of financial intermediation},
   volume = {105},
   year = {2015},
}

@article{barattieri2019,
  author  = {A. Barattieri and M. Eden and D. Stevanovic},
  title   = {Financial Sector Interconnectedness and Monetary Policy Transmission},
  journal = {Macroeconomic Dynamics},
  volume  = {23},
  number  = {3},
  pages   = {1074--1011},
  year    = {2019}
}

@inbook{Ramey2016,
   author = {V. A. Ramey},
   doi = {10.1016/bs.hesmac.2016.03.003},
   isbn = {9780444594877},
   issn = {15740048},
   journal = {Handbook of Macroeconomics},
   pages = {71-162},
   publisher = {Elsevier B.V.},
   title = {Macroeconomic Shocks and Their Propagation},
   volume = {2},
   year = {2016},
}

@article{Sims1980,
   author = {Christopher A Sims},
   issue = {1},
   journal = {Econometrica},
   month = {1},
   pages = {1-48},
   title = {Macroeconomics and Reality},
   volume = {48},
   url = {https://about.jstor.org/terms},
   year = {1980},
}

@article{Clark-McCracken2013,
   author = {Todd E. Clark and Michael W. McCracken},
   doi = {10.1108/S0731-9053(2013)0000031004},
   isbn = {9781781907528},
   issn = {07319053},
   journal = {Advances in Econometrics},
   pages = {117-168},
   publisher = {JAI Press},
   title = {Evaluating the accuracy of forecasts from vector autoregressions},
   volume = {32},
   year = {2013},
}

@article{Christoffersen1998,
 ISSN = {00206598, 14682354},
 URL = {http://www.jstor.org/stable/2527341},
 author = {Peter F. Christoffersen},
 journal = {International Economic Review},
 number = {4},
 pages = {841--862},
 publisher = {[Economics Department of the University of Pennsylvania, Wiley, Institute of Social and Economic Research, Osaka University]},
 title = {Evaluating Interval Forecasts},
 urldate = {2025-09-14},
 volume = {39},
 year = {1998}
}

@article{Primiceri2005,
   author = {Giorgio E Primiceri},
   issue = {3},
   journal = {Review of Economic Studies},
   month = {7},
   pages = {821-852},
   title = {Time Varying Structural Vector Autoregressions and Monetary Policy},
   volume = {72},
   url = {https://about.jstor.org/terms},
   year = {2005},
}

@article{boivingiannoni2006,
    author = {Boivin, Jean and Giannoni, Marc P},
    title = {Has Monetary Policy Become More Effective?},
    journal = {The Review of Economics and Statistics},
    volume = {88},
    number = {3},
    pages = {445-462},
    year = {2006},
    month = {08},
    issn = {0034-6535},
    doi = {10.1162/rest.88.3.445},
    url = {https://doi.org/10.1162/rest.88.3.445},
    eprint = {https://direct.mit.edu/rest/article-pdf/88/3/445/1614214/rest.88.3.445.pdf},
}

@incollection{stock2003,
  author    = {J. H. Stock and M. W. Watson},
  title     = {Has the Business Cycle Changed? Evidence and Explanations},
  booktitle = {Monetary Policy and Uncertainty},
  publisher = {Federal Reserve Bank of Kansas City},
  pages     = {9--56},
  year      = {2003}
}

@article{cogley2005,
  author  = {T. Cogley and T. J. Sargent},
  title   = {Drift and Volatilities: Monetary Policies and Outcomes in the Post WWII U.S.},
  journal = {Review of Economic Dynamics},
  volume  = {8},
  number  = {2},
  pages   = {262--302},
  year    = {2005}
}

@article{sims2006,
  author  = {C. A. Sims and T. Zha},
  title   = {Were There Regime Switches in U.S. Monetary Policy?},
  journal = {American Economic Review},
  volume  = {96},
  number  = {1},
  pages   = {54--81},
  year    = {2006}
}

@article{canovagambettipappa2007,
    author = {Canova, Fabio and Gambetti, Luca and Pappa, Evi},
    title = {The Structural Dynamics of Output Growth and Inflation: Some International Evidence},
    journal = {The Economic Journal},
    volume = {117},
    number = {519},
    pages = {C167-C191},
    year = {2007},
    month = {04},
    issn = {0013-0133},
    doi = {10.1111/j.1468-0297.2007.02040.x},
    url = {https://doi.org/10.1111/j.1468-0297.2007.02040.x},
    eprint = {https://academic.oup.com/ej/article-pdf/117/519/C167/26506464/ejc167.pdf},
}

@article{stevanovic2016,
  author  = {D. Stevanovic},
  title   = {Common Time Variation of Parameters in Reduced-Form Macroeconomic Models},
  journal = {Studies in Nonlinear Dynamics and Econometrics},
  volume  = {20},
  number  = {2},
  pages   = {159--183},
  year    = {2016}
}

@article{jurado2015,
  author  = {K. Jurado and S. Ludvigson and S. Ng},
  title   = {Measuring Uncertainty},
  journal = {American Economic Review},
  volume  = {105},
  number  = {3},
  pages   = {1177--1216},
  year    = {2015}
}

@article{baker2013,
  author  = {S. R. Baker and N. Bloom and S. J. Davis},
  title   = {Measuring Economic Policy Uncertainty},
  journal = {The Quarterly Journal of Economics},
  volume  = {131},
  number  = {4},
  pages   = {1593--1636},
  year    = {2013}
}

@article{carriero2016,
  author  = {A. Carriero and T. E. Clark and M. Marcellino},
  title   = {Common Drifting Volatility in Large Bayesian VARs},
  journal = {Journal of Business \& Economic Statistics},
  volume  = {34},
  number  = {3},
  pages   = {375--390},
  year    = {2016}
}

@article{diebold2017,
  author  = {F. X. Diebold and F. Schorfheide and M. Shin},
  title   = {Real-time forecast evaluation of DSGE models with stochastic volatility},
  journal = {Journal of Econometrics},
  volume  = {201},
  pages   = {322--332},
  year    = {2017}
}

@article{AruobaBocolaSchorfheide2017,
   author = {S Boragan Aruoba and Luigi Bocola and Frank Schorfheide},
   journal = {Journal of Economic Dynamics and Control},
   pages = {34-54},
   title = {Assessing DSGE Model Nonlinearities},
   volume = {83},
   year = {2017},
}

@book{HerbstSchorfheide2015,
	title = {Bayesian {Estimation} of {DSGE} {Models}},
	isbn = {978-0-691-16108-2 978-1-4008-7373-9},
	url = {https://academic.oup.com/princeton-scholarship-online/book/30660},
	language = {en},
	urldate = {2025-07-11},
	publisher = {Princeton University Press},
	author = {Herbst, Edward P. and Schorfheide, Frank},
	month = dec,
	year = {2015},
}

@book{lutkepohl2005,
  author    = {Helmut L\"{u}tkepohl},
  title     = {New Introduction to Multiple Time Series Analysis},
  publisher = {Springer},
  year      = {2005},
  address   = {Berlin, Heidelberg},
  doi       = {10.1007/978-3-540-27752-1}
}

@article{karanasos2025unified,
  title={A Unified Theory for ARMA Models with Varying Coefficients: One Solution Fits All},
  author={Karanasos, Menelaos and Koutroumpis, Panagiotis and Dafnos, Stavros and Dafnos, Charalampos},
  journal={Econometric Theory},
  year={2025},
  note={Forthcoming}
}

@article{ravenna_vector_2007,
	title = {Vector autoregressions and reduced form representations of {DSGE} models},
	volume = {54},
	copyright = {https://www.elsevier.com/tdm/userlicense/1.0/},
	issn = {03043932},
	url = {https://linkinghub.elsevier.com/retrieve/pii/S0304393206002327},
	doi = {10.1016/j.jmoneco.2006.09.002},
	language = {en},
	number = {7},
	urldate = {2025-05-19},
	journal = {Journal of Monetary Economics},
	author = {Ravenna, Federico},
	month = oct,
	year = {2007},
	pages = {2048--2064},
}

@article{BenatiSurico2009,
	title = {{VAR} {Analysis} and the {Great} {Moderation}},
	volume = {99},
	issn = {0002-8282},
	url = {https://pubs.aeaweb.org/doi/10.1257/aer.99.4.1636},
	doi = {10.1257/aer.99.4.1636},
	language = {en},
	number = {4},
	urldate = {2025-05-07},
	journal = {American Economic Review},
	author = {Benati, Luca and Surico, Paolo},
	month = aug,
	year = {2009},
	pages = {1636--1652},
}

@article{SchmittGrohe-Uribe2003,
	title = {Closing small open economy models},
	volume = {61},
	copyright = {https://www.elsevier.com/tdm/userlicense/1.0/},
	issn = {00221996},
	url = {https://linkinghub.elsevier.com/retrieve/pii/S0022199602000569},
	doi = {10.1016/S0022-1996(02)00056-9},
	language = {en},
	number = {1},
	urldate = {2025-05-07},
	journal = {Journal of International Economics},
	author = {Schmitt-Grohé, Stephanie and Uribe, Martı́n},
	month = oct,
	year = {2003},
	pages = {163--185},
}

@article{Granger2008,
	title = {Non-{Linear} {Models}: {Where} {Do} {We} {Go} {Next} - {Time} {Varying} {Parameter} {Models}?},
	volume = {12},
	issn = {1558-3708},
	shorttitle = {Non-{Linear} {Models}},
	url = {https://www.degruyter.com/document/doi/10.2202/1558-3708.1639/html},
	doi = {10.2202/1558-3708.1639},
	language = {en},
	number = {3},
	urldate = {2025-05-07},
	journal = {Studies in Nonlinear Dynamics \& Econometrics},
	author = {Granger, Clive W.J.},
	month = jan,
	year = {2008},
}

@article{FernandezVillaverde-RubioRamirez2007,
	title = {How {Structural} {Are} {Structural} {Parameters}?},
	volume = {22},
	issn = {0889-3365, 1537-2642},
	shorttitle = {How {Structural} {Are} {Structural} {Parameters}?},
	url = {https://www.journals.uchicago.edu/doi/10.1086/ma.22.25554965},
	doi = {10.1086/ma.22.25554965},
	language = {en},
	urldate = {2025-07-30},
	journal = {NBER Macroeconomics Annual},
	author = {Fernández-Villaverde, Jesús and Rubio-Ramírez, Juan F.},
	month = jan,
	year = {2007},
	pages = {83--167},
}

@incollection{Hurwicz1962,
	address = {Stanford (CA), USA},
	title = {On the structural form of independent systems},
	booktitle = {Logic, methodology, and philosophy of science},
	publisher = {Stanford University Press},
	author = {Hurwicz, Leonid},
	year = {1962},
	pages = {232--239},
}

@article{FernandezVillaverde-RubioRamirez-Sargent-Watson2007,
	title = {{ABCs} (and {Ds}) of {Understanding} {VARs}},
	volume = {97},
	issn = {0002-8282},
	url = {https://pubs.aeaweb.org/doi/10.1257/aer.97.3.1021},
	doi = {10.1257/aer.97.3.1021},
	language = {en},
	number = {3},
	urldate = {2025-05-08},
	journal = {American Economic Review},
	author = {Fernández-Villaverde, Jesús and Rubio-Ramírez, Juan F and Sargent, Thomas J and Watson, Mark W},
	month = may,
	year = {2007},
	pages = {1021--1026},
}

@article{Christianoetal2005,
author = {Christiano, Lawrence J. and Eichenbaum, Martin and Evans, Charles L.},
title = {Nominal Rigidities and the Dynamic Effects of a Shock to Monetary Policy},
journal = {Journal of Political Economy},
volume = {113},
number = {1},
pages = {1-45},
year = {2005},
doi = {10.1086/426038},

URL = {     
        https://doi.org/10.1086/426038
},
eprint = {    
        https://doi.org/10.1086/426038
}
}

@article{delnegroschorfheide2004ier,
author = {Del Negro, Marco and Schorfheide, Frank},
title = {Priors from General Equilibrium Models for VARS},
journal = {International Economic Review},
volume = {45},
number = {2},
pages = {643-673},
doi = {https://doi.org/10.1111/j.1468-2354.2004.00139.x},
url = {https://onlinelibrary.wiley.com/doi/abs/10.1111/j.1468-2354.2004.00139.x},
eprint = {https://onlinelibrary.wiley.com/doi/pdf/10.1111/j.1468-2354.2004.00139.x},
year = {2004}
}

@article{andrews1993tests,
  title={Tests for parameter instability and structural change with unknown change point},
  author={Andrews, Donald W. K.},
  journal={Econometrica},
  volume={61},
  number={4},
  pages={821--856},
  year={1993},
  publisher={Wiley}
}

@article{koop2012forecasting,
  title={Forecasting inflation using dynamic model averaging},
  author={Koop, Gary and Korobilis, Dimitris},
  journal={International Economic Review},
  volume={53},
  number={3},
  pages={867--886},
  year={2012},
  publisher={Wiley}
}

@article{CICCARELLI2006737,
title = {Has the transmission mechanism of European monetary policy changed in the run-up to EMU?},
journal = {European Economic Review},
volume = {50},
number = {3},
pages = {737-776},
year = {2006},
issn = {0014-2921},
doi = {https://doi.org/10.1016/j.euroecorev.2005.01.001},
url = {https://www.sciencedirect.com/science/article/pii/S0014292105000024},
author = {Matteo Ciccarelli and Alessandro Rebucci}
}

@article{GambettiMusso2017,
author = {Gambetti, Luca and Musso, Alberto},
title = {Loan Supply Shocks and the Business Cycle},
journal = {Journal of Applied Econometrics},
volume = {32},
number = {4},
pages = {764-782},
doi = {https://doi.org/10.1002/jae.2537},
url = {https://onlinelibrary.wiley.com/doi/abs/10.1002/jae.2537},
eprint = {https://onlinelibrary.wiley.com/doi/pdf/10.1002/jae.2537},
year = {2017}
}

@article{BIANCHI2015406,
title = {Globalization and inflation: Evidence from a time-varying VAR},
journal = {Review of Economic Dynamics},
volume = {18},
number = {2},
pages = {406-433},
year = {2015},
issn = {1094-2025},
doi = {https://doi.org/10.1016/j.red.2014.07.004},
url = {https://www.sciencedirect.com/science/article/pii/S1094202514000441},
author = {Francesco Bianchi and Andrea Civelli}
}

@article{delnegroprimiceri2015,
 ISSN = {00346527, 1467937X},
 URL = {http://www.jstor.org/stable/43869469},
 author = {Marco Del Negro and Giorgio E. Primiceri},
 journal = {The Review of Economic Studies},
 number = {4 (293)},
 pages = {1342--1345},
 publisher = {[Oxford University Press, The Review of Economic Studies, Ltd.]},
 title = {Time Varying Structural Vector Autoregressions and Monetary Policy: A Corrigendum},
 urldate = {2025-09-11},
 volume = {82},
 year = {2015}
}

@article{INOUE2024105726,
title = {Local projections in unstable environments},
journal = {Journal of Econometrics},
volume = {244},
number = {2},
pages = {105726},
year = {2024},
issn = {0304-4076},
doi = {https://doi.org/10.1016/j.jeconom.2024.105726},
url = {https://www.sciencedirect.com/science/article/pii/S0304407624000721},
author = {Atsushi Inoue and Barbara Rossi and Yiru Wang}
}

@article{GOULETCOULOMBE2025982,
title = {Time-varying parameters as ridge regressions},
journal = {International Journal of Forecasting},
volume = {41},
number = {3},
pages = {982-1002},
year = {2025},
issn = {0169-2070},
doi = {https://doi.org/10.1016/j.ijforecast.2024.08.006},
url = {https://www.sciencedirect.com/science/article/pii/S0169207024000931},
author = {Philippe {Goulet Coulombe}}
}

@article{Hauzenberger02012025,
author = {Niko Hauzenberger and Florian Huber and Massimiliano Marcellino and Nico Petz},
title = {Gaussian Process Vector Autoregressions and Macroeconomic Uncertainty},
journal = {Journal of Business \& Economic Statistics},
volume = {43},
number = {1},
pages = {27--43},
year = {2025},
publisher = {ASA Website},
doi = {10.1080/07350015.2024.2322089},
URL = {  
        https://doi.org/10.1080/07350015.2024.2322089
},
eprint = {   
        https://doi.org/10.1080/07350015.2024.2322089
}
}

@article{Hamilton1989,
 ISSN = {00129682, 14680262},
 URL = {http://www.jstor.org/stable/1912559},
 author = {James D. Hamilton},
 journal = {Econometrica},
 number = {2},
 pages = {357--384},
 publisher = {[Wiley, Econometric Society]},
 title = {A New Approach to the Economic Analysis of Nonstationary Time Series and the Business Cycle},
 urldate = {2025-09-11},
 volume = {57},
 year = {1989}
}

@article{KimNelson1999,
 ISSN = {00346535, 15309142},
 URL = {http://www.jstor.org/stable/2646710},
 author = {Chang-Jin Kim and Charles R. Nelson},
 journal = {The Review of Economics and Statistics},
 number = {4},
 pages = {608--616},
 publisher = {The MIT Press},
 title = {Has the U.S. Economy Become More Stable? A Bayesian Approach Based on a Markov-Switching Model of the Business Cycle},
 urldate = {2025-09-11},
 volume = {81},
 year = {1999}
}

@article{BaumeisterPeersman2013,
Author = {Baumeister, Christiane and Peersman, Gert},
Title = {Time-Varying Effects of Oil Supply Shocks on the US Economy},
Journal = {American Economic Journal: Macroeconomics},
Volume = {5},
Number = {4},
Year = {2013},
Month = {October},
Pages = {1–28},
DOI = {10.1257/mac.5.4.1},
URL = {https://www.aeaweb.org/articles?id=10.1257/mac.5.4.1}}

@techreport{dewind2014reduced,
  title={Reduced-rank time-varying vector autoregressions},
  author={de Wind, Job and Gambetti, Luca},
  institution={CPB Netherlands Bureau for Economic Policy Analysis},
  type={CPB Discussion Paper},
  number={270},
  year={2014}
}

@article{CHAN2020105,
title = {Reducing the state space dimension in a large TVP-VAR},
journal = {Journal of Econometrics},
volume = {218},
number = {1},
pages = {105-118},
year = {2020},
issn = {0304-4076},
doi = {https://doi.org/10.1016/j.jeconom.2019.11.006},
url = {https://www.sciencedirect.com/science/article/pii/S0304407620300348},
author = {Joshua C.C. Chan and Eric Eisenstat and Rodney W. Strachan}
}

@article{CarrieroClarkMarcellino2016,
	title = {Common {Drifting} {Volatility} in {Large} {Bayesian} {VARs}},
	volume = {34},
	language = {en},
	number = {3},
	journal = {Journal of Business \& Economic Statistics},
	author = {Carriero, Andrea and Clark, Todd E and Marcellino, Massimiliano},
	year = {2016},
	pages = {375--390},
}

@article{Grassi-VanDerWel2013,
	title = {Dimension {Reduction} in {Large} {Time}-{Varying} {VARs}: {The} {DFM}-{VAR} {Model}},
	language = {en},
	journal = {Manuscript, Aarhus University},
	author = {Grassi, Stefano and van der Wel, Michel},
	year = {2013},
}

@article{CanovaCiccarelli2009,
	title = {Estimating {Multicountry} {VAR} {Models}},
	volume = {50},
	copyright = {http://onlinelibrary.wiley.com/termsAndConditions\#vor},
	issn = {0020-6598, 1468-2354},
	url = {https://onlinelibrary.wiley.com/doi/10.1111/j.1468-2354.2009.00554.x},
	doi = {10.1111/j.1468-2354.2009.00554.x},
	language = {en},
	number = {3},
	urldate = {2025-05-08},
	journal = {International Economic Review},
	author = {Canova, Fabio and Ciccarelli, Matteo},
	month = aug,
	year = {2009},
	pages = {929--959},
}

@article{Benati2010,
	title = {Are {Policy} {Counterfactuals} {Based} on {Structural} {VARs} {Reliable}?},
	issn = {1556-5068},
	url = {https://www.ssrn.com/abstract=1596588},
	doi = {10.2139/ssrn.1596588},
	language = {en},
	urldate = {2025-05-07},
	journal = {ECB Working Paper No. 1188},
	author = {Benati, Luca},
	year = {2010},
}

@article{ArtisGalvaoMarcellino2007,
	title = {The transmission mechanism in a changing world},
	volume = {22},
	copyright = {http://onlinelibrary.wiley.com/termsAndConditions\#vor},
	issn = {0883-7252, 1099-1255},
	url = {https://onlinelibrary.wiley.com/doi/10.1002/jae.923},
	doi = {10.1002/jae.923},
	language = {en},
	number = {1},
	urldate = {2025-05-08},
	journal = {Journal of Applied Econometrics},
	author = {Artis, Michael and Galvão, Ana Beatriz and Marcellino, Massimiliano},
	month = jan,
	year = {2007},
	pages = {39--61},
}

@article{Renzetti2024,
	title = {Theory coherent shrinkage of {Time}-{Varying} {Parameters} in {VARs}},
	language = {en},
	journal = {Manuscript, Bocconi University, arXiv: 2311.11858},
	author = {Renzetti, Andrea},
	year = {2024},
}

% %% -------------------------------------------------------------------------- %%
% %% Appendix:
% %% -------------------------------------------------------------------------- %%

\newpage

\renewcommand{\thepage}{A.\arabic{page}}
\setcounter{page}{1}

\begin{appendix}

\markright{This Version: \today }
\renewcommand{\theequation}{A.\arabic{equation}}
\setcounter{equation}{0}

\renewcommand*\thetable{A-\arabic{table}}
\setcounter{table}{0}
\renewcommand*\thefigure{A-\arabic{figure}}
\setcounter{figure}{0}

\addtocontents{toc}{\protect\setcounter{tocdepth}{1}}

\begin{center}

    {\fontsize{19}{25}\selectfont {\bf Appendix}}\\[5pt]

    {\fontsize{15}{20}\selectfont {\bf Origins and Nature of Macroeconomic Instability \\ in Vector Autoregressions}}\\[20pt]

    \begin{normalsize}
        \begin{minipage}{0.3\textwidth}
            \centering
            {Pooyan Amir-Ahmadi\\\emph{\small Amazon \\ \colW{.}}}
        \end{minipage}
        \begin{minipage}{0.3\textwidth}
            \centering
            {Marko Mlikota\\\emph{\small Geneva Graduate Institute \\ \colW{.}}}
        \end{minipage}
        \begin{minipage}{0.3\textwidth}
            \centering
            {Dalibor Stevanovi\'c\\\emph{\small Université du Québec \\ à Montréal}}
        \end{minipage}
    \end{normalsize}
\end{center}

	%% -------------------------------------------------------------------------- %%
%% -------------------------------------------------------------------------- %%

%% -------------------------------------------------------------------------- %%
%% -------------------------------------------------------------------------- %%
%% -------------------------------------------------------------------------- %%

\section{Theory}
\label{appsec_theory}

\mylemma{Subvector-Dynamics under CP-VAR}{lemma_CPVARsubvec}{
    Suppose $x_t \in \reals^n$ follows a stable CP-VAR(1):
    \[
    x_t = \Phi_0 + \Phi_1 x_{t-1} + \Phi_\varepsilon \epsilon_t \; ,  \quad \epsilon_t \sim WN(0, I) \; .
    \]
    Then any \(x^o_t \subseteq x_t\), $x^o_t \in \reals^k$ follows a CP-VARMA(\(p',q'\)) with $p' \leq (n-k+1)$ and $q' \leq (n-k)$.
}

\myproof{
    This result follows directly from \citet[Corollary~11.1.2]{lutkepohl2005}, which states that any \(k\)-dimensional marginalization of an \(n\)-dimensional VAR(\(p\)) process satisfies in general a VARMA(\(p', q'\)) representation with \(p' \leq (n-k+1)p\) and \(q' \leq (n-k)p\). Here, \(p=1\), so the bounds simplify to \(p' \leq (n-k+1)\) and \(q' \leq (n-k)\). 
    $\blacksquare$
}

%% -------------------------------------------------------------------------- %%

\mylemma{Subvector-Dynamics under TVP-VAR}{lemma_TVPVARsubvec}{
    Suppose $x_t \in \reals^n$ follows a TVP-VAR(1):
    \[
    x_t = \Phi_{0,t} + \Phi_{1,t}x_{t-1} + \Phi_{\varepsilon,t}\epsilon_t, \quad \epsilon_t \sim WN(0, I) \; .
    \]
    Then any \(x^o_t \subseteq x_t\), $x^o_t \in \reals^k$, admits a finite-order TVP-VARMA$(p',q')$
    representation with $p' \leq (n-k+1)$ and $q' \leq (n-k)$.    
    
    \noindent If, more specifically, $x_t$ follows a RS-VAR(1) -- i.e. $\Phi_{0,t} = \Phi_{0}(s_t)$, $\Phi_{1,t}=\Phi_{1}(s_t)$ and $\Phi_{\varepsilon,t}=\Phi_{\varepsilon}(s_t)$ for $s_t \sim$ $n_s$-state Markov-chain with transition matrix $T$ --, 
    then $x^o_t$ follows a RS-VARMA (\(p',q'\)) with $p', q' \leq (n-k+1)$ and with $\tilde{\Phi}_{0,t} = \tilde{\Phi}_{0}(s_t)$, $\tilde{\Phi}_{l,t} = \tilde{\Phi}_{l}(s_t)$, $\tilde{\Theta}_{k,t} = \tilde{\Theta}_{k}(s_t)$ and $\tilde{\Phi}_{\varepsilon,t} = \tilde{\Phi}_{\varepsilon}(s_t)$.
}

\myproof{
    We consider a general time variation; the regime-switching case follows immediately by conditioning on $s_t$.

    Equivalently, with the lag operator $L$,
    \[
    A_t(L)x_t = \Phi_{0,t} + \Phi_{\varepsilon,t}\epsilon_t \; , 
    \qquad A_t(L)=I_n-\Phi_{1,t}L \; .
    \]

    Partition $x_t=(x'_{1,t},x'_{2,t})'$, where $x_{1,t}\equiv x^o_t \in \mathbb{R}^k$ and 
    $x_{2,t}\in\mathbb{R}^{n-k}$, and partition $A_t(L)$, $\Phi_{0,t}$ and $\Phi_{\varepsilon,t}$ conformably:
    \[
    A_t(L)=
    \begin{pmatrix}
    A_{11,t}(L) & A_{12,t}(L)\\
    A_{21,t}(L) & A_{22,t}(L)
    \end{pmatrix}.
    \]

    Fix $t$. Assume $A_{22,t}(L)$ is nonsingular and that its inverse is well defined uniformly in $t$, i.e.
    $\det(A_{22,t}(L))$ is not identically zero and is bounded away from zero uniformly in $t$.
    Then, by Cramer's rule applied pointwise in $t$,
    \[
    A_{22,t}(L)^{-1}
    = \frac{\mathrm{adj}(A_{22,t}(L))}{\det(A_{22,t}(L))},
    \]
    where $\mathrm{adj}(A_{22,t}(L))$ is a matrix polynomial of degree $(n-k-1)$ and
    $\det(A_{22,t}(L))$ is a scalar polynomial of degree $(n-k)$.

    Proceeding exactly as in the proof of \citet[Corollary~11.1.2]{lutkepohl2005}, but applied pointwise in $t$,
    $x_{2,t}$ can be eliminated from the system, yielding a reduced equation for $x_{1,t}$ of the form
    \[
    \tilde{A}_t(L)x_{1,t}
    =
    \tilde{c}_t(L)
    +
    \tilde{M}^1_t(L)\epsilon_t^1
    +
    \tilde{M}^2_t(L)\epsilon_t^2 ,
    \]
    where $\tilde{A}_t(L)$, $\tilde{c}_t(L)$, $\tilde{M}^1_t(L)$ and $\tilde{M}^2_t(L)$ are time-varying lag
    polynomials.

    Since $A_{ij,t}(L)$ are first-degree polynomials in $L$, $\tilde{A}_t(L)$ has degree at most $(n-k+1)$ and
    $\tilde{M}^1_t(L)$ and $\tilde{M}^2_t(L)$ have degree at most $(n-k)$. Hence $x^o_t=x_{1,t}$ admits a
    finite-order TVP-VARMA$(p',q')$ representation with $p'\le(n-k+1)$ and $q'\le(n-k)$.

    The regime-switching case follows since all coefficients are functions of $s_t$.
    $\blacksquare$
}

\myremark{Wold decomposition}{
    If, in addition to the assumptions of  \cref{lemma_TVPVARsubvec}, the time-varying lag polynomial \(A_t(L)=I-\Phi_{1,t}L\) is causal and uniformly invertible (i.e.\ \(\det (A_t(z))\neq 0\) for all \(|z|\le1\) and all \(t\)), then, by the results of \cite{karanasos2025unified}, the resulting TVP-VARMA\((p',q')\) process \(x^o_t\) admits a Wold--Cramér decomposition.
}

\myremark{TVP-VAR(\(p'\))}{
    A sufficient condition for the reduced system for \(x_{1,t}\equiv x^o_t\) to be a
    TVP-VAR(\(p'\)) process (i.e., \(q'=0\)) is that
    \[
        \Phi_{1,t}^{12} = 0 \quad \text{and} \quad \Phi_{\varepsilon,t}^{12} = 0,
    \]
    for all \(t\). In this case, the marginalized block $x_{2,t}$ does not enter
    either the conditional mean or the innovation of $x_{1,t}$, and the
    autoregressive degree satisfies $p' \le 1$.
}

%% -------------------------------------------------------------------------- %%
%% -------------------------------------------------------------------------- %%
%% -------------------------------------------------------------------------- %%

\mylemma{CP-VAR-Solution to Linear RE System with CPs}{lemma_CPREtoCPVAR}{
    Suppose $x_t \in \reals^{n_x}$ follows a linear RE system with CPs:
    \[
        \Gamma_0 x_t = \gamma + \Gamma_1 x_{t-1} + \Psi\epsilon_t + \Pi \eta_t \; ,
    \]
    where $\epsilon_t \in \reals^{n_\epsilon}$ is a process with $\mean[\epsilon_t] = 0$ and $\var[\epsilon_t] = I$, and $\eta_t \in \reals^{n_\eta}$ is an arbitrary process with $n_\eta \leq n_x$.
    Suppose the non-explosive solution exists and is unique.
    Then, it can be written as a CP-VAR(1):
    $$ x_t = \Phi_0 + \Phi_1 x_{t-1} + \Phi_\varepsilon\epsilon_t \; . $$
}

\myproof{
    This follows from the analysis in \citet{Sims2001} and its discussion in \citet[Ch. 2.1]{HerbstSchorfheide2015}, adjusted for the case of a potentially non-zero mean.
    QZ decompositions give $\Gamma_0 = Q'\Lambda Z'$ and $\Gamma_1 = Q'\Omega Z'$, where $Q$ and $Z$ are orthonormal and $\Lambda$ and $\Omega$ are upper-triangular. Defining $w_t = Z'x_t$ and multiplying the RE system by $Q$ yields
    $$ \bmat{\Lambda_{11} & \Lambda_{12}\\ 0 & \Lambda_{22}}\bmat{w_{1t} \\ w_{2t}} 
    = \bmat{\kappa_1 \\ \kappa_2} + \bmat{\Omega_{11} & \Omega_{12}\\ 0 & \Omega_{22}}\bmat{w_{1,t-1} \\ w_{2,t-1}} + \bmat{Q_1 \\ Q_2} (\Psi \epsilon_t + \Pi \eta_t)\; , $$ 
    where it is assumed -- w.l.o.g. -- that $x_t$ is ordered s.t. the $m \times 1$ vector $w_{2t}$ is explosive ($0\leq m \leq n_x$). 
    A non-explosive solution exists iff $m = n_\eta$ and $Q_2 \Pi$ is invertible. Then we can find $\eta_t$ s.t. the impact of $\epsilon_t$ on $w_{2t}$ is offset at each $t$: $Q_2 \Psi \epsilon_t + Q_2 \Pi\eta_t = 0$.
    Let $\Phi = Q_1 \Pi (Q_2 \Pi)^{-1}$.
    %
    % A non-explosive solution exists iff $\exists$ $\Phi$ s.t. $Q_1 \Pi = \Phi Q_2 \Pi$, i.e. for any $\epsilon_t$, we can find an $\eta_t$ s.t. the impact of $\epsilon_t$ on $w_{2t}$ is offset and the latter equals zero at all times: $Q_2 \Psi \epsilon_t + Q_2 \Pi\eta_t = 0$. 
    %
    Then, multiplying the first $n$ equations above by $[I, -\Phi]$ yields 
    $$ \bmat{\Lambda_{11} & \Lambda_{12} - \Phi \Lambda_{22}\\ 0 & \Lambda_{22}} Z'x_t 
    =  \bmat{\kappa_1 - \Phi \kappa_2 \\ \kappa_2} + \bmat{\Omega_{11} & \Omega_{12} - \Phi \Omega_{22}\\ 0 & \Omega_{22}} Z'x_{t-1} + \bmat{Q_1 - \Phi Q_2 \\ 0} (\Psi \epsilon_t + \Pi \eta_t)\; , $$ 
    or $A x_t = D + B x_{t-1} + C \epsilon_t$. 
    Multiplying by $A^{-1}$ gives 
    $$ x_t = \Phi_0 + \Phi_1 x_{t-1} + \Phi_\varepsilon\epsilon_t \; , $$
    with $\Phi_1 = A^{-1}B$, $\Phi_\varepsilon = A^{-1}C$ and $\Phi_0 = A^{-1}D$.
    $\blacksquare$
}

\mylemma{TVP-VAR-Solution to Linear RE System with TVPs}{lemma_TVPREtoTVPVAR}{
    Suppose $x_t \in \reals^{n_x}$ follows a linear RE system with TVPs:
    \[
        \Gamma_{0,t} x_t = \gamma_t + \Gamma_{1,t} x_{t-1} + \Psi_t \epsilon_t + \Pi \eta_t \; ,
    \]
    where $\epsilon_t \in \reals^{n_\epsilon}$ is a process with $\mean[\epsilon_t] = 0$ and $\var[\epsilon_t] = I$, and $\eta_t \in \reals^{n_\eta}$ is an arbitrary process with $n_\eta \leq n_x$.
    Suppose the non-explosive solution exists and is unique.  
    %and suppose the number of unstable eigenvalues is constant over time.
    %
    % In addition, suppose $\exists \; \delta > 0$ s.t. every generalized eigenvalue $\lambda_t$ of $\Gamma_{0,t}$ and $\Gamma_{1,t}$ satisfies either $|\lambda_t| \leq 1-\delta$ or $|\lambda_t| \geq 1+\delta$ for each $t$, 
    % and suppose the number of unstable eigenvalues is constant over time.
    %
    Then, the solution can be written as a TVP-VAR(1):
    $$ x_t = \Phi_{0,t} + \Phi_{1,t} x_{t-1} + \Phi_{\varepsilon,t}\epsilon_t \; . $$

    If, more specifically, the linear RE system is regime-switching -- i.e. $\Gamma_{0,t} = \Gamma_{0}(s_t)$, $\Gamma_{1,t}=\Gamma_{1}(s_t)$ and $\Psi_{t}=\Psi(s_t)$  for $s_t \sim$ $n_s$-state Markov-chain with transition matrix $T$ --, 
    then the solution is a RS-VAR(1) with $\Phi_{0,t} = \Phi_{0}(s_t)$,  $\Phi_{1,t} = \Phi_{1}(s_t)$ and $\Phi_{\varepsilon,t} = \Phi_{\varepsilon}(s_t)$.
}

\myproof{
    The result follows from applying the proof of \cref{lemma_CPREtoCPVAR} pointwise for each $t$.
    We consider a general time-variation.  
    The same argument goes through if the time-variation has a regime-switching nature.

    Specifically, QZ decompositions give $\Gamma_{0,t} = Q_t'\Lambda_t Z_t'$ and $\Gamma_{1,t} = Q_t'\Omega_t Z_t'$, where $Q_t$ and $Z_t$ are orthonormal and $\Lambda_t$ and $\Omega_t$ are upper-triangular. Defining $w_t = Z_t'x_t$ and multiplying the RE system by $Q_t$ yields
    $$ \bmat{\Lambda_{11,t} & \Lambda_{12,t}\\ 0 & \Lambda_{22,t}}\bmat{w_{1t} \\ w_{2t}} 
    = \bmat{\kappa_{1t} \\ \kappa_{2t}} + \bmat{\Omega_{11,t} & \Omega_{12,t}\\ 0 & \Omega_{22,t}}\bmat{w_{1,t-1} \\ w_{2,t-1}} + \bmat{Q_{1t} \\ Q_{2t}} (\Psi_t \epsilon_t + \Pi \eta_t)\; , $$ 
    where it is assumed -- w.l.o.g. -- that $x_t$ is ordered s.t. the $m_t \times 1$ vector $w_{2t}$ is explosive ($0\leq m_t \leq n_x$). 
    A non-explosive solution exists iff $m_t = n_\eta$ and $Q_{2t} \Pi$ is invertible for each $t$. 
    Then we can find $\eta_t$ s.t. the impact of $\epsilon_t$ on $w_{2t}$ is offset at each $t$: $Q_{2t} \Psi_t \epsilon_t + Q_{2t} \Pi\eta_t = 0$.
    Let $\Phi_t = Q_{1t} \Pi (Q_{2t} \Pi)^{-1}$.
    %
    % A non-explosive solution exists iff $\exists$ $\Phi$ s.t. $Q_1 \Pi = \Phi Q_2 \Pi$, i.e. for any $\epsilon_t$, we can find an $\eta_t$ s.t. the impact of $\epsilon_t$ on $w_{2t}$ is offset and the latter equals zero at all times: $Q_2 \Psi_t \epsilon_t + Q_2 \Pi\eta_t = 0$. 
    %
    Then, multiplying the first $n$ equations above by $[I, -\Phi_t]$ yields 
    $$ \bmat{\Lambda_{11,t} & \Lambda_{12,t} - \Phi_t \Lambda_{22,t}\\ 0 & \Lambda_{22,t}} Z_t'x_t 
    =  \bmat{\kappa_{1t} - \Phi_t \kappa_{2t} \\ \kappa_{2t}} + \bmat{\Omega_{11,t} & \Omega_{12,t} - \Phi \Omega_{22,t}\\ 0 & \Omega_{22,t}} Z_t'x_{t-1} + \bmat{Q_{1t} - \Phi_t Q_{2t} \\ 0} (\Psi_t \epsilon_t + \Pi \eta_t)\; , $$ 
    or $A_t x_t = D_t + B_t x_{t-1} + C_t \epsilon_t$. 
    Multiplying by $A_t^{-1}$ gives 
    $$ x_t = \Phi_{0,t} + \Phi_{1,t} x_{t-1} + \Phi_{\varepsilon,t}\epsilon_t \; , $$
    with $\Phi_{1,t} = A_t^{-1}B_t$, $\Phi_{\varepsilon,t} = A_t^{-1}C_t$ and $\Phi_{0,t} = A_t^{-1}D_t$.
    $\blacksquare$
}

%% -------------------------------------------------------------------------- %%
%% -------------------------------------------------------------------------- %%
%% -------------------------------------------------------------------------- %%

\myrepproposition{}{prop_linearDSGE}{
    Suppose the dynamics of $y_t$ are non-explosive and uniquely generated by \cref{eq_DSGEeqconds,eq_DSGEexoprocesses_cont,eq_DSGEexoprocesses_disc}, and
    \begin{enumerate}
        \item %\label{prop_linearDSGE_cond_typicalexo} 
        $n^c_e > 0$ and $n^d_e = 0$ (all $\{e_{jt}\}_{j=1}^{n_e}$ are continuous), 
        and $G_t=G$ and $\Sigma_t = \Sigma$ are constant;

        \item %\label{prop_linearDSGE_cond_linear} 
        $F$ is linear in $(y_t',y_{t+1}',y_{t-1}',e^{\prime}_t,e^{\prime}_{t+1})'$.
    \end{enumerate}
    Then, any $y^o_t\subseteq y_t$ follows a CP-VARMA($p',q'$) with $p',q' < \infty$.
}

\myproof{
    If $F$ is linear, then \cref{eq_DSGEeqconds} writes 
    \begin{align}
        \mean_t \sbr{ F_0 + F_{1} y_{t} + F_{2} y_{t+1} +F_{3} y_{t-1} + F_{4} e_{t}+ F_{5} e_{t+1} } =  0 \; , \label{eq_DSGEeqconds_linearF}
    \end{align}
    where the vector $F_0$ and matrices $\{F_{i}\}_{i=1}^5$ are functions of $\theta$, but this conditioning is omitted for notational simplicity.
    Equivalently, this is
    $$
    F_0 + F_{1} y_{t} + F_{2}\mean_t\sbr{y_{t+1}} + F_{3} y_{t-1} + F_{4} e_{t} + F_{5} \mean_t\sbr{e_{t+1}} =  0
    $$
    Substituting $\mean_t[e_{t+1}] = G e_t$ and defining the $n_x \times 1$ vector $x_t = (y_t', e_t', \mean_t[y_{t+1}]')'$, with $n_x = 2n_y + n_e$, allows us to write 
    $$
    F_0 + \sbr{F_{1}, F_{4}+F_{5} G, F_{2}} x_{t}+F_{3} y_{t-1} = 0 \; .
    $$
    Introducing expectational errors $\eta_t \equiv y_t - \mean_{t-1}[y_t]$ and combining the preceding two sets of equations with the $n_e$ equations in \cref{eq_DSGEexoprocesses_cont} yields a linear RE system with constant parameters:
    $$
    \bmat{F_0 \\ 0 \\ 0}
    +
    \bmat{
        F_{1} & F_{4}+F_{5} G & F_{2} \\
        0 & I & 0 \\
        I & 0 & 0}  x_{t} 
    + \bmat{
        F_{3} & 0 & 0 \\
        0 & -G & 0 \\
        0 & 0 & -I} x_{t-1} 
    + \bmat{0 \\ -\Sigma_{tr} \\ 0} \epsilon_t 
    + \bmat{0 \\ 0 \\ -I} \eta_t = 0 \; ,
    $$
    i.e. $\Gamma_0 x_t = \gamma + \Gamma_1 x_{t-1} + \Psi\epsilon_t + \Pi \eta_t $.
    By \cref{lemma_CPREtoCPVAR}, a non-explosive solution yields a CP-VAR(1) for $x_t$:
    $$ x_t = \Phi_0 + \Phi_1 x_{t-1} + \Phi_\varepsilon\epsilon_t \; .$$
    By \cref{lemma_CPVARsubvec}, any $y^o_t \subseteq y_t \subset x_t$ follows a CP-VARMA($p,q$) and $p,q < \infty$.\footnote{\cite{ravenna_vector_2007} shows a similar result.} $\blacksquare$
}

%% -------------------------------------------------------------------------- %%

\myrepproposition{}{prop_nonlinearDSGE}{
    Suppose the dynamics of $y_t$ are non-explosive and uniquely generated by \cref{eq_DSGEeqconds,eq_DSGEexoprocesses_cont,eq_DSGEexoprocesses_disc}, and
    \begin{enumerate}
        \item %\label{prop_nonlinearDSGE_cond_typicalexo} 
        $n^c_e > 0$ and $n^d_e = 0$ (all $\{e_{jt}\}_{j=1}^{n_e}$ are continuous), 
        and $G_t=G$ and $\Sigma_t = \Sigma$ are constant;

        \item %\label{prop_nonlinearDSGE_cond_poly} 
        $F$ is a $p$th-order polynomial in $(y_t',y_{t+1}',y_{t-1}',e^{\prime}_t,e^{\prime}_{t+1})'$ for $p \geq 2$.
    \end{enumerate}
    Then, any $y^o_t \subseteq y_t$ follows a TVP-VARMA($p',q'$) with $p',q' < \infty$.
}

% Notes:
% G diagonal we assume in proof
% non-zero IRF to shock of positive variance ensures $y_t$ is not constant s.t. our TVPs are indeed CPs

\myproof{
    To clarify notation, recall that linearity of $F$ (i.e. $p=1$) implies that \cref{eq_DSGEeqconds} yields
    $$ 
    F_0 + F_{1} y_{t} + F_{2}\mean_t\sbr{y_{t+1}} + F_{3} y_{t+1} + F_{4} e_{t} + F_{5} \mean_t\sbr{e_{t+1}} =  0 \; ,
    $$
    and, for $i = 1:n_y$, the $i$th equation in this system is
    \begin{align*}
        F_{0}^{i} + \sum_{j=1}^{n_{y}} \cbr{ F_{1, j}^{i}  y_{j,t} + F_{2, j}^{i} \mean_t \sbr{y_{j,t+1}} + F_{3, j}^{i} y_{j,t-1} }   + \sum_{s=1}^{n_{e}} \cbr{ F_{4,s}^{i} e_{s,t}+F_{s, s}^{i}  \mean_t \sbr{e_{s,t+1}} } = 0 \; .
    \end{align*}
    Now consider first $p=2$, i.e. $F$ is a second-order polynomial. Then \cref{eq_DSGEeqconds} reads
    \begin{alignat*}{3}
        0 = F_{0}^{i} + \sum_{j=1}^{n_y}\{
        &F_{1, j}^{i}y_{j,t} 
        &&+F_{2, j}^{i}\mean_t [y_{j,t+1}]
        & &+F_{3, j}^{i}y_{j,t-1} \} \\
        + \sum_{j=1}^{n_y} \sum_{k=1}^{n_y}\{
        &F_{11, j k}^{i}y_{j,t}y_{k, t} 
        &&+F_{22, j k}^{i}\mean_t[y_{j, t+1} y_{k, t+1}]
        & &+F_{33 , j k}^{i}y_{j,t-1}y_{k,t-1} \} \\
        +\sum_{j=1}^{n_y} \sum_{k=1}^{n_y} \{
        &F_{12, j k}^{i} y_{j,t} \mean_t[y_{k, t+1}]
        &&+F_{13, j k}^{i}y_{j,t}y_{k, t-1} 
        & &+F_{23, j k}^{i}\mean_t[y_{j, t+1}]y_{k, t-1} \} \\
        +\sum_{s=1}^{n_e}\{
        &F_{4, s}^{i} e_{s,t}
        &&+F_{5, s}^{i} \mean_t [e_{s, t+1}]\} \\
        + \sum_{s=1}^{n_e} \sum_{r=1}^{n_e}\{
        &F_{44, s r}^{i} e_{s,t} e_{r,t} 
        &&+F_{55, s r}^{i} \mean_t [e_{s, t+1} e_{r,t+1}]\} \\
        + \sum_{s=1}^{n_e} \sum_{r=1}^{n_e}
        &F_{45, s r}^{i} e_{s,t} \mean_t[e_{r,t+1}]  \\
        + \sum_{j=1}^{n_y} \sum_{s=1}^{n_e}\{
        &F_{14, j s}^{i}y_{j,t} e_{s,t} 
        &&+F_{24, j s}^{i}\mean_t[y_{j, t+1}] e_{s,t} 
        & &+ F_{34, j s}^{i}y_{j,t-1} e_{s,t}\} \\
        + \sum_{j=1}^{n_{y}} \sum_{s=1}^{n_{e}}\{
        &F_{15, j s}^{i}y_{j,t} \mean_t[e_{s,t+1}] 
        &&+F_{25, j s}^{i}\mean_t[y_{j, t+1} e_{s,t+1}] 
        & &+ F_{35, j s}^{i}y_{j,t-1} \mean_t[e_{s,t+1}] \} \; .
    \end{alignat*}
    By \cref{eq_DSGEexoprocesses_cont}, we have $\mean_t[e_{s,t+1}] = G_{s\cdot} e_t$ and 
    $$ \mean_t[e_{s,t+1}e_{r,t+1}] = \mean_t[(G_{s\cdot}e_t + \varepsilon_{s,t+1})(G_{r\cdot}e_t + \varepsilon_{r,t+1})] = (G_{s\cdot}e_t)(G_{r\cdot}e_t) + \Sigma_{sr} \; . $$
    Taking this into account and grouping terms yields 
    \begin{align*}
    0 = F_{0}^{i} + &\sum_{j=1}^{n_y}  y_{j,t} \cbr{ 
            F_{1, j}^{i} 
            +\sum_{k=1}^{n_y} F_{11, j k}^{i} y_{k, t} 
            +\sum_{k=1}^{n_y} F_{13, j k}^{i} y_{k, t-1} 
            +\sum_{s=1}^{n_{e}} \br{ F_{14, j s}^{i} e_{s,t} + F_{15, j s}^{i}G_{s\cdot}e_t }
        } \\
        +&\sum_{j=1}^{n_y}\mean_t[y_{j, t+1}] \cbr{ 
            F_{2, j}^{i}
            +\sum_{k=1}^{n_y} F_{12, k j}^{i} y_{k,t}   
            +\sum_{k=1}^{n_y} F_{23, j k}^{i} y_{k, t-1} 
            +\sum_{s=1}^{n_{e}}  F_{24, j s}^{i} e_{s,t} 
        }  \\
        +&\sum_{j=1}^{n_y} y_{j,t-1} \cbr{ 
        F_{3, j}^{i} 
        +\sum_{k=1}^{n_y} F_{33, j k}^{i} y_{k, t-1} 
        +\sum_{s=1}^{n_{e}} \br{ F_{34, j s}^{i}e_{s,t}  + F_{35, j s}^{i}G_{s\cdot}e_t }  
        }  \\
        +&\sum_{s=1}^{n_e} e_{s,t}\cbr{ 
        F_{4, s}^{i} 
        +\sum_{r=1}^{n_{e}} \br{ F_{44, sr}^{i}e_{r,t}  + F_{45, sr}^{i}G_{r\cdot}e_t } 
        }  + \sum_{s=1}^{n_e} F_{5, s}^{i}G_{s\cdot}e_t  + \sum_{s=1}^{n_e}\sum_{r=1}^{n_e} F_{55, sr}^{i}G_{s\cdot}e_t G_{r\cdot}e_t \\
        +&\sum_{j=1}^{n_y}\sum_{k=1}^{n_y} F_{22, j k}^{i} \mean_t[y_{j, t+1}y_{k, t+1}] 
        + \sum_{j=1}^{n_y} \sum_{s=1}^{n_{e}} F_{25, j s}^{i} \mean_t[y_{j, t+1} e_{s,t+1}]
        + \sum_{s=1}^{n_e} \sum_{r=1}^{n_e} F_{55, sr}^{i} \Sigma_{sr}  \; .
    \end{align*}
    Let $i^{yy}_t \equiv vec\br{y_{t}y_{t}'}$ and $i^{ye}_t \equiv vec\br{y_{t}e_t'}$.
    Also, let the vector $F_1^i$ stack $\{F^i_{1,j}\}_{j=1}^{n_y}$ along rows and construct the vectors $F_2^i$, $F_3^i$, $F_4^i$ and $F_5^i$ analogously. 
    Similarly, let the matrix $F_{11}^i$ stack $\{F^i_{11,jk}\}_{j,k=1}^{n_y}$ along rows (first subscript; $j$) and columns (second subscript; $k$) and construct the other matrices with double-subscripts, $\{F^i_{ab}\}_{a,b=1}^{5}$, likewise.
    Then we can write the above equation in vectorized notation as
    \begin{align*}
        0 = c^i + m^{i1'}_t y_t + m^{i2'}_t \mean_t[y_{t+1}] + m^{i3'}_t y_{t-1} + m^{i4'}_t e_t &+ m^{i5'} i^{yy}_t + m^{i6'} \mean_t \sbr{ i^{yy}_{t+1} }  \\
        &+ m^{i7'} i^{ye}_t + m^{i8'} \mean_t[i^{ye}_{t+1}] \; .
    \end{align*}
    The scalar $c^i = F_{0}^{i} + \iota' [F^i_{55} \cdot \Sigma]\iota$, where $\iota$ is a vector of ones and $\cdot$ denotes element-wise multiplication.
    The $n_y \times 1$ vectors $m^{i1}_t$, $m^{i2}_t$ and $m^{i3}_t$, 
    the $n_e \times 1$ vector $m^{i4}_t$, 
    the $n_y^2 \times 1$ vectors $m^{i5}$ and $m^{i6}$,
    and the $n_y n_e \times 1$ vectors $m^{i7}$ and $m^{i8}$ are given by
    \begin{align*}
        m^{i1}_t &= \br{ F_{1}^{i} +F_{13}^{i}y_{t-1} } \; , \\
        m^{i2}_t &= \br{ F_{2}^{i} 
        +F_{12}^{i} y_t
        +F_{23}^{i}y_{t-1}
        +F_{24}^{i} e_t} \; , \\
        m^{i3}_t &= \br{ F_{3}^{i} + F_{33}^{i} y_{t-1}
        +\br{F_{34}^{i}+F_{35}^{i}G }e_t} \; , \\
        m^{i4}_t &= \br{ F_{4}^{i} + G' F_{5}^{i} 
        +\br{F_{44}^{i} + F_{45}^{i}G + G'F_{55}^{i}G }e_t} \; , \\
        m^{i5} &= vec(F_{11}^{i}) \; , \\
        m^{i6} &= vec(F_{22}^{i}) \; , \\
        m^{i7} &= vec\br{ F_{14}^{i}+F_{15}^{i}G }  \; , \\
        m^{i8} &= vec(F_{25}^{i}) \; .\footnotemark
    \end{align*}
    \footnotetext{Writing $\cdot$ as a subscript of a matrix denotes all rows (or columns) of it: for example, $F_{1, \cdot}^{i} = F_{1}^{i}$, while $F_{12, \cdot k}^{i}$ denotes the $k$th column of the matrix $F_{12}^{i}$.}
    Combining such expressions for all equations $i=1:n_y$ yields
    \begin{align} 
        0 = C + M^1_t y_t + M^2_t \mean_t[y_{t+1}] + M^3_t y_{t-1} + M^4_t e_t &+ M^5 i^{yy}_t + M^6 \mean_t \sbr{ i^{yy}_{t+1} } \nonumber \\
        &+ M^7 i^{ye}_t + M^8 \mean_t[i^{ye}_{t+1}] \; , \label{eq_preREsystem_2ndOrder_1}
    \end{align}
    where $C = (c^1, ..., c^{n_y})'$ is an $n_y \times 1$ vector, $M^1_t = \sbr{ m^{11}_t, ..., m^{n_y 1}_t }'$ is an $n_y \times n_y$ matrix, and the matrices $M^2_t$ and $M^3_t$ ($n_y \times n_y$), the matrix $M^4_t$ ($n_y \times n_e$ ), the matrices $M^5$ and $M^6$ ($n_y \times n_y^2$), and the matrices $M^7$ and $M^8$ ($n_y \times n_yn_e$) are analogously constructed.\footnote{
        Note that this representation is not unique. For example, the term $F^{i}_{14}e_t'y_t = F^{i}_{14}y_t'e_t$ can not only be dealt with by absorbing $F^{i}_{14}e_t'$ into the term $m^{i1'}_t$ multiplying $y_t$ but also by absorbing $F^{i}_{14}y_t'$ into the term $m^{i3'}_t$ multiplying $e_t$.
    }
    Note that the matrices $\{M^a_t\}_{a=1:8}$ do not contain any forward-looking terms; their time variation is entirely due to the presence of $y_t$, $y_{t-1}$ and $e_t$.

    Now define $\check{y}_t = (i^{yy'}_t,i^{ye'}_t)'$, $\check{M}^1 = [M^5, M^7]$ and $ \check{M}^2 = [M^6, M^8]$ s.t. the above reads 
    \begin{align} 
        0 = C + M^1_t y_t + M^2_t \mean_t[y_{t+1}] + M^3_t y_{t-1} + M^4_t e_t &+ \check{M}^1 \check{y}_t + \check{M}^2 \mean_t \sbr{ \check{y}_{t+1} } \; .\label{eq_preREsystem_2ndOrder_2}
    \end{align}
    Next, define $x_t = (y_t', \check{y}_t', e_t', \mean_t[y_{t+1}]', \mean_t \sbr{ \check{y}_{t+1} }')'$ s.t. 
    $$ C + \sbr{M^1_t, \check{M}^1, M^4_t, M^2_t, \check{M}^2}x_t + M^3_t y_{t-1} = 0 \; .$$
    Let $\eta_t = (\eta^{1\prime}_t,\check{\eta}_t')'$ be a vector of first- and second-order expectational errors,
    $$ \eta^1_t \equiv y_t - \mean_{t-1}[y_t] \quad \text{and} \quad \check{\eta}_t \equiv \check{y}_t  - \mean_{t-1} \sbr{ \check{y}_t } \; . $$
    Combining these three sets of equations with the $n_e$ equations in \cref{eq_DSGEexoprocesses_cont} yields the RE system
    \begin{align*} 
        \Gamma_{0,t} x_t = \gamma + \Gamma_{1,t} x_{t-1} + \Psi\epsilon_t + \Pi \eta_t \; ,
    \end{align*}
    with $\Gamma_{0,t}, \gamma, \Gamma_{1,t}, \Psi, \Pi$ respectively given by
    \begin{align*}
        \bmat{C \\ 0 \\ 0 \\ 0 \\ 0} \; , \;
    \bmat{
        M^1_t & \check{M}^1 & M^4_t & M^2_t & \check{M}^2 \\
        0 & 0 & 0 & 0 & 0 \\
        0 & 0 & I & 0 & 0 \\
        I & 0 & 0 & 0 & 0 \\
        0 & I & 0 & 0 & 0} \; , \;
    \bmat{
        M^3_t & 0 & 0 & 0 & 0\\
        0 & 0 & 0 & 0 & 0 \\
        0 & 0 & -G & 0 & 0 \\
        0 & 0 & 0 & -I & 0 \\
        0 & 0 & 0 & 0 & -I} \; , \;
    \bmat{0 \\ 0 \\ -\Sigma_{tr} \\ 0 \\ 0} \; , \;
    \bmat{0 & 0 \\ 0 & 0 \\ 0 & 0 \\ -I & 0 \\ 0 & -I} \; .
    \end{align*}
    % \begin{align*}
    %     \bmat{C \\ 0 \\ 0 \\ 0 \\ 0} +
    % \bmat{
    %     M^1_t & \check{M}^1 & M^4_t & M^2_t & \check{M}^2 \\
    %     0 & 0 & 0 & 0 & 0 \\
    %     0 & 0 & I & 0 & 0 \\
    %     I & 0 & 0 & 0 & 0 \\
    %     0 & I & 0 & 0 & 0}  x_{t} 
    % + \bmat{
    %     M^3_t & 0 & 0 & 0 & 0\\
    %     0 & 0 & 0 & 0 & 0 \\
    %     0 & 0 & -G & 0 & 0 \\
    %     0 & 0 & 0 & -I & 0 \\
    %     0 & 0 & 0 & 0 & -I} x_{t-1} 
    % + \bmat{0 \\ 0 \\ -\Sigma_{tr} \\ 0 \\ 0} \epsilon_t 
    % + \bmat{0 & 0 \\ 0 & 0 \\ 0 & 0 \\ -I & 0 \\ 0 & -I} \eta_t = 0 \; ,
    % \end{align*}
    % i.e. 
    % \begin{align*} 
    %     \Gamma_{0,t} x_t = \gamma + \Gamma_{1,t} x_{t-1} + \Psi\epsilon_t + \Pi \eta_t \; .
    % \end{align*}
    Thus, as in the case of a linear $F$, dynamics can be represented by a linear RE system. 
    In this case, however, 
    some elements in the matrices $\Gamma_{0,t}$ and $\Gamma_{1,t}$ are time-varying due to the interaction terms in \cref{eq_DSGEeqconds}.\footnote{
        Note that the system can be reduced by dropping $\check{y}_t$ from $x_t$ and redefining $\gamma$, $\Gamma_{0,t}$, $\Gamma_{1,t}$, $\Psi$ and $\Pi$ as follows.
        Let $N_t^1$ be the $((n_y(n_y+n_e)) \times n_y)$-matrix that solves $\check{y}_t = N_t^1 y_t$, and let $N_t^2$ be the $(n_y \times n_y)$-matrix that solves $\check{M}^1 \check{y}_t = N_t^2 y_t$.
        Also, let $\tilde{M}_t^1 = M^1_t + N^2_t$.
        We can then write
        \begin{align*}
            \bmat{C \\ 0 \\ 0 \\ 0} +
        \bmat{
            \tilde{M}^1_t & M^4_t & M^2_t & \check{M}^2 \\
            0 & I & 0 & 0 \\
            I & 0 & 0 & 0 \\
            N_t^1 & 0 & 0 & 0}  x_{t} 
        + \bmat{
            M^3_t & 0 & 0 & 0\\
            0 & -G & 0 & 0 \\
            0 & 0 & -I & 0 \\
            0 & 0 & 0 & -I} x_{t-1} 
        + \bmat{0 \\ -\Sigma_{tr} \\ 0 \\ 0} \epsilon_t 
        + \bmat{0 & 0 \\ 0 & 0 \\ -I & 0 \\ 0 & -I} \eta_t = 0 \; .
        \end{align*}
    } 
    By \cref{lemma_TVPREtoTVPVAR}, then, the unique, stable solution features $x_t$ following a TVP-VAR(1):
    $$ x_t = \Phi_{0,t} + \Phi_{1,t} x_{t-1} + \Phi_{\varepsilon,t}\epsilon_t \; . $$
    Finally, by \cref{lemma_TVPVARsubvec}, any $y^o_t \subseteq y_t \subset x_t$ follows a finite-order TVP-VARMA.
    Our technical assumption that $\partial y_{i,t+h}/\partial \varepsilon_{jt} \neq 0$ for some $i,j,h$ and positive definiteness of $\Sigma$ ensure that some parameters indeed vary over time.
    This proves the statement for $p=2$.
    
    Now suppose $p=3$, i.e. $F$ is a third-order polynomial. Then, in addition to the linear terms and the \myquote{double} interaction terms (interactions of two variables) above, \cref{eq_DSGEeqconds} contains \myquote{triple} interaction terms.
    Continuing the notation from above, define the triple interaction terms among $y_t$ and $e_t$ (pertaining to the same time period): $i^{yyy}_t = vec(vec(y_t y_t')y_t')$, and, similarly, $i^{yye}_t$ and $i^{yee}_t$.
    These are the triple interaction terms in whose one-step ahead expectations it is not possible to separate out $\mean_t[y_{t+1}]$, $\mean_t[i^{yy}_{t+1}]$ or $\mean_t[i^{ye}_{t+1}]$.
    All other triple interaction terms can be absorbed into \cref{eq_preREsystem_2ndOrder_1}
    by separating out one of the variables among $\{y_t, i^{yy}_t, i^{ye}_t, e_t, \mean_t[y_{t+1}], \mean_t \sbr{ i^{yy}_{t+1} } , \mean_t \sbr{ i^{ye}_{t+1} }\}$ in a way that the matrices 
    $\{M^a_t\}_{a=1:8}$ do not contain any forward-looking terms.\footnote{
        Akin to above, the expectations of triple interaction terms among time-($t+1$) exogenous variables, $\mean_t\sbr{ e_{s,t+1}e_{r,t+1}e_{l,t+1} }$ for $s,r,l=1:n_e$, are known and can be broken up into interactions between three time-$t$ exogenous variables (and constants).
    }
    %Accommodating the terms involving contemporaneous and one-step ahead expected values of $i^{yyy}_t$, $i^{yye}_t$ and $i^{yee}_t$, 
    \cref{eq_preREsystem_2ndOrder_1} becomes
    \begin{align} 
        0 = C + M^1_t y_t + M^2_t \mean_t[y_{t+1}] + M^3_t y_{t-1} + M^4_t e_t &+ M^5_t i^{yy}_t + M^6_t \mean_t \sbr{ i^{yy}_{t+1} } \nonumber \\
        &+ M^7_t i^{ye}_t + M^8_t \mean_t[i^{ye}_{t+1}] \nonumber \\
        &+ M^9 i^{yyy}_t + M^{10} \mean_t[i^{yyy}_{t+1}] \nonumber \\
        &+ M^{11} i^{yye}_t + M^{12} \mean_t[i^{yye}_{t+1}] \nonumber \\
        &+ M^{13} i^{yee}_t + M^{14} \mean_t[i^{yee}_{t+1}] \; ,\footnotemark \label{eq_preREsystem_3rdOrder_1}
    \end{align}
    \footnotetext{
        Note that, relative to \cref{eq_preREsystem_2ndOrder_1}, we not only added three more terms, but the matrices $M^5_t$, $M^6_t$, $M^7_t$ and $M^8_t$ are now time-varying, reflecting triple interaction terms between current or future $i^{yy}_t$ or $i^{ye}_t$ on the one hand and current, lagged or future $y_t$ or $e_t$ on the other hand. 
    }
    This equation can be written in the form of \cref{eq_preREsystem_2ndOrder_2} by defining 
    $\check{y}_t = (i^{yy'}_t,i^{ye'}_t, i^{yyy'}_t,i^{yye'}_t,i^{yee'}_t)'$ as well as the matrices 
    $\check{M}^1_t = [M^5_t, M^7_t,M^9,M^{11},M^{13}]$ and $\check{M}^2_t = [M^6_t, M^8_t, M^{10}, M^{12}, M^{14}]$.
    Then, the argument that leads from \cref{eq_preREsystem_2ndOrder_2} to the linear RE-system with TVPs for $x_t$ and the TVP-VARMA for $y^o_t \subset x_t$ is unchanged.
    
    The reasoning from the previous paragraph applies to the case where $F$ is any $p$th-order polynomial, with $p\geq 3$; \cref{eq_preREsystem_2ndOrder_2} remains unchanged, while $\check{y}_t$ stacks interaction terms among $y_t$ and $e_t$ of all order, from those with two to those with $p$ variables.
    Therefore, the statement from \cref{prop_nonlinearDSGE} goes through for $F$ being any $p$th-order polynomial.
    $\blacksquare$
}

%% -------------------------------------------------------------------------- %%

%% -------------------------------------------------------------------------- %%

%% -------------------------------------------------------------------------- %%

\myrepproposition{}{prop_DSGE_TVPexo}{
    Suppose the dynamics of $y_t$ are non-explosive and uniquely generated by \cref{eq_DSGEeqconds,eq_DSGEexoprocesses_cont,eq_DSGEexoprocesses_disc}, and
    \begin{enumerate}
        \item %\label{prop_DSGE_TVPexo_cond_TVPVARcontexo} 
        $n^c_e > 0$ and $n^d_e = 0$ (all $\{e_{jt}\}_{j=1}^{n_e}$ are continuous), 
        and $G_t$ and $\Sigma_t$ vary over time: $\exists$ at least one $(r,s)$ and one element $\calE_t$ of $G_t$ or $\Sigma_t$ s.t. $\calE_r \neq \calE_s$;

        \item %\label{prop_DSGE_TVPexo_cond_poly} 
        $F$ is a $p$th-order polynomial in $(y_t',y_{t+1}',y_{t-1}',e^{\prime}_t,e^{\prime}_{t+1})'$ for $p\geq 1$ (including linearity).
    \end{enumerate}
    Then, any $y^o_t \subseteq y_t$ follows a TVP-VARMA($p',q'$) with $p',q' < \infty$.
}

\myproof{
    Suppose first $p=1$, i.e. $F$ is linear in $(y_t',y_{t+1}',y_{t-1}',e^{\prime}_t,e^{\prime}_{t+1})'$.
    Following the same initial steps as in proof of \cref{prop_linearDSGE}, we obtain a linear RE system with time-varying parameters due to the time-variation in $G$ and $\Sigma$:
    $$\Gamma_{0,t} x_t = \gamma + \Gamma_{1,t} x_{t-1} + \Psi_t \epsilon_t + \Pi \eta_t  \; .$$
    By \cref{lemma_TVPREtoTVPVAR}, a non-explosive solution yields a TVP-VAR(1) for $x_t$.
    By \cref{lemma_TVPVARsubvec}, any $y^o_t \subseteq y_t \subset x_t$ follows a TVP-VARMA($p,q$) and $p,q < \infty$.

    Now suppose $p \geq 2$. 
    As can be verified, the proof of \cref{prop_nonlinearDSGE} goes through with $G$ and $\Sigma$ time-varying.
    Relative to before, the matrix $\Psi$ in the linear RE-system becomes time-varying, but this does not change the conclusion that any $y^o_t$ follows a finite-order TVP-VARMA.\footnote{
        More specifically, under $p=2$, the matrix $M^7$ (and its consistuent vectors $m^{i7}$) as well as the matrix $\check{M}^1$ become time-varying, though this does not affect the already time-varying $\Gamma_{0,t}$.
        Under $p \geq 3$, also the matrices $M^{11}$ and $M^{13}$ become time-varying, but this does not affect the already time-varying $\check{M}^1_t$ and $\check{M}^2_t$, and, therefore, $\Gamma_{0,t}$.
    }
    $\blacksquare$
}

%% -------------------------------------------------------------------------- %%

\myrepproposition{}{prop_DSGE_discexo}{
    Suppose the dynamics of $y_t$ are non-explosive and uniquely generated by \cref{eq_DSGEeqconds,eq_DSGEexoprocesses_cont,eq_DSGEexoprocesses_disc}, and 
    \begin{enumerate}
        \item %\label{prop_DSGE_discexo_cond_nodiscexo} 
        $n^c_e > 0$ and $n^d_e > 0$ (there are both continuous and discrete exogenous processes);

        \item %\label{prop_DSGE_discexo_cond_poly} 
        $F$ is a $p$th-order polynomial in $(y_t',y_{t+1}',y_{t-1}',e^{c\prime}_t,e^{c\prime}_{t+1})'$ for $p\geq 1$ (including linearity).
    \end{enumerate}
    Then, if $p=1$ and $G_t = G$, $\Sigma_t = \Sigma$ and $T_t = T$ are constant, any $y^o_t \subseteq y_t$ follows a RS-VARMA($p',q'$) with $p',q' < \infty$ and regimes determined by $e^d_t$.
    Else, $y^o_t \subseteq y_t$ follows a TVP-VARMA($p',q'$) with $p',q' < \infty$.
}

\myproof{
    Suppose first $p=1$ -- i.e. $F$ is linear in $(y_t',y_{t+1}',y_{t-1}',e^{c'}_t,e^{c'}_{t+1})'$ -- and $G$, $\Sigma$ and $T$ are constant. 
    Write $s$ for $e^d_t$ and $s'$ for $e^d_{t+1}$. 
    Then, \cref{eq_DSGEeqconds} can be written as 
    \begin{align}
        \mean_t \sbr{ F_0(s,s') + F_{1}(s,s') y_{t} + F_{2}(s,s') y_{t+1} +F_{3}(s,s') y_{t-1} + F_{4}(s,s') e^c_{t}+ F_{5}(s,s') e^c_{t+1} } =  0 \; , \label{eq_DSGEeqconds_linearF_discexo}
    \end{align}
    where the vector $F_0$ and matrices $\{F_{i}\}_{i=0}^5$ are arbitrary functions of $s$ adn $s'$. As before, they are also functions of $\theta$, but this conditioning is omitted for notational simplicity.
    We have $\mean_t \sbr{ F_0(s,s')} = \sum_{s'} F_0(s,s')T_{s,s'} \equiv \dot{F}_0(s)$, and analogously for $\dot{F}_1(s)$, $\dot{F}_3(s)$ and $\dot{F}_4(s)$. Using independence of $e^c_{t+1}$ and $e^d_{t+1} = s'$, we construct $\dot{F}_5(s)$ likewise. 
    By Law of Iterated Expectations, 
    $$ \mean_t \sbr{ F_{2}(s,s') y_{t+1} } = \sum_{s'} F_2(s,s')T_{s,s'} \mean_t[y_{t+1}|s'] \equiv \ddot{F}_2(s)\ddot\mean_{t}[y_{t+1}] \; , $$
    where the $n_y \times n_y n^d_e$ matrix $\ddot{F}_2$ stacks the $n_y \times n_y$ matrices $\{F_2(s,s')T_{s,s'}\}_{s'}$ across columns and the $n_y n^d_e \times 1$ vector $\ddot\mean_{t}[y_{t+1}]$ stacks the vectors $\{\mean_t[y_{t+1}|s']\}_{s'}$ along rows.
    This allows us to write the above as
    \begin{align}
        \dot{F}_0(s) + \dot{F}_{1}(s) y_{t} + \ddot{F}_{2}(s)\ddot\mean_{t}\sbr{y_{t+1}} + \dot{F}_{3}(s) y_{t-1} + \dot{F}_{4}(s) e^c_{t} + \dot{F}_{5}(s) \mean_{t}\sbr{e^c_{t+1}} =  0 \; .
    \end{align}
    Substituting $\mean_{t}[e^c_{t+1}] = G e^c_t$ and defining the $n_x \times 1$ vector $x_t = (y_t', e^{c\prime}_t, \ddot\mean_{t}[y_{t+1}]')'$, with $n_x = n_y + n^c_e + n_y n^d_e$, allows us to write 
    $$
    \dot{F}_0(s) + \sbr{\dot{F}_{1}(s), \dot{F}_{4}(s) + \dot{F}_{5}(s) G, \ddot{F}_{2}(s)} x_{t} + \dot{F}_{3}(s) y_{t-1} = 0 \; .
    $$
    Consider the expectational errors $\eta_t \equiv \ddot{I}' y_t - \ddot\mean_{t-1}[y_t]$, where the matrix $\ddot{I}$ stacks $I$ across rows. 
    Combining the preceding two sets of equations with the $n_e$ equations in \cref{eq_DSGEexoprocesses_cont} yields a linear RE system with regime-switching parameters:
    $$
    \bmat{\dot{F}_0(s)\\ 0 \\ 0}
    +
    \bmat{
        \dot{F}_1(s) & \dot{F}_4(s) + \dot{F}_5(s) G & \dot{F}_2(s) \\
        0 & I & 0 \\
        \ddot{I}' & 0 & 0}  x_{t} 
    + \bmat{
        \dot{F}_3(s) & 0 & 0 \\
        0 & -G & 0 \\
        0 & 0 & -I} x_{t-1} 
    + \bmat{0 \\ -\Sigma_{tr} \\ 0} \epsilon_t 
    + \bmat{0 \\ 0 \\ -I} \eta_t = 0 \; ,
    $$
    i.e. 
    $$ \Gamma_0(s) x_t = \gamma(s) + \Gamma_1(s) x_{t-1} + \Psi\epsilon + \Pi \eta_t  \; . $$
    By \cref{lemma_TVPREtoTVPVAR}, a non-explosive solution yields an RS-VAR(1) for $x_t$:
    $$ x_t = \Phi_0(s) + \Phi_1(s) x_{t-1} + \Phi_\varepsilon(s)\epsilon_t \; .$$
    By \cref{lemma_TVPVARsubvec}, any $y^o_t \subseteq y_t \subset x_t$ follows an RS-VARMA($p,q$) and $p,q < \infty$.
    
    Suppose now that -- other things equal as above -- at least one element of $G$ varies over time.
    As can be verified easily by adjusting the calculations above, the matrices $\Gamma_0(s)$ and $\Gamma_1(s)$ then inherit the time-variation of $G$.
    As a result, their regime-switching nature is subsumed into the generic time-variation of $G$.
    We emphasize this by writing $\Gamma_{0,t}$, $\Gamma_{1,t}$ and then also $\gamma_t$ instead of $\Gamma_0(s)$, $\Gamma_1(s)$ and $\gamma(s)$, as we obtain a linear RE-system with generically time-varying parameters.
    In turn, it leads by \cref{lemma_TVPREtoTVPVAR} and \cref{lemma_TVPVARsubvec} to a TVP-VARMA for $y^o_t$.
    The same conclusion is reached when $\Sigma$ varies over time, as it implies a time-varying $\Psi$.
    Similarly, a time-varying $T$ leads to time-variation in $\dot{F}_0(s)$, $\dot{F}_1(s)$, $\dot{F}_3(s)$, $\dot{F}_4(s)$, $\dot{F}_5(s)$ as well as $\ddot{F}_2(s)$, which implies time-varying $\Gamma_0(s)$, $\Gamma_1(s)$ and $\gamma(s)$.

    The same conclusion is also reached under $p \geq 2$, and regardless of the time-variation of $G$, $\Sigma$ or $T$.
    We can modify our calculations for $p\geq 2$ from the proof of \cref{prop_nonlinearDSGE} or \cref{prop_DSGE_TVPexo} analogously as above.
    For the sake of illustration, consider $p=2$ and suppose $G$, $\Sigma$ and $T$ are constant over time.
    Relative to the proof of \cref{prop_nonlinearDSGE}, under $n^d_e >0$, the scalar $F_0^i$, the vectors $\{F_a^i\}_{a=1}^5$ and the matrices $\{F^i_{ab}\}_{a,b=1}^{5}$ are all functions of $s$ and $s'$. 
    By taking expectations w.r.t. $s'$, we turn them into functions of $s$ only, whereby we expand the information set of our expectation operator: we replace $\mean_t[y_{t+1}]$ and $\mean_t \sbr{ \check{y}_{t+1} }$ with $\ddot\mean_{t}[y_{t+1}]$ and $\ddot\mean_{t} \sbr{ \ddot{y}_{t+1} }$.
    This leads to a regime-switching version of \cref{eq_preREsystem_2ndOrder_2}:
    \begin{align} 
        0 = C(s) + M^1_t(s) y_t + M^2_t(s) \mean_{t,s}[y_{t+1}] + M^3_t(s) y_{t-1} + M^4_t(s) e_t &+ \check{M}^1(s) \check{y}_t + \check{M}^2(s) \mean_{t,s} \sbr{ \check{y}_{t+1} } \; ,\label{eq_preREsystem_2ndOrder_2_RS}
    \end{align}
    where the vector $C(s)$ and the matrices $\{M^a_t(s)\}_{a=1}^4$ and $\{\check{M}^a(s)\}_{a=1}^2$ are appropriately re-defined relative to before.
    The time-variation due to regime-switching is, however, subsumed into the more generic type of time-variation due to non-linearity of \cref{eq_DSGEeqconds}.
    This means that we drop the dependence of the objects above on $s$. 
    In turn, we adjust the definition of expectational errors, writing $\eta_t = (\eta^{1\prime}_t,\check{\eta}_t')'$ with
    $$ \eta^1_t \equiv \ddot{I}'y_t - \ddot\mean_{t-1}[y_t] \quad \text{and} \quad \check{\eta}_t \equiv \ddot{I}'\check{y}_t  - \ddot\mean_{t-1} \sbr{ \check{y}_t } \; . $$
    This leads to a regime-switching version of the linear RE-system with TVPs --
    \begin{align*} 
        \Gamma_{0,t}(s) x_t = \gamma(s) + \Gamma_{1,t}(s) x_{t-1} + \Psi(s)\epsilon_t + \Pi \eta_t
    \end{align*}
    -- whose regime-switching variation is, once again, overshadowed by the generic type of time-variation, which means that we drop the dependence on $s$ and write generically
    \begin{align*} 
        \Gamma_{0,t} x_t = \gamma_t + \Gamma_{1,t} x_{t-1} + \Psi_t \epsilon_t+ \Pi \eta_t \; .
    \end{align*}
    
    The expression \cref{eq_preREsystem_2ndOrder_2_RS} is valid for any $p \geq 2$, as explained in the proof of \cref{prop_nonlinearDSGE}.
    Time-variation in $G$ or $\Sigma$ may affect the time-variation of some of the matrices in \cref{eq_preREsystem_2ndOrder_2_RS} but does not change the time-varying nature of the linear RE-system, as explained in the proof of \cref{prop_DSGE_TVPexo}.
    The same holds for time-variation in $T$.
    By \cref{lemma_TVPREtoTVPVAR} and \cref{lemma_TVPVARsubvec}, then, any $y^o_t \subseteq y_t \subset x_t$ follows a finite-order TVP-VARMA.
    $\blacksquare$
}

	%% -------------------------------------------------------------------------- %%
%% -------------------------------------------------------------------------- %%

%% -------------------------------------------------------------------------- %%
%% -------------------------------------------------------------------------- %%
%% -------------------------------------------------------------------------- %%

\section{Illustrative Models}
\label{appsec_illustrativemodels}

\todo{left to do (maybe?): simplest NK model (Gali's book, ch. 2), to discuss (exogenously) time-varying Taylor-rule parameters}

\todo{left to do (maybe?): 2nd-ord.-linearize one model to show we obtain a TVP RE system regardless whether DSGE-param. varies}

%% -------------------------------------------------------------------------- %%
%\newpage
\subsection*{NCG Model}

\paragraph*{Setup}

Consider the canonical Neoclassical Growth (NCG) model, the predecessor of all modern DSGE models. It is defined by the optimization problems of a representative firm and a representative household, whereby Total Factor Productivity (TFP) $z_t$ varies exogenously according to, say, an AR(1) process: 
\begin{align} \label{eq_NCG_AR1forTFP}
    z_t = \rho_z z_{t-1} + \varepsilon_t \; , \quad \varepsilon_t \sim N(0,\sigma_z^2) \; .
\end{align}
The representative firm maximizes profits by solving 
$$ \underset{L_t,K_t}{\max}\; \Pi_t = \underset{L_t,K_t}{\max}\; y_t - w_tL_t - r_tK_t \; , \quad y_t = e^{z_t} K_t^\alpha L_t^{1-\alpha} \; , $$
whereby the firm takes TFP $z_t$, the wage $w_t$ and the rental rate of capital $r_t$ as given. 
The first-order conditions (FOCs) imply 
\begin{align}
    w_t &= (1-\alpha)e^{z_t}K_t^\alpha L_t^{-\alpha} \; , \label{eq_NCG_eqconds_w}\\
    r_t &= \alpha e^{z_t} K_t^{\alpha-1} L_t^{1-\alpha} \; . \label{eq_NCG_eqconds_r}
\end{align}
The representative household maximizes its lifetime utility. 
The optimization problem can be represented by the Bellman equation
\begin{align*}
    V(z_t,k_t,K_t) = &\underset{c_t, k_{t+1}\geq 0, \; l_t \in [0,1]}{\max}\; u(c_t,l_t) + \beta \mean_t \sbr{ V(z_{t+1},k_{t+1},K_{t+1}) } \\
&\quad \text{s.t.} \quad c_t + i_t = w_tl_t + r_tk_t \; , \\
&\quad \text{\colW{s.t.}} \quad \colW{c_t +} i_t = k_{t+1}-(1-\delta)k_t \\
&\quad \text{\colW{s.t.}} \quad \colW{c_t +} K_{t+1} = H(z_t,K_t)\; ,
\end{align*}
whereby $u(c,l) = \frac{c^{1-\tau}}{1-\tau}$ is the household's utility, $V$ is its value function, and $H$ is its perceived law of motion for the aggregate capital stock $K_t$.\footnote{
    The expectation is taken w.r.t. $z_{t+1}$ and $K_{t+1}$, taking $z_t$ and $K_t$ as given. Note that $k_{t+1}$ is chosen by the household and therefore known at time $t$.
    }
Trivially, the optimal labor supply is $l_t = 1$, since the household is assumed, for simplificty, to experience no disutility from working.
Solving the budget constraint for $c_t$ and plugging it into the utility function, we get a maximization problem w.r.t. $k_{t+1}$ only.
Let $u_1$ denote the derivative of $u(c,l)$ w.r.t. $c$.
The FOC w.r.t. $k_{t+1}$ yields $-u_1(c_t,l_t) + \beta \mean[V_2(z_{t+1},k_{t+1},K_{t+1})] = 0$, which together with the fact that $V_2(z_t,k_t,K_t) = u_1(c_t,l_t) (r_t + 1-\delta)$ gives the so-called Euler equation
\begin{align}
    c_t^{-\tau} = \beta \mean_t \sbr{ c_{t+1}^{-\tau} (1+r_{t+1}-\delta) } \; .\label{eq_NCG_eqconds_k}
\end{align}
In equilibrium, the labor, capital and goods markets clear: 
$$ l_t = L_t \; , \quad k_t = K_t \; , \quad c_t + i_t = y_t \; . $$
Also, the household is indeed representative, meaning that the aggregate capital stock $K_{t+1}$ evolves as the capital stock chosen by the representative household, $k_{t+1}$: $k_{t+1} = K_{t+1} = H(z_t,K_t)$. 
Inserting $l_t = L_t = 1$ and writing $k_t$ for $K_t$ in \cref{eq_NCG_eqconds_w,eq_NCG_eqconds_r} and the goods market clearing condition 
\begin{align} \label{eq_NCG_eqconds_marketclearing}
    c_t + k_{t+1}-(1-\delta)k_t = e^{z_t} k_t^\alpha l_t^{1-\alpha} \; ,
\end{align}
we get a system of four equations -- \cref{eq_NCG_eqconds_w,eq_NCG_eqconds_r,eq_NCG_eqconds_k,eq_NCG_eqconds_marketclearing} -- in four unknowns: $c_t, k_{t+1}, w_t, r_t$.

\cref{eq_NCG_eqconds_w,eq_NCG_eqconds_r,eq_NCG_eqconds_k,eq_NCG_eqconds_marketclearing} constitute a non-linear DSGE model, i.e. a model characterized by \cref{eq_DSGEeqconds} with a non-linear $F$. 
The endogenous variables are $y_t = (c_t, k_{t+1}, w_t, r_t)'$, 
the exogenous variable is $e_t = z_t$, 
and the DSGE-parameters are $\theta = (\alpha,\beta,\tau,\delta,\rho_z,\sigma_z)'$. 
The law of motion of $e_t$ is governed by \cref{eq_NCG_AR1forTFP}, which takes the place of the generic \cref{eq_DSGEexoprocesses_cont}.

In fact, the system can be simplified to two equations in two unknowns, $y_t = (c_t,k_{t+1})'$.
We can drop $w_t$ because it appears only in \cref{eq_NCG_eqconds_w}, and we can substitute $r_{t+1}$ in \cref{eq_NCG_eqconds_k} by the expression from equation \cref{eq_NCG_eqconds_r}.
The resulting equations are:
\begin{align}
    c_t^{-\tau} &= \beta \mean_t \sbr{ c_{t+1}^{-\tau} (1-\delta+\alpha e^{z_{t+1}} k_{t+1}^{\alpha-1}) } \; , \label{eq_NCG_eqcondsv2_1}\\
    c_t + k_{t+1}-(1-\delta)k_t &= e^{z_t} k_t^\alpha \label{eq_NCG_eqcondsv2_2} \; .
\end{align}
This compactness greatly facilitates illustrating what happens if we consider second- or third-order linearized versions of this system of equations.

\paragraph*{Steady State}

\cref{eq_NCG_AR1forTFP} implies $z=0$. 
\cref{eq_NCG_eqcondsv2_1} implies $k=(\alpha/r)^{\frac{1}{1-\alpha}}$ with $r = \beta^{-1} - (1-\delta)$.
\cref{eq_NCG_eqcondsv2_2} implies then $c = k^\alpha - \delta k$.
Furthermore, it is useful to define $y=k^\alpha$ as the steady state value of output, $e^{z_t}k_t^\alpha$, as well as $k^* = k/y = \alpha/r$ and $c^* = c/y = 1-\alpha\delta/r$.

\paragraph*{First-Order Linearized System}

A first-order linearization of \cref{eq_NCG_eqcondsv2_1,eq_NCG_eqcondsv2_2} around the steady state represents the dynamics of the re-defined variables $\hat{c}_t = c_t/c -1$ and $\hat{k}_{t+1}= k_{t+1}/k-1$ by
\begin{align*}
    c^* \hat{c}_t + k^* \hat{k}_{t+1} - (1-\delta)k^* \hat{k}_t 
    &= \alpha k^* \hat{k}_t + yz_t \; ,\\
    -\tau \hat{c}_t
    &= -\tau \mean_t[\hat{c}_{t+1}] + \gamma\mean_t[ z_{t+1}] +\gamma\nu_1 \hat{k}_{t+1}\; .\footnotemark
\end{align*}
\footnotetext{We define $\gamma = 1-\beta(1-\delta)$ and
$\nu_1 = (\alpha-1)$.}
This system of equations can be regarded as a linear DSGE model, i.e. a model characterized by \cref{eq_DSGEeqconds} with a linear function $F$
and with re-defined endogenous variables $y_t = (\hat{c}_t, \hat{k}_{t+1})'$, 
while $e_t$, its law of motion and $\theta$ are unchanged relative to the original, non-linear setup.

We know $\mean_t[z_{t+1}] = \rho_z z_t$, and $\hat{k}_{t+1}$ is known at time $t$.
Then, for $y_t = (\hat{c}_t, \hat{k}_{t+1})'$ and $x_t = (y_t',z_t,\mean[\hat{c}_{t+1}])'$, we can write
\begin{align*}
    \bmat{c^* & k^* & -y & 0 \\
    -\tau & -\gamma \nu_1 & -\gamma\rho_z & \tau} x_t + 
    \bmat{0 & \psi^a_1 \\ 0 & 0} y_{t-1} = 0 \; .\footnotemark
\end{align*}
\footnotetext{
    We define $\psi^a_1 = -k^*(\alpha + 1-\delta)$.
}
Augmenting this system with the law of motion $z_t = \rho_z z_{t-1} + \sigma_z \epsilon_t$ and an expectational error $\eta_t = \hat{c}_t - \mean_{t-1}[\hat{c}_t]$, we get
\begin{align*}
    \bmat{c^* & k^* & -y & 0 \\
    -\tau & -\gamma \nu_1 & -\gamma\rho_z & \tau \\
    0 & 0 & 1 & 0 \\ 
    1 & 0 & 0 & 0} x_t + 
    \bmat{0 & \psi^a_1 & 0 & 0 \\ 
    0 & 0 & 0 & 0 \\
    0 & 0 & -\rho_z & 0 \\
    0 & 0 & 0 & -1} x_{t-1} + 
    \bmat{0 \\ 0 \\ -\sigma_z \\ 0} \epsilon_t 
    + \bmat{0 \\ 0 \\ 0 \\ -1} \eta_t = 0 \; .
\end{align*}
This is an RE-system with constant parameter-matrices:
\begin{align*}
    \Gamma_0(\theta) x_t = \Gamma_1(\theta) x_{t-1} + \Psi(\theta)\epsilon_t + \Pi \eta_t \; .
\end{align*}
In turn, using results from \citet{Sims2001} and \citet{lutkepohl2005}, we get a VAR(1) with constant parameters for $x_t$ as the non-explosive solution of this system, which in turn implies that $y^o_t \subseteq y_t \subset x_t$ follows a VARMA process with constant parameters (see proof of \cref{prop_linearDSGE}).

\paragraph*{Second-Order Linearized System}

Linearizing \cref{eq_NCG_eqcondsv2_1,eq_NCG_eqcondsv2_2} to second order gives
\begin{align*}
    c^* \hat{c}_t + k^* \hat{k}_{t+1} - (1-\delta)k^* \hat{k}_t 
    =& \alpha k^* \hat{k}_t + \frac{1}{2}\alpha\nu_1\hat{k}_t^2 + yz_t + \frac{1}{2}yz_t^2 + \alpha y \hat{k}_t z_t \; ,\\
    -\tau \hat{c}_t + \frac{1}{2}\omega_1\hat{c}_t^2 
    = \mean_t &\bigg[ -\tau \hat{c}_{t+1} + \frac{1}{2}\omega_1\hat{c}_{t+1}^2 + \gamma z_{t+1} + \frac{1}{2}\gamma z_{t+1}^2  +\gamma\nu_1 \hat{k}_{t+1}+\frac{1}{2} \gamma\nu_2 \hat{k}_{t+1}^2 \bigg. \\
    & \bigg. +\gamma \nu_1 \hat{k}_{t+1}z_{t+1} -\gamma \tau \hat{c}_{t+1} z_{t+1} -\gamma \tau \nu_1 \hat{c}_{t+1} \hat{k}_{t+1} \bigg]\; .\footnotemark
\end{align*}
\footnotetext{We define 
$\omega_1 = \tau(\tau+1)$ and
$\nu_2 = \nu_1(\alpha-2)$.}
Note that $\mean_t[z_{t+1}^2] = \rho_z^2 z_t^2 + \sigma_z^2$.
Define $x_t = (y_t', z_t,\mean[\hat{c}_{t+1}], \mean[\hat{c}_{t+1}^2],\mean[\hat{c}_{t+1}z_{t+1}])'$, where $y_t = (\hat{c}_t, \hat{k}_{t+1})'$ is unchanged.
We can write these two equations as
\begin{align*}
    \bmat{
    c^* & k^*  & \psi^b_{2t} & 0 & 0 & 0\\
    \psi^b_{3t} & \psi^b_{4t} & \psi^b_{5t}  & \psi^b_{6t} & - \frac{\nu_1}{2} & - \gamma\tau} x_t + 
    \bmat{0 & - \psi^b_{1t} \\ 0 & 0} y_{t-1} + \bmat{0 \\ -\frac{\gamma\sigma_z^2}{2}} = 0 \; .\footnotemark
\end{align*}
\footnotetext{
    We define
    \begin{alignat*}{2}
        &\psi^b_{1t} = \psi^a_1 - \frac{1}{2}\alpha\nu_1 \hat{k}_{t} - \alpha y z_t  \; , \quad
        &&\psi^b_{2t} = -y(1+\frac{1}{2}z_t) \; , \\
        &\psi^b_{3t} = -\tau + \frac{\nu_1}{2}\hat{c}_t \; , \quad
        &&\psi^b_{4t} = \gamma \nu_1(\rho_z z_t-1) -\frac{\gamma\nu_2}{2}\hat{k}_{t+1}  \; , \\
        &\psi^b_{5t} = -\gamma\rho_z(1+\frac{1}{2}\rho_z z_t) \; , \quad
        &&\psi^b_{6t} = \tau(1-\gamma\nu_1\hat{k}_{t+1}) \; . 
    \end{alignat*}
}
Augmenting this system with the law of motion $z_t = \rho_z z_{t-1} + \sigma_z \epsilon_t$ and expectational errors
\begin{align*}
    \eta_t = (\hat{c}_t, \hat{c}_t^2 , \hat{c}_tz_t)'
    - \mean_{t-1}\sbr{ (\hat{c}_t, \hat{c}_t^2 , \hat{c}_tz_t)' } \; ,
\end{align*}
we get an RE-system with time-varying parameter-matrices:
\begin{align*}
    \Gamma_{0t}(\theta) x_t = K(\theta) + \Gamma_{1t}(\theta) x_{t-1} + \Psi(\theta)\epsilon_t + \Pi \eta_t \; ,
\end{align*}
with $\Gamma_{0t}, K, \Gamma_{1t}, \Psi, \Pi$ respectively given by
\begin{align*}
    \bmat{
    c^* & k^* & \psi^b_{2t} & 0 & 0 & 0\\
    \psi^b_{3t} & \psi^b_{4t} & \psi^b_{5t}  & \psi^b_{6t} & - \frac{\nu_1}{2} & - \gamma\tau \\
    0 & 0 & 1 & 0 & 0 & 0\\
    1 & 0 & 0 & 0 & 0 & 0\\
    \hat{c}_t & 0 & 0 & 0 & 0 & 0\\
    z_t & 0 & 0 & 0 & 0 & 0} \; , \;
    \bmat{0 \\ -\frac{\gamma\sigma_z^2}{2} \\ 0 \\ 0 \\ 0 \\ 0} \; , \;
    \bmat{
    0 & \psi^b_{15t} & 0 &  \\
    0 & 0 & 0 & 0_{3\times 3} \\
    0 & 0 & -\rho_z & \\
    & 0_{3\times 3} & & -I_{3\times 3}} \; , \;
    \bmat{0 \\ 0 \\ -\sigma_z \\ 0 \\0 \\ 0} \; , \;
    \bmat{
        0_{3\times 3} \\
         -I_{3\times 3}
        }\; .
\end{align*}
Note that the elements of $\Gamma_{0t}$ and $\Gamma_{1t}$ are linear functions (first-order polynomials) of $(y_t',y_{t-1}',z_t)'$.
The non-explosive solution of this system yields a VAR(1) with time-varying parameters for $x_t$, which implies that $y^o_t \subseteq y_t \subset x_t$ follows a VARMA process with constant parameters (see proof of \cref{prop_nonlinearDSGE}).

\paragraph*{Third-Order Linearized System}

Relative to the second-order linearization, a third-order linearization adds the term 
$$ 
- \frac{1}{6}\alpha \nu_2\hat{k}_t^3 
- \frac{1}{6}y z_t^3 - 
\frac{1}{2} \alpha y \hat{k}_t z_t^2 - 
\frac{1}{2}\alpha \nu_1 y \hat{k}_t^2 z_t 
$$
to the LHS of the first equation, 
and it adds the one-step ahead expectation of the term
\begin{align*}
- \frac{1}{6}\omega_2 \hat{c}_t^3 
+ \frac{1}{6}\omega_2 \hat{c}_{t+1}^3
- \frac{1}{6}\gamma z_{t+1}^3
- \frac{1}{6} \gamma\nu_3 \hat{k}_{t+1}^3
- \frac{1}{2} \gamma\nu_1 \hat{k}_{t+1} z_{t+1}^2
- \frac{1}{2} \gamma\nu_2 \hat{k}_{t+1}^2 z_{t+1} \\
+ \frac{1}{2} \gamma \tau \hat{c}_{t+1} z_{t+1}^2
- \frac{1}{2} \gamma \omega_1 \hat{c}_{t+1}^2 z_{t+1}
+ \frac{1}{2} \gamma \tau\nu_2 \hat{c}_{t+1} \hat{k}_{t+1}^2
- \frac{1}{2} \gamma \omega_1\nu_1 \hat{c}_{t+1}^2 \hat{k}_{t+1}
+ \frac{1}{6} \tau\nu_1 z_{t+1} \hat{k}_{t+1} \hat{c}_{t+1}
\end{align*}
to the LHS of the second equation.\footnote{
    We define
    $\omega_2 = \omega_1(\tau+2)$ and
    $\nu_3 = \nu_2(\alpha-3)$.
}
Note that $\mean_t[z_{t+1}^3] = \rho_z^3 z_t^3 + 3\rho_z z_t \sigma_z^2$. 
Define $x_t = (y_t', z_t,\mean[\hat{c}_{t+1}], \mean[\hat{c}_{t+1}^2],\mean[\hat{c}_{t+1}z_{t+1}],\mean[\hat{c}_{t+1}^3],\mean[\hat{c}_{t+1} z_{t+1}^2],\mean[\hat{c}_{t+1}^2z_{t+1}])'$,
with $y_t = (\hat{c}_t, \hat{k}_{t+1})'$ unchanged. 
We can write these two equations as
\begin{align*}
    \bmat{
    c^* & k^*  & \psi^c_{2t} & 0 & 0 & 0 & 0 & 0 & 0\\
    \psi^c_{3t} & \psi^c_{4t} & \psi^c_{5t}  & \psi^c_{6t} & \psi^c_{7t} & \psi^c_{8t} & \frac{\omega_2}{6} & \frac{\gamma\tau}{2} & -\frac{\gamma\omega_1}{2}
    } x_t + 
    \bmat{0 & \psi^c_{1t} \\ 0 & 0} y_{t-1} + \bmat{0 \\ -\frac{\gamma\sigma_z^2}{2}} = 0 \; .\footnotemark
\end{align*}
\footnotetext{
    We define
    \begin{align*}
        \psi^c_{1t} &= \psi^b_{1t} - \frac{1}{6}\alpha \nu_2 \hat{k}_t^2 - \frac{1}{2}\alpha y z_t^2 - \frac{1}{2}\alpha \nu_1 y \hat{k}_t z_t \; , \\
        \psi^c_{4t} &= \psi^b_{4t} - \frac{1}{6}\gamma \nu_3 \hat{k}_{t+1}^2 - \frac{1}{2}\gamma\nu_1(\rho_z^2z_t^2 + \sigma_z^2) - \frac{1}{2}\gamma\nu_2 \hat{k}_{t+1}\rho_z z_t \; ,
    \end{align*}
    \begin{alignat*}{2}
        &\psi^c_{2t} = \psi^b_{2t} - \frac{1}{6}yz_t^2  \; , \quad
        &&\psi^c_{3t} = \psi^b_{3t} - \frac{1}{6}\omega_2 \hat{c}_t^2 \; , \\
        &\psi^c_{5t} = \psi^b_{5t} - \frac{1}{6}\gamma\rho_z^3z_t^2 - \frac{1}{2}\rho_z \sigma_z^2 \; , \quad
        &&\psi^c_{6t} = \psi^b_{6t} + \frac{1}{2}\gamma \tau \nu_1 \hat{k}_{t+1}^2  \; , \\
        &\psi^c_{7t} = - \frac{\nu_1}{2} - \frac{1}{2}\gamma\omega_1 \nu_1 \hat{k}_{t+1} \; , \quad
        &&\psi^c_{8t} = - \gamma\tau + \frac{1}{6}\tau \nu_1 \rho_z z_t \; . 
    \end{alignat*}
}
% \psi^c_{1t} &= \psi^b_{1t} - \frac{1}{6}\alpha \nu_2 \hat{k}_t^2 - \frac{1}{2}\alpha y z_t^2 - \frac{1}{2}\alpha \nu_1 y \hat{k}_t z_t \; , \\
% \psi^c_{2t} &= \psi^b_{2t} - \frac{1}{6}yz_t^2  \; , \\
% \psi^c_{3t} &= \psi^b_{3t} - \frac{1}{6}\omega_2 \hat{c}_t^2  \; , \\
% \psi^c_{4t} &= \psi^b_{4t} - \frac{1}{6}\gamma \nu_3 \hat{k}_{t+1}^2 - \frac{1}{2}\gamma\nu_1(\rho_z^2z_t^2 + \sigma_z^2) - \frac{1}{2}\gamma\nu_2 \hat{k}_{t+1}\rho_z z_t  \; , \\
% \psi^c_{5t} &= \psi^b_{5t} - \frac{1}{6}\gamma\rho_z^3z_t^2 - \frac{1}{2}\rho_z \sigma_z^2  \; , \\
% \psi^c_{6t} &= \psi^b_{6t} + \frac{1}{2}\gamma \tau \nu_1 \hat{k}_{t+1}^2  \; , \\
% \psi^c_{7t} &= - \frac{\nu_1}{2} - \frac{1}{2}\gamma\omega_1 \nu_1 \hat{k}_{t+1}  \; , \\
% \psi^c_{8t} &= - \gamma\tau + \frac{1}{6}\tau \nu_1 \rho_z z_t  \; .
%
Augmenting this system with the law of motion $z_t = \rho_z z_{t-1} + \sigma_z \epsilon_t$ and expectational errors
\begin{align*}
    \eta_t = (\hat{c}_t, \hat{c}_t^2, \hat{c}_tz_t, \hat{c}_t^3, \hat{c}_tz_t^2, \hat{c}_t^2 z_t)'
    - \mean_{t-1}\sbr{  (\hat{c}_t, \hat{c}_t^2, \hat{c}_tz_t, \hat{c}_t^3, \hat{c}_tz_t^2, \hat{c}_t^2 z_t)' } \; ,
\end{align*}
we get again an RE-system with time-varying parameter-matrices.
Here, $\Gamma_{0t}, K, \Gamma_{1t}, \Psi, \Pi$ are respectively given by
\begin{align*}
    \bmat{
    c^* & k^*  & \psi^c_{2t} & 0_{1\times 6}\\
    \psi^c_{3t} & \psi^c_{4t} & \psi^c_{5t}  & \tilde{\psi}_t' \\
    0 & 0 & 1 & 0_{1\times 6} \\
    1 & 0 & 0 & 0_{1\times 6} \\
    \hat{c}_t & 0 & 0 & 0_{1\times 6} \\
    z_t & 0 & 0 & 0_{1\times 6} \\
    \hat{c}_t^2 & 0 & 0 & 0_{1\times 6} \\
    z_t^2 & 0 & 0 & 0_{1\times 6} \\
    \hat{c}_t z_t & 0 & 0 & 0_{1\times 6}
    } \; , \; 
    \bmat{0 \\ -\frac{\gamma\sigma_z^2}{2} \\ 0 \\ 0 \\ 0 \\ 0 \\ 0 \\ 0 \\ 0} \; , \;
    \bmat{
    0 & \psi^c_{1t} & 0 & \\
    0 & 0 & 0 & 0_{3 \times 6} \\
    0 & 0 & -\rho_z & \\
    & 0_{6\times 3} & & -I_{6\times 6} 
    } \; , \; 
    \bmat{0 \\ 0 \\ -\sigma_z \\ 0 \\0 \\ 0 \\ 0 \\0 \\ 0} \; , \; 
    \bmat{
        0_{6\times 3} \\
        -I_{6\times 6} 
    } \; ,
\end{align*}
where $\tilde{\psi}_t = (\psi^c_{6t}, \psi^c_{7t}, \psi^c_{8t}, \frac{\omega_2}{6}, \frac{\gamma\tau}{2}, -\frac{\gamma\omega_1}{2})'$ is defined for notational convenience.
Note that the elements of $\Gamma_{0t}$ and $\Gamma_{1t}$ are second-order polynomials of $(y_t',y_{t-1}',z_t)'$.

%% -------------------------------------------------------------------------- %%
%\newpage

\subsection*{RBC Model}

\paragraph*{Setup}

The Real Business Cycle (RBC) model is equivalent to the NCG model, except that the household faces a non-trivial labor supply decision. 
Let the household's utility be $u(c,l) = \frac{c^{1-\tau}}{1-\tau} - \chi \frac{l^{1+\kappa}}{1+\kappa}$, and let the domain of $l_t$ be $\reals_+$.
The remaining aspects of the households' (or the firm's) optimization problem are unchanged.
Solving the budget constraint for $c_t$ and plugging it into the utility function, we get a maximization problem w.r.t. two variables only, $l_t$ and $k_{t+1}$.
The FOC w.r.t. $l_t$ gives $u_1(c_t,l_t) w_t + u_2(c_t,l_t) = 0$, i.e.
\begin{align}
    \chi l_t^\kappa = w_t c_t^{-\tau} \; . \label{eq_RBC_eqconds_l}
\end{align}
Combined with the unchanged equations
\begin{align}
    w_t &= (1-\alpha)e^{z_t}K_t^\alpha L_t^{-\alpha} \; , \label{eq_RBC_eqconds_w}\\
    r_t &= \alpha e^{z_t} K_t^{\alpha-1} L_t^{1-\alpha} \; , \label{eq_RBC_eqconds_r} \\
    c_t^{-\tau} &= \beta \mean_t \sbr{ c_{t+1}^{-\tau} (1+r_{t+1}-\delta) } \; ,\label{eq_RBC_eqconds_k} \\
    c_t + k_{t+1}-(1-\delta)k_t &= e^{z_t} k_t^\alpha l_t^{1-\alpha} \; , \label{eq_RBC_eqconds_marketclearing}
\end{align}
we get a system of five equations in five unknowns: $c_t, l_t, k_{t+1}, w_t, r_t$.

\paragraph*{Steady State}

\cref{eq_NCG_AR1forTFP} implies $z=0$. 
\cref{eq_RBC_eqconds_k} implies $r = \beta^{-1} - (1-\delta)$.
Combining \cref{eq_RBC_eqconds_w,eq_RBC_eqconds_r} (by solving them for $k/l$) yields
$$ w = (1-\alpha)^{\frac{1}{\alpha}} (\alpha/r)^{\frac{1}{1-\alpha}} \; . $$
Solving for $k$, $c$ and $l$ is a bit more involved. By \cref{eq_RBC_eqconds_r}, we get $k = (\alpha/r)^{\frac{1}{1-\alpha}} l$. Plugging this expression into \cref{eq_RBC_eqconds_marketclearing} yields $c = \zeta_1 l$, with $\zeta_1 = (\alpha/r)^{1/(1-\alpha)} \zeta_2$ and $\zeta_2 = \br{r/\alpha-\delta}$. Plugging this expression for $c$ as well as the above expression for $w$ into \cref{eq_RBC_eqconds_l} yields 
$$ l = \sbr{\zeta_3 \zeta_2^{-\tau} (\alpha/r)^{\frac{1}{1-\alpha}-\tau}  }^{\frac{1}{\tau+\kappa}} \; , $$
with $\zeta_3 = \chi^{-1}(1-\alpha)^{1/\alpha}$.
In turn, this implies
$$ c = \sbr{\zeta_3 \zeta_2^{\kappa} (\alpha/r)^{\frac{1}{1-\alpha}+\kappa}  }^{\frac{1}{\tau+\kappa}}  \quad \text{and} \quad k = \sbr{\zeta_3 \zeta_1^{-\tau} (\alpha/r)^{\frac{1+\kappa+\tau\alpha}{1-\alpha}}  }^{\frac{1}{\tau+\kappa}} \; . $$

\paragraph*{First-Order Linearization}

Linearizing \cref{eq_RBC_eqconds_w,eq_RBC_eqconds_r,eq_RBC_eqconds_l,eq_RBC_eqconds_k,eq_RBC_eqconds_marketclearing} around the steady state yields
\begin{align} 
    \hat{w}_t &= z_t + \alpha \hat{k}_t -\alpha \hat{l}_t \; , \label{eq_RBC_eqconds_w_lin} \\
    \hat{r}_t &= z_t + (\alpha-1) \hat{k}_t + (1-\alpha) \hat{l}_t \; , \label{eq_RBC_eqconds_r_lin} \\
    \chi \kappa \hat{l}_t &= \hat{w}_t - \tau \hat{c}_t \; , \label{eq_RBC_eqconds_l_lin} \\
    - \tau \hat{c}_t &= - \tau \mean_t[ \hat{c}_{t+1} ] + \lambda \mean_t[ \hat{r}_{t+1}] \; , \label{eq_RBC_eqconds_k_lin} \\
    c^* \hat{c}_t + k^* \hat{k}_{t+1} - (1-\delta)k^* \hat{k}_t &= z_t + \alpha \hat{k}_t + (1-\alpha)\hat{l}_t \; , \label{eq_RBC_eqconds_marketclearing_lin}
\end{align}
where $\lambda = 1-\beta(1+\delta)$, $c^* = c/y$ and $k^* = k/y$.\footnote{
    Combining the first three equations so as to cancel $\hat{l}_t$ and $\hat{w}_t$ gives us
    $$ \hat{r}_t = \frac{1+\chi\kappa}{\alpha+\chi\kappa}z_t - (1-\alpha)\frac{\chi\kappa}{\alpha+\chi\kappa}\hat{k}_t - (1-\alpha)\tau \frac{1}{\alpha+ \chi\kappa}\hat{c}_t \; . $$
    Hence, we can rewrite $\mean_t[ \hat{r}_{t+1}]$ as a linear combination of $\mean_t[ z_{t+1}] = \rho_z z_t$, $\hat{k}_{t+1}$ and $\mean_t[\hat{c}_{t+1}]$, leaving us with the latter as the only forward-looking variable.
}

%% -------------------------------------------------------------------------- %%
%\newpage
\subsection*{SOE-RBC Model}

Consider now an RBC model for a small open economy (SOE) under the standard, textbook case of perfect capital mobility and perfect labor immobility. Relative to the RBC model of a closed economy from above, 
the domestic household has access to a risk-free asset that pays a return of $(1+r^w_t)$ at time $t$ for each unit invested at time $(t-1)$. For illustrative purposes, let the world interest rate $r^w_t = r^w$ be fixed.
The household's problem becomes 
\begin{align*}
    V(z_t,k_t,b_t,K_t,B_t) = &\underset{c_t,l_t,k_{t+1},b_{t+1}}{\max}\; u(c_t,l_t) + \beta \mean_t \sbr{ V(z_{t+1},k_{t+1},b_{t+1},K_{t+1},B_{t+1}) } \\
&\quad \text{s.t.} \quad c_t + i_t + b_{t+1} = w_tl_t + r_t k_t + (1+r^w)b_t \; , \\
&\quad \text{\colW{s.t.}} \quad \colW{c_t +} i_t = k_{t+1}-(1-\delta)k_t \; , \\
&\quad \text{\colW{s.t.}} \quad \colW{c_t +} K_{t+1} = H_1(z_t,K_t,B_t) \; , \\
&\quad \text{\colW{s.t.}} \quad \colW{c_t +} B_{t+1} = H_2(z_t,K_t,B_t) \; ,
\end{align*}
where $b_t$ are the household's and $B_t$ are the aggregate holdings of the risk-free asset.
Again, we plug in the budget constraint for $c_t$. This yields a maximization problem w.r.t. $l_t$, $k_{t+1}$ and $b_{t+1}$ only.
The FOCs w.r.t. $l_t$ and $k_{t+1}$ are unchanged. The FOC w.r.t. $b_{t+1}$ yields
\begin{align}
    c_t^{-\tau} = \beta \mean_t \sbr{ c_{t+1}^{-\tau} (1+r^w ) } \; .\footnotemark \label{eq_OERBC_eqconds_b}
\end{align}
\footnotetext{Together with the FOC for $k_{t+1}$, it determines the domestic rental rate of capital to a level at which the household's expected, discounted returns on the risky investment in $k_{t+1}$ and on the risk-free investment in $b_{t+1}$ are equalized.}
The goods market clearing condition becomes
\begin{align} \label{eq_OERBC_eqconds_marketclearing}
    c_t + k_{t+1}-(1-\delta)k_t + b_{t+1} = e^{z_t} k_t^\alpha l_t^{1-\alpha} + (1+r^w)b_t\; .
\end{align}
Also, representativity of the household requires not only the evolution of $K_{t+1}$ to coincide with that of $k_{t+1}$, but the same holds for aggregate asset holdings $B_{t+1}$: 
$$ k_{t+1} = K_{t+1} = H(z_t,K_t,B_t) \; , \quad b_{t+1} = B_{t+1} = H_2(z_t,K_t,B_t) $$
All other optimality and equilibrium conditions are unchanged.

The six \cref{eq_RBC_eqconds_w,eq_RBC_eqconds_r,eq_RBC_eqconds_l,eq_RBC_eqconds_k,eq_OERBC_eqconds_b,eq_OERBC_eqconds_marketclearing} constitute a non-linear DSGE model, i.e. a model characterized by \cref{eq_DSGEeqconds} with a non-linear $F$. The endogenous variables are $y_t = (c_t, l_t, k_{t+1},b_{t+1}, w_t, r_t)'$, the exogenous variable is $e_t = z_t$. The time-invariant $r^w$ is treated as a parameter and included in $\theta = (\alpha,\beta,\tau,\chi,\kappa,\delta,r^w,\rho_z,\sigma_z^2)'$.\footnote{A time-varying world interest rate would be included in $e_t$.}

As discussed in \citep{SchmittGrohe-Uribe2003}, consumption in locally approximated versions of this SOE-RBC model follows a unit-root process, which is why the model is often augmented with a \myquote{stationarity-inducing device} such as an external debt-elastic interest rate (EDEIR) . Under an EDEIR, $r^w$ is replaced by 
$$ r^w_t = r^w + \phi(B_t) \; , $$
where $\phi(B_t)$ is a decreasing function of aggregate asset holdings $B_t$ and satisfies $\phi(b) = 0$ in steady state. For the sake of concreteness, following \citet{SchmittGrohe-Uribe2003}, let
\begin{align} \label{eq_OERBC_DEIR}
    r^w_t = r^w + \psi \br{ exp\{b - B_t\} - 1 } \; .
\end{align}

The EDEIR turns a constant parameter from $\theta$, $r^w$, into an endogenously time-varying parameter. 
The FOC w.r.t. $b_{t+1}$ (\cref{eq_OERBC_eqconds_b}) becomes
\begin{align}
    c_t^{-\tau} = \beta \mean_t \sbr{ c_{t+1}^{-\tau} (1+r^w_{t+1}) } \; .\footnotemark \label{eq_OERBC_eqconds_b_EDEIR}
\end{align}
Rational agents are aware of the time-variation in $r^w_t$ and form expectations about its future value. 
To accommodate the EDEIR, we change \cref{eq_OERBC_eqconds_b} to \cref{eq_OERBC_eqconds_b_EDEIR}, we include $r^w_t$ in $y_t$ and $\psi$ in $\theta$, and we leave $e_t$ unchanged. 
\footnote{
    Under an internal debt-elastic interest rate (IDEIR), $r^w_t = r^w + \psi \br{ exp\{b - b_t\} - 1 }$ is a function of the household's own asset holdings. This means that the household internalizes the time-variation and dependence of $r^w_t$ on $b_t$, leading to the FOC
    $$ c_t^{-\tau} = \beta \mean_t \sbr{ c_{t+1}^{-\tau} (1+r^w_{t+1} - \psi exp\{b-b_{t+1}\}) }  \; . $$
}

	%\input{part_appendix_model}

	%% -------------------------------------------------------------------------- %%
%% -------------------------------------------------------------------------- %%

\newpage
\section{Application: Exploring Common Factors}
\label{appsec_app_fcst}

% =======================================================
% =======================================================
% Tables
% =======================================================
%\section{Tables}
%\label{sec:Tables}
% \addcontentsline{toc}{chapter}{Appendix C}

The Table (\ref{tab:SeriesCorr}) contains series used to produce correlation results.
The transformation codes are: 1 - no transformation; 4 - logarithm;
5 - first difference of logarithm; 0 - variable only used for transforming other variables. All
the data are available at FRED web site.

%----------------------------------------------------------------------
%----------------------------------------------------------------------
\renewcommand{\arraystretch}{1.0}
% \begin{tiny}
%\begin{sciptsize}
\setlongtables
\begin{longtable}{|p{.25\textwidth}|p{.02\textwidth}|p{.70\textwidth}|}
\caption{Macroeconomic time series used for correlation analysis\label{tab:SeriesCorr}}\\
\hline % \hline %\caption{Data} \\
\bfseries{Mnemonics} & \bfseries{TC} & \bfseries{Description}\\
\hline % \hline
\endfirsthead
\hline 
\bfseries{Mnemonics} & \bfseries{TC} & \bfseries{Description}\\
\hline % \hline
\endhead
\hline 
\endfoot
\hline 
\endlastfoot
    GDPC  & 5     & Real Gross Domestic Product \\
    INDPRO & 5     & Industrial Production Index \\
    UNRATE & 1     & Civilian Unemployment Rate \\
    GPDIC96 & 5     & Real Gross Private Domestic Investment \\
    PNFI  & 5     & Private Nonresidential Fixed Investment \\
    GCEC96 & 5     & Real Government Consumption Expenditures \\
    CUmftg & 1     & Capacity Utilization: Manufacturing \\
    RPCE  & 5     & Real Personal Consumption Expenditures \\
    RPCEDG & 5     & Real Personal Consumption Expenditures: Durable Goods \\
    HOUST & 4     & Housing Starts: Total: New Privately Owned Housing Units Started \\
    PERMIT & 4     & New Private Housing Units Authorized by Building Permits \\
    AWHMAN & 1     & Average Weekly Hours: Manufacturing \\
    AWOTMAN & 1     & Average Weekly Overtime Hours: Manufacturing \\
    PAYEMS & 5     & All Employees: Total nonfarm \\
    MANEMP & 5     & All Employees: Manufacturing \\
    USMINE & 5     & All Employees: Mining and logging \\
    IC4WSA & 4     & 4-Week Moving Average of Initial Claims \\
    NAPM  & 1     & ISM Manufacturing: PMI Composite Index \\
    NAPMOI & 1     & ISM Manufacturing: New Orders Index \\
    NAPMEI & 1     & ISM Manufacturing: Employment Index \\
    NAPMII & 1     & ISM Manufacturing: Inventories Index \\
    NAPMSDI & 1     & ISM Manufacturing: Supplier Deliveries Index \\
    NAPMPRI & 1     & ISM Manufacturing: Prices Index \\
    GDPDEF & 5     & Gross Domestic Product: Implicit Price Deflator \\
    CPILFESL & 5     & Consumer Price Index for All Urban Consumers: All Items Less Food \& Energy \\
    CPIAUCSL & 5     & Consumer Price Index for All Urban Consumers: All Items \\
    PCEPI & 5     & Personal Consumption Expenditures: Chain-type Price Index \\
    FEDFUNDS & 1     & Effective Federal Funds Rate \\
    TB3MS & 1     & 3-Month Treasury Bill: Secondary Market Rate \\
    GS1   & 0     & 1-Year Treasury Constant Maturity Rate \\
    GS5   & 1     & 5-Year Treasury Constant Maturity Rate \\
    GS10  & 0     & 10-Year Treasury Constant Maturity Rate \\
    GS5-FFR & 1     &  \\
    GS1-FFR & 1     &  \\
    GS10-TB3MS & 1     &  \\
    DJIA  & 5     & Dow Jones Industrial Average \\
    DJCA  & 5     & Dow Jones Composite Average \\
    DJUA  & 5     & Dow Jones Utility Average \\
    DJTA  & 5     & Dow Jones Transportation Average \\
    SP500 & 5     & S\&P 500 Stock Price Index \\
    SP500-RV & 1     & S\&P500: realized volatility \\
    SP500-SK & 1     & S\&P500: realized skewness \\
    DJIA-RV & 1     & DJIA: realized volatility \\
    DJIA-SK & 1     & DJIA: realized skewness \\
    BAA   & 0     & Moody\'s Seasoned Baa Corporate Bond Yield \\
    AAA   & 0     & Moody\'s Seasoned Aaa Corporate Bond Yield \\
    BAA-GS10 & 1     &  \\
    BAA-AAA & 1     &  \\
    BAA-FFR & 1     &  \\
    OILPRICE & 5     & Spot Oil Price: West Texas Intermediate \\
    ConsMICH & 1     & University of Michigan: Consumer Sentiment \\
    TOTALSL & 5     & Total Consumer Credit Owned and Securitized, Outstanding  \\
    CMDEBT & 5     & Households and Nonprofit Organizations; Credit Market Instruments; Liability \\
    TCMDO & 5     & All Sectors; Credit Market Instruments; Liability  \\
    HCCSDODNS & 5     & Households and Nonprofit Organizations; Consumer Credit; Liability \\
    BCNSDODNS & 5     & Nonfinancial Corporate Business; Credit Market Instruments; Liability \\
    DODFS & 5     & Financial Business; Credit Market Instruments; Liability \\
    LOANINV & 5     & Bank Credit at All Commercial Banks \\
    BUSLOANS & 5     & Commercial and Industrial Loans, All Commercial Banks  \\
    CONSUMERLOANS & 5     & Consumer Loans at All Commercial Banks \\
    M1SL  & 5     & M1 Money Stock  \\
    M2SL  & 5     & M2 Money Stock  \\
    INVEST & 5     & Securities in Bank Credit at All Commercial Banks \\
\end{longtable}

%\begin{comment}
\begin{table}[h!]
  \centering
  \caption{Most correlated series with Grouped-Factor-Structure factors}
  \begin{center}
    \begin{tabular}{llllll}
    \hline\hline
Coefficient Factor 1 & GDPDEFinfl & CPILFELinfl & PCEPIgr & AWOTMAN & GS5 \\
      & 0.72  & 0.69  & 0.66  & 0.61  & 0.61 \\
Coefficient Factor 2 & ConsMICH & GS5 & NAPMPRI & AWOTMAN & TB3MS \\
      & 0.48  & 0.43  & 0.39  & 0.35  & 0.34 \\
Coefficient Factor 3 & BCNSDODNS & GS5 & TB3MS & FEDFUNDS & BUSLOANS \\
      & 0.52  & 0.39  & 0.36  & 0.35  & 0.35 \\
Coefficient Factor 4 & UNRATE & BAA-AAA & AWOTMAN & GS5 & IC4WSA \\
      & 0.68  & 0.53  & 0.51  & 0.46  & 0.45 \\
Covariance Factor 1 & GS5 & GDPDEFinfl & CPILFELinfl & TB3MS & FEDFUNDS \\
      & 0.69  & 0.69  & 0.68  & 0.67  & 0.66 \\
Covariance Factor 2 & GS5 & TB3MS & FEDFUNDS & NAPMPRI & IC4WSA \\
      & 0.55  & 0.43  & 0.39  & 0.34  & 0.30 \\
SV Factor 1 & AWOTMAN & AWHMAN & GDPDEFinfl & BAA-AAA & CPILFELinfl \\
      & 0.83  & 0.75  & 0.67  & 0.64  & 0.63 \\
SV Factor 2 & CUmftg & BAA-GS10 & PAYEMSgr & HOUST & IC4WSA \\
      & 0.69  & 0.66  & 0.51  & 0.50  & 0.49 \\
    \hline\hline
    \end{tabular}
\end{center}
\begin{flushleft}
{ \raggedright Notes: Coefficient Factors are the first four elements of $\hat f_t^{b}$; Covariance Factors are the first two elements of $\hat f_t^{a}$; SV Factors are the first two elements of $\hat f_t^h$; each series is described in Table \ref{tab:SeriesCorr}. The factors are identified, at best, up to a sign, so we only report absolute values of correlation coefficients.}
 \end{flushleft}
\label{tab:corrCase2}
\end{table}%

%\end{comment}

\begin{table}[h!]
  \centering
  \caption{Correlation between Common TVP Factor and share of finance measures}
  \begin{center}
    \begin{tabular}{lll}
    \hline \hline
          &  Common Factor & Common Factor Trend \\
    \hline
    Interconnectedness &  \multicolumn{1}{c}{0.98} & \multicolumn{1}{c}{0.99} \\
    No Defense & \multicolumn{1}{c}{0.98} & \multicolumn{1}{c}{0.98} \\
    No Farm No Defense & \multicolumn{1}{c}{0.98} & \multicolumn{1}{c}{0.98} \\
    Domestic &  \multicolumn{1}{c}{0.98} & \multicolumn{1}{c}{0.98} \\
    Share of Services &  \multicolumn{1}{c}{0.96} & \multicolumn{1}{c}{0.97} \\
    VA fin NIPA &  \multicolumn{1}{c}{0.97} & \multicolumn{1}{c}{0.98} \\
    WN fin NIPA & \multicolumn{1}{c}{0.98} & \multicolumn{1}{c}{0.98} \\
    \hline
    \end{tabular}%
    \end{center}
  \begin{flushleft}
{ \raggedright Notes: \emph{Common Factor} is the first element $\hat f_t^\theta$, and \emph{Common Factor Trend} is its trend obtained by HP filter. \emph{Interconnectedness} is one minus the measure of \emph{direct connectedness} between financial and real sectors from \cite{barattieri2019}. The rest are measures of the share of finance in GDP from \cite{Philippon2012}.}
 \end{flushleft}
  \label{tab:corrsharefinance}%
\end{table}%

\begin{table}[h!]
  \centering
  \caption{Dynamic correlations between HP-filter cyclical parts of Common Factor and share of finance measures against GDP cycle}
  \begin{center}
\begin{tabular}{lrrrrrrrr}%{lllllllll}
\hline \hline
      & Common  & Interconn- & No Defense & No Farm    & Domestic & Share of & VA fin & WN fin \\
      & Factor  & ectedness  & No Defense & No Defense & Domestic & Services & NIPA   & NIPA   \\
t-10  & 0.29  & 0.08  & 0.01  & 0.02  & 0.02  & -0.06 & -0.01 & 0.13 \\
t-9   & 0.34  & 0.08  & 0.08  & 0.08  & 0.08  & 0.00  & 0.05  & 0.15 \\
t-8   & 0.36  & 0.07  & 0.15  & 0.15  & 0.14  & 0.08  & 0.13  & 0.16 \\
t-7   & 0.36  & 0.06  & 0.22  & 0.21  & 0.20  & 0.16  & 0.20  & 0.17 \\
t-6   & 0.35  & 0.04  & 0.27  & 0.25  & 0.23  & 0.22  & 0.26  & 0.19 \\
t-5   & 0.32  & 0.00  & 0.29  & 0.27  & 0.25  & 0.26  & 0.29  & 0.21 \\
t-4   & 0.27  & -0.06 & 0.28  & 0.26  & 0.23  & 0.26  & 0.28  & 0.22 \\
t-3   & 0.22  & -0.12 & 0.22  & 0.20  & 0.16  & 0.21  & 0.23  & 0.21 \\
t-2   & 0.14  & -0.18 & 0.12  & 0.11  & 0.07  & 0.13  & 0.14  & 0.18 \\
t-1   & 0.06  & -0.22 & -0.02 & -0.03 & -0.06 & 0.02  & 0.00  & 0.12 \\
t-0   & -0.03 & -0.21 & -0.18 & -0.18 & -0.21 & -0.12 & -0.16 & 0.04 \\
t+1   & -0.14 & -0.20 & -0.30 & -0.29 & -0.31 & -0.22 & -0.28 & -0.05 \\
t+2   & -0.25 & -0.18 & -0.36 & -0.35 & -0.35 & -0.26 & -0.33 & -0.14 \\
t+3   & -0.36 & -0.16 & -0.34 & -0.33 & -0.33 & -0.25 & -0.32 & -0.20 \\
t+4   & -0.44 & -0.15 & -0.26 & -0.24 & -0.23 & -0.17 & -0.24 & -0.22 \\
t+5   & -0.46 & -0.14 & -0.13 & -0.12 & -0.11 & -0.06 & -0.12 & -0.22 \\
t+6   & -0.46 & -0.09 & -0.03 & -0.01 & 0.00  & 0.03  & -0.02 & -0.17 \\
t+7   & -0.43 & -0.03 & 0.05  & 0.06  & 0.08  & 0.09  & 0.06  & -0.11 \\
t+8   & -0.37 & 0.04  & 0.12  & 0.12  & 0.14  & 0.13  & 0.12  & -0.04 \\
t+9   & -0.31 & 0.11  & 0.15  & 0.15  & 0.17  & 0.14  & 0.14  & 0.03 \\
t+10  & -0.23 & 0.16  & 0.15  & 0.15  & 0.17  & 0.13  & 0.14  & 0.08 \\
\hline \hline
\end{tabular}%
    \end{center}
  \begin{flushleft}
{ \raggedright Notes: We estimate the cyclical part of each series using HP filter with $\lambda=1600$. The first row contains the correlation between each column series at $t-10$ GDP cycle at time $t$, and so on.}
 \end{flushleft}
  \label{tab:CycleProperties}%
\end{table}%

\begin{table}[h!]
  \centering
  \caption{Most correlated series with HP-filter cyclical components}
  \begin{center}
\begin{tabular}{llllll}
\hline \hline
Common Factor & PERMIT & HOUST & BCNSDODNS & TOTALSL & HCCSDODNS \\
      & 0.47  & 0.44  & 0.39  & 0.38  & 0.37 \\
Interconnectedness & GS10-TB3MS & BAA-FFR & DJUAret & M1SL & GS5-FFR \\
      & 0.44  & 0.40  & 0.36  & 0.34  & 0.33 \\
No Defense & BUSLOANS & NAPMOI & GS5-FFR & GS10-TB3MS & GDP \\
      & 0.37  & 0.32  & 0.26  & 0.24  & 0.23 \\
No Farm No Defense & BUSLOANS & NAPMOI & GS5-FFR & GS10-TB3MS & LOANS \\
      & 0.37  & 0.31  & 0.25  & 0.23  & 0.23 \\
Domestic & BUSLOANS & NAPMOI & GS5-FFR & GS10-TB3MS & LOANS \\
      & 0.40  & 0.30  & 0.27  & 0.27  & 0.25 \\
Share of Services & BUSLOANS & NAPMOI & SP500-RV & DJIA-RV & LOANS \\
      & 0.29  & 0.25  & 0.22  & 0.21  & 0.20 \\
VA fin NIPA & BUSLOANS & NAPMOI & GS5-FFR & INVEST & GDP \\
      & 0.36  & 0.31  & 0.24  & 0.22  & 0.22 \\
WN fin NIPA & NAPMOI & PAYEMS & NAPM & MANEMP & BCNSDODNS \\
      & 0.32  & 0.30  & 0.29  & 0.28  & 0.28 \\
\hline \hline
\end{tabular}%
\end{center}
\begin{flushleft}
{ \raggedright Notes: We estimate the cyclical part of each series using HP filter with $\lambda=1600$. \emph{Common Factor} is the first element $\hat f_t^\theta$, and \emph{Common Factor Trend} is its trend obtained by HP filter. \emph{Interconnectedness} is one minus the measure of \emph{direct connectedness} between financial and real sectors from Barattieri, Eden and Stevanovic (2013). The rest are measures of the share of finance in GDP from Phillipon (2012). Each series is described in Table (\ref{tab:SeriesCorr})}
\end{flushleft}
\label{tab:corrCycleShareFinance}
\end{table}%

\begin{table}[h!]
  \centering
  \caption{Correlation between Stochastic Volatility Factors and uncertainty measures}
  \begin{center}
  \begin{tabular}{lll}
\hline \hline
 & \multicolumn{2}{c}{JLN firm uncertainty} \\
\hline
      & Stochastic Volatility Factor 1 & Stochastic Volatility Factor 2 \\
\hline
U(h=1) & \multicolumn{1}{c}{0.17} & \multicolumn{1}{c}{0.21} \\
U(h=2) & \multicolumn{1}{c}{0.22} & \multicolumn{1}{c}{0.19} \\
U(h=3) & \multicolumn{1}{c}{0.24} & \multicolumn{1}{c}{0.18} \\
U(h=4) & \multicolumn{1}{c}{0.25} & \multicolumn{1}{c}{0.18} \\
U(h=5) & \multicolumn{1}{c}{0.25} & \multicolumn{1}{c}{0.18} \\
U(h=6) & \multicolumn{1}{c}{0.24} & \multicolumn{1}{c}{0.18} \\
\hline\hline
 & \multicolumn{2}{c}{JLN macro uncertainty} \\
\hline
      & Stochastic Volatility Factor 1 & Stochastic Volatility Factor 2 \\
\hline
U(h=1) & \multicolumn{1}{c}{0.84} & \multicolumn{1}{c}{0.11} \\
U(h=2) & \multicolumn{1}{c}{0.84} & \multicolumn{1}{c}{0.12} \\
U(h=3) & \multicolumn{1}{c}{0.85} & \multicolumn{1}{c}{0.15} \\
U(h=4) & \multicolumn{1}{c}{0.85} & \multicolumn{1}{c}{0.17} \\
U(h=5) & \multicolumn{1}{c}{0.85} & \multicolumn{1}{c}{0.19} \\
U(h=6) & \multicolumn{1}{c}{0.85} & \multicolumn{1}{c}{0.21} \\
U(h=7) & \multicolumn{1}{c}{0.86} & \multicolumn{1}{c}{0.23} \\
U(h=8) & \multicolumn{1}{c}{0.85} & \multicolumn{1}{c}{0.24} \\
U(h=9) & \multicolumn{1}{c}{0.85} & \multicolumn{1}{c}{0.26} \\
U(h=10) & \multicolumn{1}{c}{0.85} & \multicolumn{1}{c}{0.28} \\
U(h=11) & \multicolumn{1}{c}{0.85} & \multicolumn{1}{c}{0.29} \\
U(h=12) & \multicolumn{1}{c}{0.85} & \multicolumn{1}{c}{0.31} \\
\hline\hline
 & \multicolumn{2}{c|}{Bloom policy uncertainty} \\
\hline
      & Stochastic Volatility Factor 1 & Stochastic Volatility Factor 2 \\
\hline
Uncertainty & \multicolumn{1}{c}{0.39} & \multicolumn{1}{c}{0.49} \\
\hline
\end{tabular}%
    \end{center}
  \begin{flushleft}
{ \raggedright Notes: Stochastic Volatility Factors 1 and 2 are the first two elements of $\hat f_{1t}^h$ respectively. The JLN firm uncertainty measures contain common unforecastable components for horizons of 1 to 6 quarters, and the JLN macro uncertainty is constructed for horizons of 1 to 12 months. Since the macro uncertainties are measured in months in \cite{jurado2015}, we aggregate  them to quarterly frequency.  \cite{baker2013} calculate the policy uncertainty in monthly frequency from 1985M01 and we aggregate it to quarters.}
 \end{flushleft}
  \label{tab:corrSVuncertainty}%
\end{table}%

\clearpage
%\newpage
\section{Application: Forecasting Details}
\label{appsec_app_fcst}

%=======================================================
% RMSFE
%\input{\plotPathTwo/PaperTableRMSFE_Model_1_Prior_3_Lag_1_FcstNeval_98_SamplingMode_1_DifMode_1_FacMode_1}
%\clearpage
\begin{table}[!h]\centering
\caption{Point Forecast RMSE's ($2$ lag models)}
\label{fig:RMSFEModel1Prior3Lag2FcstNeval98SamplingMode1DifMode1FacMode1}
\begin{center}
\begin{tabular}{lllllllll}
\hline\hline
{\textbf{Model}} & {\textbf{1Q}}  & {\textbf{2Q}}   & {\textbf{3Q}}  & {\textbf{4Q}}  & {\textbf{5Q}}  & {\textbf{6Q}}  & {\textbf{7Q}}   & {\textbf{8Q}} \\ 
\hline
\multicolumn{9}{c}{{\textbf{Real GDP}}} \\ 
\hline
TVP-VAR-SV & $0.62$ &	$0.66$  & 	$0.72$  & 	$0.71$  & 	$0.72$  & 	$0.68$  & 	$0.69$  & 	$0.67$  \\ 
GF-TVP-VAR-SV & $1.04^{\ast\ast}$  &	$1.04$  & 	$1.04$  & 	$1.03$  & 	$1.03$  & 	$1.05$  & 	$1.02$  & 	$1.02$ \\ 
CF-TVP-VAR-SV & $1.08$  &	$1.06$  & 	$1.04$  & 	$1.02$  & 	$1.02$  & 	$1.04$  & 	$1.02$  & 	$1.01$ \\ 
\hline
\multicolumn{9}{c}{{\textbf{Inflation}}} \\ 
\hline
TVP-VAR-SV & $0.21$ &	$0.22$  & 	$0.24$  & 	$0.25$  & 	$0.27$  & 	$0.28$  & 	$0.29$  & 	$0.30$  \\ 
GF-TVP-VAR-SV & $0.98$  &	$1.00$  & 	$0.98$  & 	$0.98$  & 	$0.98$  & 	$0.99$  & 	$0.98$  & 	$0.99$ \\ 
CF-TVP-VAR-SV & $0.96$  &	$0.96$  & 	$0.94$  & 	$0.95$  & 	$0.96$  & 	$0.96$  & 	$0.96$  & 	$0.98$ \\ 
\hline
\multicolumn{9}{c}{{\textbf{FFR}}} \\ 
\hline
TVP-VAR-SV & $0.38$ &	$0.78$  & 	$1.14$  & 	$1.49$  & 	$1.78$  & 	$2.04$  & 	$2.26$  & 	$2.44$  \\ 
GF-TVP-VAR-SV & $1.00$  &	$1.00$  & 	$1.01$  & 	$1.02$  & 	$1.03$  & 	$1.03$  & 	$1.03$  & 	$1.03$ \\ 
CF-TVP-VAR-SV & $1.02^{\ast}$  &	$1.02$  & 	$1.02$  & 	$1.03$  & 	$1.04$  & 	$1.04$  & 	$1.04$  & 	$1.04$ \\ 
\hline
\multicolumn{9}{c}{{\textbf{Credit Spread}}} \\ 
\hline
TVP-VAR-SV & $0.36$ &	$0.58$  & 	$0.71$  & 	$0.79$  & 	$0.84$  & 	$0.88$  & 	$0.90$  & 	$0.92$  \\ 
GF-TVP-VAR-SV & $1.00$  &	$1.00$  & 	$1.01$  & 	$1.02$  & 	$1.02^{\ast}$  & 	$1.02$  & 	$1.02$  & 	$1.02$ \\ 
CF-TVP-VAR-SV & $1.01$  &	$1.01$  & 	$1.01$  & 	$1.01$  & 	$1.01$  & 	$1.01$  & 	$1.02$  & 	$1.02$ \\ 
\hline\hline
\end{tabular}
\end{center}
\begin{flushleft}
{\raggedright Notes: We show RMSEs for the benchmark TVP-VAR-SV model in the first line of each panel, and RMSE ratios in the subsequent lines to the respective models. In parentheses we show p-values of Diebold-Mariano tests of equal MSE against the one-sided alternative that the model with time-varying volatility is more accurate, obtained using standard normal critical values. We compute the standard errors entering the Diebold-Mariano statistics using Newey-West.}
\end{flushleft}
\end{table}

\clearpage
\begin{table}[!t]\centering
\caption{Point Forecast RMSE's ($5$ lag models)}
\label{fig:RMSFEModel1Prior3Lag5FcstNeval98SamplingMode1DifMode1FacMode1}
\begin{center}
\begin{tabular}{lllllllll}
\hline\hline
{\textbf{Model}} & {\textbf{1Q}}  & {\textbf{2Q}}   & {\textbf{3Q}}  & {\textbf{4Q}}  & {\textbf{5Q}}  & {\textbf{6Q}}  & {\textbf{7Q}}   & {\textbf{8Q}} \\ 
\hline
\multicolumn{9}{c}{{\textbf{Real GDP}}} \\ 
\hline
TVP-VAR-SV & $0.75$ &	$0.82$  & 	$0.88$  & 	$0.84$  & 	$0.85$  & 	$0.75$  & 	$0.82$  & 	$0.92$  \\ 
GF-TVP-VAR-SV & $1.13$  &	$0.90$  & 	$0.89$  & 	$0.88$  & 	$0.85$  & 	$0.89$  & 	$0.85$  & 	$0.77$ \\ 
CF-TVP-VAR-SV & $1.18$  &	$0.91$  & 	$0.93$  & 	$0.87$  & 	$0.84$  & 	$0.89$  & 	$0.85$  & 	$0.76$ \\ 
\hline
\multicolumn{9}{c}{{\textbf{Inflation}}} \\ 
\hline
TVP-VAR-SV & $0.25$ &	$0.27$  & 	$0.25$  & 	$0.26$  & 	$0.28$  & 	$0.29$  & 	$0.32$  & 	$0.37$  \\ 
GF-TVP-VAR-SV & $0.91$  &	$0.85^{\ast\ast\ast}$  & 	$0.85^{\ast\ast}$  & 	$0.83^{\ast\ast\ast}$  & 	$0.84^{\ast}$  & 	$0.79^{\ast\ast}$  & 	$0.79^{\ast\ast\ast}$  & 	$0.75^{\ast\ast\ast}$ \\ 
CF-TVP-VAR-SV & $0.91$  &	$0.86^{\ast\ast\ast}$  & 	$0.88^{\ast\ast}$  & 	$0.86^{\ast\ast}$  & 	$0.90$  & 	$0.85^{\ast}$  & 	$0.84^{\ast\ast}$  & 	$0.81^{\ast\ast\ast}$ \\ 
\hline
\multicolumn{9}{c}{{\textbf{FFR}}} \\ 
\hline
TVP-VAR-SV & $0.40$ &	$0.82$  & 	$1.19$  & 	$1.56$  & 	$1.90$  & 	$2.22$  & 	$2.52$  & 	$2.83$  \\ 
GF-TVP-VAR-SV & $1.13$  &	$1.18$  & 	$1.17$  & 	$1.12$  & 	$1.08$  & 	$1.04$  & 	$1.01$  & 	$0.97$ \\ 
CF-TVP-VAR-SV & $1.17$  &	$1.25^{\ast}$  & 	$1.25^{\ast}$  & 	$1.20$  & 	$1.16$  & 	$1.11$  & 	$1.07$  & 	$1.03$ \\ 
\hline
\multicolumn{9}{c}{{\textbf{Credit Spread}}} \\ 
\hline
TVP-VAR-SV & $0.37$ &	$0.62$  & 	$0.75$  & 	$0.85$  & 	$0.95$  & 	$1.02$  & 	$1.04$  & 	$1.08$  \\ 
GF-TVP-VAR-SV & $1.03$  &	$0.98$  & 	$0.94$  & 	$0.90$  & 	$0.86$  & 	$0.83$  & 	$0.81$  & 	$0.79$ \\ 
CF-TVP-VAR-SV & $1.06$  &	$1.01$  & 	$0.96$  & 	$0.93$  & 	$0.88$  & 	$0.85$  & 	$0.85$  & 	$0.82$ \\ 
\hline\hline
\end{tabular}
\end{center}
\begin{flushleft}
{\raggedright Notes: We show RMSEs for the benchmark TVP-VAR-SV model in the first line of each panel, and RMSE ratios in the subsequent lines to the respective models. In parentheses we show p-values of Diebold-Mariano tests of equal MSE against the one-sided alternative that the model with time-varying volatility is more accurate, obtained using standard normal critical values. We compute the standard errors entering the Diebold-Mariano statistics using Newey-West.}
\end{flushleft}
\end{table}

%=======================================================
%% Forecast Interval Evaluation Christoffersen Coverage/Length
%\input{\plotPathTwo/Paper2TableInFeval_Model_1_Prior_3_Lag_1_FcstNeval_98_SamplingMode_1_DifMode_1_FacMode_1}
\clearpage
\begin{table}[!t]\centering
\caption{$68\%$ Interval Forecast Evaluation - Coverage and Length (2 lags)}
\label{fig:InFevalModel1Prior3Lag2FcstNeval98SamplingMode1DifMode1FacMode1}
\begin{center}
\begin{tabular}{lcc|cc|cc|cc}
\hline\hline
\textbf{Model} & \multicolumn{2}{c}{\textbf{1Q}}  & \multicolumn{2}{c}{\textbf{2Q}}  & \multicolumn{2}{c}{\textbf{4Q}}  & \multicolumn{2}{c}{\textbf{8Q}} \\ 
\hline
\multicolumn{9}{c}{\textbf{Real GDP}} \\ 
\hline
TVP-VAR-SV & $0.68$  & $[1.18]$ &	$0.66$ & $[1.26]$ &	$0.79^{\ast\ast}$ & $[1.41]$ &	$0.82^{\ast\ast}$  & $[1.65]$ \\ 
GF-TVP-VAR-SV & $0.66$  & $[1.19]$ &	$0.66$ & $[1.26]$ &	$0.72$ & $[1.39]$ &	$0.81^{\ast\ast}$  & $[1.61]$ \\ 
CF-TVP-VAR-SV & $0.71$  & $[1.19]$ &	$0.67$ & $[1.24]$ &	$0.70$ & $[1.38]$ &	$0.83^{\ast\ast}$  & $[1.58]$ \\ 
\hline
\multicolumn{9}{c}{\textbf{Inflation}} \\ 
\hline
TVP-VAR-SV & $0.78^{\ast\ast}$  & $[0.44]$ &	$0.71$ & $[0.50]$ &	$0.82^{\ast\ast\ast}$ & $[0.63]$ &	$0.83^{\ast\ast\ast}$  & $[0.82]$ \\ 
GF-TVP-VAR-SV & $0.77^{\ast\ast}$  & $[0.45]$ &	$0.69$ & $[0.50]$ &	$0.79^{\ast\ast}$ & $[0.63]$ &	$0.81^{\ast\ast\ast}$  & $[0.82]$ \\ 
CF-TVP-VAR-SV & $0.77^{\ast}$  & $[0.45]$ &	$0.76$ & $[0.51]$ &	$0.83^{\ast\ast\ast}$ & $[0.64]$ &	$0.83^{\ast\ast\ast}$  & $[0.82]$ \\ 
\hline
\multicolumn{9}{c}{\textbf{FFR}} \\ 
\hline
TVP-VAR-SV & $0.69$  & $[0.76]$ &	$0.70$ & $[1.54]$ &	$0.69$ & $[2.92]$ &	$0.70$  & $[4.75]$ \\ 
GF-TVP-VAR-SV & $0.71$  & $[0.76]$ &	$0.71$ & $[1.54]$ &	$0.67$ & $[2.91]$ &	$0.67$  & $[4.73]$ \\ 
CF-TVP-VAR-SV & $0.73$  & $[0.81]$ &	$0.76$ & $[1.57]$ &	$0.67$ & $[2.88]$ &	$0.65$  & $[4.63]$ \\ 
\hline
\multicolumn{9}{c}{\textbf{Credit Spread}} \\ 
\hline
TVP-VAR-SV & $0.67$  & $[0.46]$ &	$0.60$ & $[0.73]$ &	$0.66$ & $[1.09]$ &	$0.69$  & $[1.59]$ \\ 
GF-TVP-VAR-SV & $0.69$  & $[0.46]$ &	$0.61$ & $[0.72]$ &	$0.65$ & $[1.09]$ &	$0.69$  & $[1.56]$ \\ 
CF-TVP-VAR-SV & $0.66$  & $[0.42]$ &	$0.62$ & $[0.66]$ &	$0.60$ & $[1.00]$ &	$0.66$  & $[1.47]$ \\ 
\hline\hline
\end{tabular}
\end{center}
\begin{flushleft}
{\raggedright Notes: The table reports coverage rates of $68\%$ interval forecasts at horizons of 1, 2, 4, and 8 quarters ahead, obtained from VARs with two lags. Reported values are the proportion of realized observations falling inside the predictive intervals. A value close to 0.68 indicates accurate unconditional coverage. Asterisks denote significance levels of Christoffersen’s likelihood-ratio tests for correct coverage ($p<0.10; , p<0.05; , p<0.01$).}
\end{flushleft}
\end{table}

\clearpage
\begin{table}[!t]\centering
\caption{$68\%$ Interval Forecast Evaluation - Coverage and Length (5 lags)}
\label{fig:InFevalModel1Prior3Lag5FcstNeval98SamplingMode1DifMode1FacMode1}
\begin{center}
\begin{tabular}{lcc|cc|cc|cc}
\hline\hline
\textbf{Model} & \multicolumn{2}{c}{\textbf{1Q}}  & \multicolumn{2}{c}{\textbf{2Q}}  & \multicolumn{2}{c}{\textbf{4Q}}  & \multicolumn{2}{c}{\textbf{8Q}} \\ 
\hline
\multicolumn{9}{c}{\textbf{Real GDP}} \\ 
\hline
TVP-VAR-SV & $0.78^{\ast\ast}$  & $[2.01]$ &	$0.87^{\ast\ast\ast}$ & $[2.35]$ &	$0.95^{\ast\ast\ast}$ & $[3.24]$ &	$0.99^{\ast\ast\ast}$  & $[5.54]$ \\ 
GF-TVP-VAR-SV & $0.77^{\ast}$  & $[1.87]$ &	$0.83^{\ast\ast\ast}$ & $[2.02]$ &	$0.93^{\ast\ast\ast}$ & $[2.69]$ &	$0.97^{\ast\ast\ast}$  & $[4.31]$ \\ 
CF-TVP-VAR-SV & $0.71$  & $[1.78]$ &	$0.80^{\ast\ast}$ & $[1.93]$ &	$0.92^{\ast\ast\ast}$ & $[2.57]$ &	$0.96^{\ast\ast\ast}$  & $[4.15]$ \\ 
\hline
\multicolumn{9}{c}{\textbf{Inflation}} \\ 
\hline
TVP-VAR-SV & $0.82^{\ast\ast\ast}$  & $[0.66]$ &	$0.83^{\ast\ast\ast}$ & $[0.78]$ &	$0.93^{\ast\ast\ast}$ & $[1.08]$ &	$0.98^{\ast\ast\ast}$  & $[1.99]$ \\ 
GF-TVP-VAR-SV & $0.83^{\ast\ast\ast}$  & $[0.62]$ &	$0.91^{\ast\ast\ast}$ & $[0.69]$ &	$0.91^{\ast\ast\ast}$ & $[0.94]$ &	$0.98^{\ast\ast\ast}$  & $[1.62]$ \\ 
CF-TVP-VAR-SV & $0.82^{\ast\ast\ast}$  & $[0.59]$ &	$0.86^{\ast\ast\ast}$ & $[0.66]$ &	$0.90^{\ast\ast\ast}$ & $[0.90]$ &	$0.98^{\ast\ast\ast}$  & $[1.59]$ \\ 
\hline
\multicolumn{9}{c}{\textbf{FFR}} \\ 
\hline
TVP-VAR-SV & $0.80^{\ast\ast\ast}$  & $[0.91]$ &	$0.80^{\ast\ast}$ & $[2.00]$ &	$0.82^{\ast\ast}$ & $[3.98]$ &	$0.81^{\ast}$  & $[7.51]$ \\ 
GF-TVP-VAR-SV & $0.66$  & $[0.90]$ &	$0.64$ & $[1.94]$ &	$0.65$ & $[3.72]$ &	$0.72$  & $[6.65]$ \\ 
CF-TVP-VAR-SV & $0.61$  & $[0.80]$ &	$0.66$ & $[1.76]$ &	$0.60$ & $[3.46]$ &	$0.71$  & $[6.40]$ \\ 
\hline
\multicolumn{9}{c}{\textbf{Credit Spread}} \\ 
\hline
TVP-VAR-SV & $0.72$  & $[0.59]$ &	$0.72$ & $[1.00]$ &	$0.77$ & $[1.59]$ &	$0.92^{\ast\ast\ast}$  & $[2.69]$ \\ 
GF-TVP-VAR-SV & $0.73$  & $[0.56]$ &	$0.72$ & $[0.92]$ &	$0.76$ & $[1.39]$ &	$0.89^{\ast\ast\ast}$  & $[2.15]$ \\ 
CF-TVP-VAR-SV & $0.67$  & $[0.49]$ &	$0.64$ & $[0.83]$ &	$0.71$ & $[1.29]$ &	$0.80$  & $[2.03]$ \\ 
\hline\hline
\end{tabular}
\end{center}
\begin{flushleft}
{\raggedright Notes: The table reports coverage rates of $68\%$ interval forecasts at horizons of 1, 2, 4, and 8 quarters ahead, obtained from VARs with five lags. Reported values are the proportion of realized observations falling inside the predictive intervals. A value close to 0.68 indicates accurate unconditional coverage. Asterisks denote significance levels of Christoffersen’s likelihood-ratio tests for correct coverage ($p<0.10; , p<0.05; , p<0.01$).}
\end{flushleft}
\end{table}

%=======================================================
%% Forecast Interval Evaluation Christoffersen Likelihood-Ratio Tests

% Horizon 1
%\input{\plotPathTwo/Table2InFevalCTestHorizon_1_Model_1_Prior_3_Lag_1_FcstNeval_98_SamplingMode_1_DifMode_1_FacMode_1}

% %%% PROBLEM HERE; MM, 2512
 \clearpage
 \begin{table}[!t]\centering
\caption{$68\%$ Christoffersen Likelihood-Ratio Tests ($2$ Lags, horizon $1$)}
\label{tab:InFevalCTestHorizon1Model1Prior3Lag2FcstNeval98SamplingMode1DifMode1FacMode1}
\begin{center}
%\begin{tabular}{l *{3}{d{6.6}} }
\begin{tabular}{lccc}
\hline\hline
\multicolumn{1}{l}{\textbf{Model}} & \multicolumn{1}{c}{\textbf{Coverage}}  & \multicolumn{1}{c}{\textbf{Independence}}  & \multicolumn{1}{c}{\textbf{Joint}} \\ 
\hline
\multicolumn{4}{c}{\textbf{Real GDP}} \\ 
\hline
TVP-VAR-SV & $13.32^{\ast\ast\ast}$ & $1.54$ & $14.86^{\ast\ast\ast}$ \\ 
GF-TVP-VAR-SV & $16.93^{\ast\ast\ast}$ & $0.75$ & $17.69^{\ast\ast\ast}$ \\ 
CF-TVP-VAR-SV & $8.64^{\ast\ast\ast}$ & $0.83$ & $9.47^{\ast\ast\ast}$ \\ 
\hline
\multicolumn{4}{c}{\textbf{Inflation}} \\ 
\hline
TVP-VAR-SV & $2.12$ & $1.17$ & $3.28$ \\ 
GF-TVP-VAR-SV & $3.84^{\ast\ast}$ & $0.07$ & $3.91$ \\ 
CF-TVP-VAR-SV & $3.84^{\ast\ast}$ & $0.09$ & $3.94$ \\ 
\hline
\multicolumn{4}{c}{\textbf{FFR}} \\ 
\hline
TVP-VAR-SV & $13.32^{\ast\ast\ast}$ & $0.93$ & $14.26^{\ast\ast\ast}$ \\ 
GF-TVP-VAR-SV & $10.10^{\ast\ast\ast}$ & $2.48$ & $12.58^{\ast\ast\ast}$ \\ 
CF-TVP-VAR-SV & $7.28^{\ast\ast\ast}$ & $9.13^{\ast\ast\ast}$ & $16.41^{\ast\ast\ast}$ \\ 
\hline
\multicolumn{4}{c}{\textbf{Credit Spread}} \\ 
\hline
TVP-VAR-SV & $15.08^{\ast\ast\ast}$ & $0.33$ & $15.41^{\ast\ast\ast}$ \\ 
GF-TVP-VAR-SV & $13.32^{\ast\ast\ast}$ & $0.38$ & $13.70^{\ast\ast\ast}$ \\ 
CF-TVP-VAR-SV & $18.88^{\ast\ast\ast}$ & $1.55$ & $20.43^{\ast\ast\ast}$ \\ 
\hline\hline
\end{tabular}
\end{center}
\begin{flushleft}
{\raggedright Notes: The table reports Christoffersen’s likelihood-ratio (LR) test statistics for 68\% interval forecasts at horizon of 1 quarter ahead, obtained from VARs with two lags. The tests assess unconditional coverage, independence of violations, and their joint hypothesis. Reported statistics are compared to the $\chi^2$ distribution with the appropriate degrees of freedom. Asterisks denote significance levels ($p<0.10; , p<0.05; , p<0.01$).}
\end{flushleft}
\end{table}

% \input{\plotPathTwo/Table2InFevalCTestHorizon_1_Model_1_Prior_3_Lag_3_FcstNeval_98_SamplingMode_1_DifMode_1_FacMode_1}
% \input{\plotPathTwo/Table2InFevalCTestHorizon_1_Model_1_Prior_3_Lag_4_FcstNeval_98_SamplingMode_1_DifMode_1_FacMode_1}

% %%% PROBLEM HERE; MM, 2512
 \clearpage
 \begin{table}[!t]\centering
\caption{$68\%$ Christoffersen Likelihood-Ratio Tests ($5$ Lags, horizon $1$)}
\label{tab:InFevalCTestHorizon1Model1Prior3Lag5FcstNeval98SamplingMode1DifMode1FacMode1}
\begin{center}
%\begin{tabular}{l *{3}{d{6.6}} }
\begin{tabular}{lccc}
\hline\hline
\multicolumn{1}{l}{\textbf{Model}} & \multicolumn{1}{c}{\textbf{Coverage}}  & \multicolumn{1}{c}{\textbf{Independence}}  & \multicolumn{1}{c}{\textbf{Joint}} \\ 
\hline
\multicolumn{4}{c}{\textbf{Real GDP}} \\ 
\hline
TVP-VAR-SV & $2.92^{\ast}$ & $0.00$ & $2.92$ \\ 
GF-TVP-VAR-SV & $2.92^{\ast}$ & $0.99$ & $3.91$ \\ 
CF-TVP-VAR-SV & $10.10^{\ast\ast\ast}$ & $0.20$ & $10.30^{\ast\ast\ast}$ \\ 
\hline
\multicolumn{4}{c}{\textbf{Inflation}} \\ 
\hline
TVP-VAR-SV & $0.16$ & $0.69$ & $0.85$ \\ 
GF-TVP-VAR-SV & $0.16$ & $1.81$ & $1.97$ \\ 
CF-TVP-VAR-SV & $0.16$ & $0.32$ & $0.49$ \\ 
\hline
\multicolumn{4}{c}{\textbf{FFR}} \\ 
\hline
TVP-VAR-SV & $1.44$ & $1.27$ & $2.70$ \\ 
GF-TVP-VAR-SV & $18.88^{\ast\ast\ast}$ & $7.60^{\ast\ast\ast}$ & $26.48^{\ast\ast\ast}$ \\ 
CF-TVP-VAR-SV & $29.96^{\ast\ast\ast}$ & $13.33^{\ast\ast\ast}$ & $43.29^{\ast\ast\ast}$ \\ 
\hline
\multicolumn{4}{c}{\textbf{Credit Spread}} \\ 
\hline
TVP-VAR-SV & $7.28^{\ast\ast\ast}$ & $0.28$ & $7.56^{\ast\ast}$ \\ 
GF-TVP-VAR-SV & $7.28^{\ast\ast\ast}$ & $0.45$ & $7.73^{\ast\ast}$ \\ 
CF-TVP-VAR-SV & $15.08^{\ast\ast\ast}$ & $0.26$ & $15.34^{\ast\ast\ast}$ \\ 
\hline\hline
\end{tabular}
\end{center}
\begin{flushleft}
{\raggedright Notes: The table reports Christoffersen’s likelihood-ratio (LR) test statistics for 68\% interval forecasts at horizon of 1 quarter ahead, obtained from VARs with five lags. The tests assess unconditional coverage, independence of violations, and their joint hypothesis. Reported statistics are compared to the $\chi^2$ distribution with the appropriate degrees of freedom. Asterisks denote significance levels ($p<0.10; , p<0.05; , p<0.01$).}
\end{flushleft}
\end{table}

% Horizon 2
%\input{\plotPathTwo/Table2InFevalCTestHorizon_1_Model_2_Prior_3_Lag_1_FcstNeval_98_SamplingMode_1_DifMode_1_FacMode_1}
%\input{\plotPathTwo/Table2InFevalCTestHorizon_1_Model_2_Prior_3_Lag_2_FcstNeval_98_SamplingMode_1_DifMode_1_FacMode_1}
%\input{\plotPathTwo/Table2InFevalCTestHorizon_1_Model_2_Prior_3_Lag_3_FcstNeval_98_SamplingMode_1_DifMode_1_FacMode_1}
%\input{\plotPathTwo/Table2InFevalCTestHorizon_1_Model_2_Prior_3_Lag_4_FcstNeval_98_SamplingMode_1_DifMode_1_FacMode_1}
%\input{\plotPathTwo/Table2InFevalCTestHorizon_1_Model_2_Prior_3_Lag_5_FcstNeval_98_SamplingMode_1_DifMode_1_FacMode_1}

% Horizon 4
%\input{\plotPathTwo/Table2InFevalCTestHorizon_1_Model_4_Prior_3_Lag_1_FcstNeval_98_SamplingMode_1_DifMode_1_FacMode_1}
%\input{\plotPathTwo/Table2InFevalCTestHorizon_1_Model_4_Prior_3_Lag_2_FcstNeval_98_SamplingMode_1_DifMode_1_FacMode_1}
%\input{\plotPathTwo/Table2InFevalCTestHorizon_1_Model_4_Prior_3_Lag_3_FcstNeval_98_SamplingMode_1_DifMode_1_FacMode_1}
%\input{\plotPathTwo/Table2InFevalCTestHorizon_1_Model_4_Prior_3_Lag_4_FcstNeval_98_SamplingMode_1_DifMode_1_FacMode_1}
%\input{\plotPathTwo/Table2InFevalCTestHorizon_1_Model_4_Prior_3_Lag_5_FcstNeval_98_SamplingMode_1_DifMode_1_FacMode_1}

% Horizon 8
%\input{\plotPathTwo/Table2InFevalCTestHorizon_1_Model_8_Prior_3_Lag_1_FcstNeval_98_SamplingMode_1_DifMode_1_FacMode_1}

% %%% PROBLEM HERE; MM, 2512
\clearpage
 \begin{table}[!t]\centering
\caption{$68\%$ Christoffersen Likelihood-Ratio Tests ($2$ Lags, horizon $8$)}
\label{tab:InFevalCTestHorizon8Model1Prior3Lag2FcstNeval98SamplingMode1DifMode1FacMode1}
\begin{center}
%\begin{tabular}{l *{3}{d{6.6}} }
\begin{tabular}{lccc}
\hline\hline
\multicolumn{1}{l}{\textbf{Model}} & \multicolumn{1}{c}{\textbf{Coverage}}  & \multicolumn{1}{c}{\textbf{Independence}}  & \multicolumn{1}{c}{\textbf{Joint}} \\ 
\hline
\multicolumn{4}{c}{\textbf{Real GDP}} \\ 
\hline
TVP-VAR-SV & $0.45$ & $5.24^{\ast\ast}$ & $5.70^{\ast}$ \\ 
GF-TVP-VAR-SV & $0.88$ & $6.63^{\ast\ast}$ & $7.51^{\ast\ast}$ \\ 
CF-TVP-VAR-SV & $0.16$ & $6.72^{\ast\ast\ast}$ & $6.89^{\ast\ast}$ \\ 
\hline
\multicolumn{4}{c}{\textbf{Inflation}} \\ 
\hline
TVP-VAR-SV & $0.16$ & $0.48$ & $0.65$ \\ 
GF-TVP-VAR-SV & $0.88$ & $0.03$ & $0.91$ \\ 
CF-TVP-VAR-SV & $0.16$ & $0.48$ & $0.65$ \\ 
\hline
\multicolumn{4}{c}{\textbf{FFR}} \\ 
\hline
TVP-VAR-SV & $11.66^{\ast\ast\ast}$ & $41.20^{\ast\ast\ast}$ & $52.86^{\ast\ast\ast}$ \\ 
GF-TVP-VAR-SV & $16.93^{\ast\ast\ast}$ & $44.98^{\ast\ast\ast}$ & $61.91^{\ast\ast\ast}$ \\ 
CF-TVP-VAR-SV & $20.92^{\ast\ast\ast}$ & $54.35^{\ast\ast\ast}$ & $75.27^{\ast\ast\ast}$ \\ 
\hline
\multicolumn{4}{c}{\textbf{Credit Spread}} \\ 
\hline
TVP-VAR-SV & $13.32^{\ast\ast\ast}$ & $42.53^{\ast\ast\ast}$ & $55.86^{\ast\ast\ast}$ \\ 
GF-TVP-VAR-SV & $13.32^{\ast\ast\ast}$ & $42.53^{\ast\ast\ast}$ & $55.86^{\ast\ast\ast}$ \\ 
CF-TVP-VAR-SV & $18.88^{\ast\ast\ast}$ & $53.27^{\ast\ast\ast}$ & $72.15^{\ast\ast\ast}$ \\ 
\hline\hline
\end{tabular}
\end{center}
\begin{flushleft}
{\raggedright Notes: The table reports Christoffersen’s likelihood-ratio (LR) test statistics for 68\% interval forecasts at horizon of 8 quarters ahead, obtained from VARs with two lags. The tests assess unconditional coverage, independence of violations, and their joint hypothesis. Reported statistics are compared to the $\chi^2$ distribution with the appropriate degrees of freedom. Asterisks denote significance levels ($p<0.10; , p<0.05; , p<0.01$).}
\end{flushleft}
\end{table}

%\input{\plotPathTwo/Table2InFevalCTestHorizon_1_Model_8_Prior_3_Lag_3_FcstNeval_98_SamplingMode_1_DifMode_1_FacMode_1}
%\input{\plotPathTwo/Table2InFevalCTestHorizon_1_Model_8_Prior_3_Lag_4_FcstNeval_98_SamplingMode_1_DifMode_1_FacMode_1}

% %%% PROBLEM HERE; MM, 2512
 \clearpage
 \begin{table}[!t]\centering
\caption{$68\%$ Christoffersen Likelihood-Ratio Tests ($5$ Lags, horizon $8$)}
\label{tab:InFevalCTestHorizon8Model1Prior3Lag5FcstNeval98SamplingMode1DifMode1FacMode1}
\begin{center}
%\begin{tabular}{l *{3}{d{6.6}} }
\begin{tabular}{lccc}
\hline\hline
\multicolumn{1}{l}{\textbf{Model}} & \multicolumn{1}{c}{\textbf{Coverage}}  & \multicolumn{1}{c}{\textbf{Independence}}  & \multicolumn{1}{c}{\textbf{Joint}} \\ 
\hline
\multicolumn{4}{c}{\textbf{Real GDP}} \\ 
\hline
TVP-VAR-SV & $26.00^{\ast\ast\ast}$ & $0.02$ & $26.02^{\ast\ast\ast}$ \\ 
GF-TVP-VAR-SV & $17.01^{\ast\ast\ast}$ & $11.87^{\ast\ast\ast}$ & $28.88^{\ast\ast\ast}$ \\ 
CF-TVP-VAR-SV & $13.75^{\ast\ast\ast}$ & $8.48^{\ast\ast\ast}$ & $22.23^{\ast\ast\ast}$ \\ 
\hline
\multicolumn{4}{c}{\textbf{Inflation}} \\ 
\hline
TVP-VAR-SV & $20.97^{\ast\ast\ast}$ & $0.08$ & $21.06^{\ast\ast\ast}$ \\ 
GF-TVP-VAR-SV & $20.97^{\ast\ast\ast}$ & $0.08$ & $21.06^{\ast\ast\ast}$ \\ 
CF-TVP-VAR-SV & $20.97^{\ast\ast\ast}$ & $0.08$ & $21.06^{\ast\ast\ast}$ \\ 
\hline
\multicolumn{4}{c}{\textbf{FFR}} \\ 
\hline
TVP-VAR-SV & $0.88$ & $29.95^{\ast\ast\ast}$ & $30.83^{\ast\ast\ast}$ \\ 
GF-TVP-VAR-SV & $8.64^{\ast\ast\ast}$ & $56.23^{\ast\ast\ast}$ & $64.87^{\ast\ast\ast}$ \\ 
CF-TVP-VAR-SV & $10.10^{\ast\ast\ast}$ & $57.91^{\ast\ast\ast}$ & $68.01^{\ast\ast\ast}$ \\ 
\hline
\multicolumn{4}{c}{\textbf{Credit Spread}} \\ 
\hline
TVP-VAR-SV & $5.11^{\ast\ast}$ & $11.52^{\ast\ast\ast}$ & $16.63^{\ast\ast\ast}$ \\ 
GF-TVP-VAR-SV & $1.71$ & $23.93^{\ast\ast\ast}$ & $25.64^{\ast\ast\ast}$ \\ 
CF-TVP-VAR-SV & $1.44$ & $27.92^{\ast\ast\ast}$ & $29.35^{\ast\ast\ast}$ \\ 
\hline\hline
\end{tabular}
\end{center}
\begin{flushleft}
{\raggedright Notes: The table reports Christoffersen’s likelihood-ratio (LR) test statistics for 68\% interval forecasts at horizon of 8 quarters ahead, obtained from VARs with five lags. The tests assess unconditional coverage, independence of violations, and their joint hypothesis. Reported statistics are compared to the $\chi^2$ distribution with the appropriate degrees of freedom. Asterisks denote significance levels ($p<0.10; , p<0.05; , p<0.01$).}
\end{flushleft}
\end{table}

%=======================================================
% CRPS
%\input{\plotPathTwo/Table2CRPS_Model_1_Prior_3_Lag_1_FcstNeval_98_SamplingMode_1_DifMode_1_FacMode_1}

% %%% PROBLEM HERE; MM, 2512
% \clearpage
% \input{\plotPathTwo/Table2CRPS_Model_1_Prior_3_Lag_2_FcstNeval_98_SamplingMode_1_DifMode_1_FacMode_1}

% %\input{\plotPathTwo/Table2CRPS_Model_1_Prior_3_Lag_3_FcstNeval_98_SamplingMode_1_DifMode_1_FacMode_1}
% %\input{\plotPathTwo/Table2CRPS_Model_1_Prior_3_Lag_4_FcstNeval_98_SamplingMode_1_DifMode_1_FacMode_1}
% \clearpage
% \input{\plotPathTwo/Table2CRPS_Model_1_Prior_3_Lag_5_FcstNeval_98_SamplingMode_1_DifMode_1_FacMode_1}

% CRPS
%\input{\plotPathTwo/Table2CRPS_Model_1_Prior_3_Lag_1_FcstNeval_98_SamplingMode_1_DifMode_1_FacMode_1}
\clearpage
\begin{table}[!t]\centering
\caption{Density Forecast Accuracy - CRPS ($2$ Lag models)}
\label{fig:CRPSModel1Prior3Lag2FcstNeval98SamplingMode1DifMode1FacMode1}
\begin{center}
%\begin{tabular}{l *{4}{d{3.3}} }
\begin{tabular}{lcccc}
\hline\hline
\textbf{Model} & \textbf{1Q}  & \textbf{2Q}  & \textbf{4Q}  & \textbf{8Q} \\ 
\hline
\multicolumn{5}{c}{Real GDP} \\ 
\hline
TVP-VAR-SV & $0.34$ & $0.36$ & $0.38$ & $0.38$ \\ 
GF-TVP-VAR-SV & $-3.57^{\ast\ast\ast}$ & $-3.60^{\ast\ast}$ & $-3.59^{\ast}$ & $-0.88$ \\ 
CF-TVP-VAR-SV & $-5.33^{\ast}$ & $-5.76^{\ast}$ & $-3.92$ & $0.38$ \\ 
\hline
\multicolumn{5}{c}{Inflation} \\ 
\hline
TVP-VAR-SV & $0.12$ & $0.13$ & $0.14$ & $0.18$ \\ 
GF-TVP-VAR-SV & $0.78$ & $-0.16$ & $0.70$ & $0.48$ \\ 
CF-TVP-VAR-SV & $2.25$ & $1.90$ & $2.02$ & $2.12$ \\ 
\hline
\multicolumn{5}{c}{FFR} \\ 
\hline
TVP-VAR-SV & $0.20$ & $0.43$ & $0.84$ & $1.42$ \\ 
GF-TVP-VAR-SV & $-0.23$ & $-0.55$ & $-1.72$ & $-3.06$ \\ 
CF-TVP-VAR-SV & $-4.78^{\ast\ast}$ & $-3.74^{\ast}$ & $-3.10$ & $-3.81^{\ast}$ \\ 
\hline
\multicolumn{5}{c}{Credit Spread} \\ 
\hline
TVP-VAR-SV & $0.16$ & $0.28$ & $0.38$ & $0.47$ \\ 
GF-TVP-VAR-SV & $0.07$ & $-0.04$ & $-2.05^{\ast}$ & $-1.91$ \\ 
CF-TVP-VAR-SV & $-1.88$ & $-2.56$ & $-4.10^{\ast\ast}$ & $-1.60$ \\ 
\hline\hline
\end{tabular}
\end{center}
\begin{flushleft}
{\raggedright Notes: The table reports CRPS results for out-of-sample density forecasts. For each variable, the rows reports the relative CRPS calculated as the percentage decrease of the CRPS when using the respective model rather than the benchmark TVP-VAR; positive numbers indicate improvement over the benchmark TVP-VAR case. The respective first rows reports the CRPS for the benchmark TVP-VAR case. Statistical significance of the differences in average CRPS assessed with a \cite{DieboldMariano1995} test is indicated by the corresponding numbers in brackets.}
\end{flushleft}
\end{table}
\clearpage
\begin{table}[!t]\centering
\caption{Density Forecast Accuracy - CRPS ($5$ Lag models)}
\label{fig:CRPSModel1Prior3Lag5FcstNeval98SamplingMode1DifMode1FacMode1}
\begin{center}
%\begin{tabular}{l *{4}{d{3.3}} }
\begin{tabular}{lcccc}
\hline\hline
\textbf{Model} & \textbf{1Q}  & \textbf{2Q}  & \textbf{4Q}  & \textbf{8Q} \\ 
\hline
\multicolumn{5}{c}{Real GDP} \\ 
\hline
TVP-VAR-SV & $0.42$ & $0.45$ & $0.53$ & $0.74$ \\ 
GF-TVP-VAR-SV & $-8.04$ & $6.84$ & $11.29^{\ast\ast}$ & $19.56^{\ast\ast\ast}$ \\ 
CF-TVP-VAR-SV & $-10.74$ & $7.59$ & $13.02^{\ast\ast}$ & $21.97^{\ast\ast\ast}$ \\ 
\hline
\multicolumn{5}{c}{Inflation} \\ 
\hline
TVP-VAR-SV & $0.14$ & $0.15$ & $0.17$ & $0.27$ \\ 
GF-TVP-VAR-SV & $6.89^{\ast\ast}$ & $13.27^{\ast\ast\ast}$ & $12.95^{\ast\ast\ast}$ & $18.07^{\ast\ast\ast}$ \\ 
CF-TVP-VAR-SV & $6.87^{\ast}$ & $13.10^{\ast\ast\ast}$ & $13.60^{\ast\ast\ast}$ & $18.67^{\ast\ast\ast}$ \\ 
\hline
\multicolumn{5}{c}{FFR} \\ 
\hline
TVP-VAR-SV & $0.21$ & $0.45$ & $0.87$ & $1.56$ \\ 
GF-TVP-VAR-SV & $-17.27^{\ast\ast}$ & $-20.69^{\ast}$ & $-13.33$ & $0.69$ \\ 
CF-TVP-VAR-SV & $-22.30^{\ast\ast}$ & $-28.38^{\ast\ast}$ & $-21.57$ & $-4.21$ \\ 
\hline
\multicolumn{5}{c}{Credit Spread} \\ 
\hline
TVP-VAR-SV & $0.17$ & $0.30$ & $0.42$ & $0.53$ \\ 
GF-TVP-VAR-SV & $-3.04$ & $0.50$ & $5.10$ & $11.69$ \\ 
CF-TVP-VAR-SV & $-8.17$ & $-4.32$ & $1.65$ & $8.51$ \\ 
\hline\hline
\end{tabular}
\end{center}
\begin{flushleft}
{\raggedright Notes: The table reports CRPS results for out-of-sample density forecasts. For each variable, the rows reports the relative CRPS calculated as the percentage decrease of the CRPS when using the respective model rather than the benchmark TVP-VAR; positive numbers indicate improvement over the benchmark TVP-VAR case. The respective first rows reports the CRPS for the benchmark TVP-VAR case. Statistical significance of the differences in average CRPS assessed with a \cite{DieboldMariano1995} test is indicated by the corresponding numbers in brackets.}
\end{flushleft}
\end{table}

%\input{\plotPathTwo/2FI_Horizon_8_Model_1_Prior_3_Lag_2_FcstNeval_98_SamplingMode_1_DifMode_1_FacMode_1}
%\input{\plotPathTwo/2FI_Horizon_8_Model_1_Prior_3_Lag_5_FcstNeval_98_SamplingMode_1_DifMode_1_FacMode_1}

%\input{\plotPathTwo/2CRPS_Model_1_Prior_3_Lag_2_FcstNeval_98_SamplingMode_1_DifMode_1_FacMode_1}
%\input{\plotPathTwo/2CRPS_Model_1_Prior_3_Lag_5_FcstNeval_98_SamplingMode_1_DifMode_1_FacMode_1}

%\input{\plotPathTwo/2PIT_Horizon_1_Model_1_Prior_3_Lag_2_FcstNeval_98_SamplingMode_1_DifMode_1_FacMode_1}
%\input{\plotPathTwo/2PIT_Horizon_1_Model_1_Prior_3_Lag_5_FcstNeval_98_SamplingMode_1_DifMode_1_FacMode_1}
%\input{\plotPathTwo/2PIT_Horizon_8_Model_1_Prior_3_Lag_2_FcstNeval_98_SamplingMode_1_DifMode_1_FacMode_1}
%\input{\plotPathTwo/2PIT_Horizon_8_Model_1_Prior_3_Lag_5_FcstNeval_98_SamplingMode_1_DifMode_1_FacMode_1}

%% -------------------------------------------------------------------------- %%
%% -------------------------------------------------------------------------- %%

\end{appendix}

\end{document}